\documentclass[fleqn,usenatbib]{mnras}
\newcommand{\persqdegrees}{deg$^{-2}$~}
\usepackage{tabularray}
\usepackage{ulem}
\newcommand{\sqdegrees}{deg$^{2}$~}

\usepackage{lineno}
\usepackage{newtxtext,newtxmath}

\newcommand{\eqreff}{Eq.~\eqref}
\usepackage{arydshln}
\newcommand{\secreff}{Section~\ref}
\newcommand{\appreff}{Appendix~\ref}
\newcommand{\figreff}{Figure~\ref}
\newcommand{\tabreff}{Table~\ref}

\usepackage{soul}
\usepackage[T1]{fontenc}
\DeclareRobustCommand{\VAN}[3]{#2}
\let\VANthebibliography\thebibliography
\def\thebibliography{\DeclareRobustCommand{\VAN}[3]{##3}\VANthebibliography}
\def\fnl{f_\text{\rm NL}^{\rm loc}}

\usepackage{graphicx}	
\usepackage{amsmath}	

\title[Targeting the future of spectroscopic surveys with WST]{Targeting the future of spectroscopic galaxy surveys: A cosmological study case with the Wide-field Spectroscopic Telescope}

\author[WST Cosmology WG -- Target Selection Team]{
\parbox{\textwidth}{
Constantin~Payerne,$^{1}$\thanks{E-mail: constantin.payerne@gmail.com}
Antoine~Rocher,$^{2}$
Aurélien~Verdier,$^{2}$
William~d'Assignies,$^{3, 4}$
Behnood~Bandi,$^{5}$\thanks{Corresponding author. E-mail: b.bandi@sussex.ac.uk}
Sofia~Contarini,$^{6}$
Matteo~Ferro,$^{7}$
Jean-Paul~Kneib,$^{2}$
Vincenzo~Mainieri,$^{8}$
Michele~Moresco,$^{9, 10}$
Marco~Palla,$^{9, 10, 11}$
Maxime~Rombach,$^{2}$
Christophe~Yèche,$^{1}$
Furkan~Akbaba,$^{12}$
Hélène~Courtois,$^{13}$
Darko~Donevski,$^{14, 15}$
Massimo~Guidi,$^{9, 10}$
Valentin~Ivanov,$^{8}$
Roelof~S.~de~Jong,$^{16}$
Konstantinos~Migkas,$^{17}$
Ivan~Minchev,$^{16}$
Ignacio~del~Moral-Castro,$^{18}$
Olcay~Plevne,$^{19}$
Mojtaba~Raouf,$^{17, 20, 21}$
Laurence~Tresse,$^{22}$
and the WST Collaboration  (affiliations are listed at the end of the paper)}
}
\date{Accepted --. Received --; in original form --}

\pubyear{\the\year{}}

\begin{document}
\label{firstpage}
\pagerange{\pageref{firstpage}--\pageref{lastpage}}
\maketitle

\begin{abstract}
The next generation of Stage-V spectroscopic surveys are designed to measure hundreds of millions of galaxy and quasar spectra, extending precision large-scale-structure cosmology to redshifts and volumes well beyond the reach of current Stage-IV facilities. Realising this potential requires revisiting the target selection strategies and designing them explicitly around the depth, multiplexing, and survey time of the new instruments, rather than a direct extrapolation of existing recipes. This work develops and characterises new strategies for the Wide-field Spectroscopic Telescope (WST), a proposed 12-metre, highly multiplexed facility expected to operate in the 2040s. Using deep multi-band photometry, including COSMOS-field data from CLAUDS and HSC, we revisit low-redshift $z\lesssim 1.6$ tracer selection of SDSS, BOSS/eBOSS, DESI and 4MOST-CRS -- Bright Galaxies, Luminous Red Galaxies, Emission Line Galaxies and quasars -- to photometric depth achievable in the WST era, and introduce a purely magnitude-limited sample (MagMax) exploiting three-dimensional colour-colour star-galaxy separation, and a set of Lyman-Break Galaxy dropout samples extending target selection to $z\gtrsim2$. Using Fisher-matrix forecasts, we showcase the constraining power of a WST-like cosmological survey on BAO distance parameters $\alpha_\parallel$ and $\alpha_\perp$, the growth rate $f$, the evolving dark energy equation $w_0w_a$, the local-type primordial non-Gaussianity $f_{\rm NL}^{\rm loc}$, and the sum of the neutrino mass $\Sigma m_\nu$. The main outcome of this work is an optimisation framework that directly links instrumental capabilities and observing time to cosmological constraining power, providing a science-driven template applicable to the design of future generations of Stage-V spectroscopic surveys.
\end{abstract}

\section{Introduction}

Optical spectroscopy has been a cornerstone of modern cosmology for the past three decades, with the advent of multi-fibre instruments. The Sloan Digital Sky Survey (SDSS;~\citealt{York2000sdsstechnicalreport}) and the 2 degree Field Galaxy Redshift Survey (2dFGRS;~\citealt{colless_2df_2001}) revolutionized wide-area 3D mapping of the galaxy distribution and measurements of galaxy clustering. In parallel, deep, high-density galaxy surveys but limited in sky area, like the VIMOS Public Extragalactic Redshift Survey (VIPERS;~\citealt{guzzo2014vimos}) extending the accessible redshift range, provided dedicated study of galaxy clustering, redshift-space distortions, growth of structure, and the cosmic web~\citep{bel_vimos_2014}. For two decades, these surveys and there extensions (e.g. 6dFGRS~\citealt{beutler_6df_2012}, eBOSS~\citealt{dawson_sdss-iv_2016}) map millions of extra-galactic objects including bright galaxies (BGs), luminous red galaxies (LRGs), emission-line galaxies (ELGs), and quasi-stellar objects (QSOs), to trace the expansion history of the Universe out to redshift $z=2.2$ and $z<3.5$ using the Lyman-alpha forest of QSOs, achieving percent-level precision on cosmological parameters~\citep{Tegmark2006sdsslrg, Alam2021sdsseboss, cole_2df_2005}.
Today, the Dark Energy Spectroscopic Instrument (DESI, \citealt{Levi2013desiwhitepaper}) observes a third of the sky with a target density at least 2.5 times higher than SDSS, allowing it to reach sub-per cent-precision cosmological measurements of the Baryon Acoustic Oscillation (BAO) distances~\citep{desi2025baodr2}. DESI has already completed its nominal five-year survey and has begun a three-year extension, increasing both target density and sky area (DESI Collaboration, in prep.). Other forthcoming Stage-IV redshift surveys —the 4-metre Multi-Object Spectroscopic Telescope (4MOST;~\citealt{deJong20124most} with the cosmology redshift survey (CRS;~\citealt{richard_4most_2019, Verdier_2025}), the Subaru Prime Focus Spectrograph (PFS; \citealt{Tamura2016PFS}), and the \textit{Euclid} space mission~\citep{Mellier2025euclid}—will complement DESI in redshift range and sky coverage. 

The recent results from DESI, combined with other cosmological probes, have revealed hints of an evolving dark energy model that challenge our current $\Lambda$CDM paradigm \citep{Adame2025desiY1BAO,desi2025baodr2}, well established since the 1980s~\citep{lahav_cdm_2026}.
Beyond differentiating between dark energy models, spectroscopic surveys also probe more recent questions in cosmology: is the assumption of a Gaussian primordial Universe valid? What is the neutrino mass hierarchy? To address these questions, the redshift-survey community is advocating for Stage-V spectroscopic surveys. Several projects have been proposed, including the MUltiplexed Survey Telescope (MUST;~\citealt{mZhao2024must}), Spec-S5~\citep{Besuner2025specs5}, and the Wide-field Spectroscopic Telescope\footnote{\url{https://wstelescope.eu}} (WST;~\citealt{Bacon2024whitepaper,Mainieri2024wst}). Using large-aperture telescopes with highly multiplexed spectroscopic instruments, these surveys plan to reinforce the density of low-redshift tracers at redshift $z<1.6$, along with targeting high-redshift galaxies such as Lyman-break galaxies (LBGs;~\citealt{Steidel1996LBG}) and Lyman-$\alpha$ emitters (LAEs;~\citealt{Ouchi2020lae}), extending precise mapping of the Universe out to redshift 5 over the coming decades, enabling observation of the linear modes in the primordial Universe. As a pilot for these future projects, the DESI collaboration is preparing DESI Run 2~\citep{Schlegel2022desi2}, an extension of DESI Run 1, aiming to start in 2029 and to significantly extend the science reach of its predecessor, collecting 40 million spectra at higher density and higher redshift than DESI Run 1, while serving as a pathfinder for a future Stage-V survey.
Stage-V surveys (Spec-S5, WST) expect to measure hundreds of millions of redshifts, opening a new and largely uncharted discovery space. Executing such large-scale spectroscopic campaigns depends critically on available imaging surveys with sufficient sky coverage, wavelength range, and depth to identify suitable targets. Over the coming decade, large photometric surveys such as the Rubin Observatory's Legacy Survey of Space and Time (LSST; \citealt{LSST2009whitepaper}), the \textit{Euclid} satellite \citep{Mellier2025euclid}, the Nancy Grace Roman Space Telescope~\citep{Akeson2019Romanwhitepaper}, and the China Space Station Telescope (CSST;~\citealt{Gong2026csst}) will cover a wide spectral range from the optical to the infrared, enabling hundreds of millions of tracers to be selected for Stage-V experiments. Together, these new datasets will allow us to probe the mechanisms driving the accelerated expansion of the Universe—both in its earliest moments (inflation) and at late times (dark energy)—while also addressing the neutrino mass problem and the internal tensions within the $\Lambda$CDM paradigm, which have been brought into sharper focus by recent DESI results.
A key challenge in cosmology with redshift surveys is to provide galaxy or quasar samples with both high density and high purity, in order to optimise the scientific return of these experiments. In this paper, we explore new target-selection strategies for future Stage-V spectroscopic surveys that extend current tracer-selection approaches. At redshift $z<1.6$, we propose an alternative to the traditional, DESI-like tracer-based strategy, selecting magnitude-limited samples instead. In a second part, we optimise these selections for the specific case of a WST-like instrument. We examine the survey design of WST and its implications for the selection of WST tracers, and we present cosmological forecasts to quantify what WST could deliver. For the high-redshift survey, given by LBGs primarily, we propose to define the target selection using a science-driven approach, in which the magnitude limits are directly constrained by the scientific return of the survey. This cosmology-oriented paper is part of the wider effort of the WST community to provide recommendations to maximize the scientific outcome of beyond Stage-IV spectroscopic surveys, in the particular case of the WST (see e.g. \citealt{Tsaprazi2025gravitational_redshift_clusters_WST_white_paper,Ivanov2025wstalert,Tomasetti2025cosmicchronometer_eso,Battaglia2025localgroup_eso,Guidi2025clustering_eso,Milakovic2026fundamentalconstant,PC_voids,Loveday2026mapping_dark_matter_eso,Montano2026GR_eso}).

The paper is organised as follows: In \secreff{sec:WST}, we present the main features of the WST. In \secreff{sec:methodology_paper}, we make a summary of the different steps we have to go through to be able to estimate the cosmological constraints for a WST-like spectroscopic survey properly. In \secreff{sec:target_selection}, we present the methods for selecting targets in the WST era, after enumerating the different photometric datasets to be available in the 2040s. In \secreff{sec:modelling_wst_cosmo_survey}, in light of the WST design, we explore the spectroscopic redshift success rate as well as the completeness and galaxy bias of the aforementioned target selection. In \secreff{sec:forecast}, we present possible survey designs of the WST, based on the target densities and their respective constraining power on different cosmological parameters and models. In \secreff{sec:discussion_limitation}, we discuss the limitations and further improvements of this work, and we conclude in \secreff{sec:conclusions}.

\section{Cosmology with Stage-V spectroscopic experiments: the WST case}

\label{sec:WST}

The WST is designed to deliver major advances across a broad range of astrophysical topics, from the formation of the first stars and galaxies to the mapping of dark and baryonic matter in the cosmic web, the reconstruction of the formation history of the Milky Way and nearby dwarf galaxies, and the characterisation of transient events. Substantial progress is expected in cosmology, in particular regarding the nature and expansion history of the dark Universe. A comprehensive description of the WST facility and its scientific objectives are presented in \citet{Bacon2024whitepaper} and in \citet{Mainieri2024wst}.

\subsection{Overview of the WST cosmology survey}
\label{sec:WST overview}

\begin{figure}
    \centering
    \includegraphics[width=\linewidth]{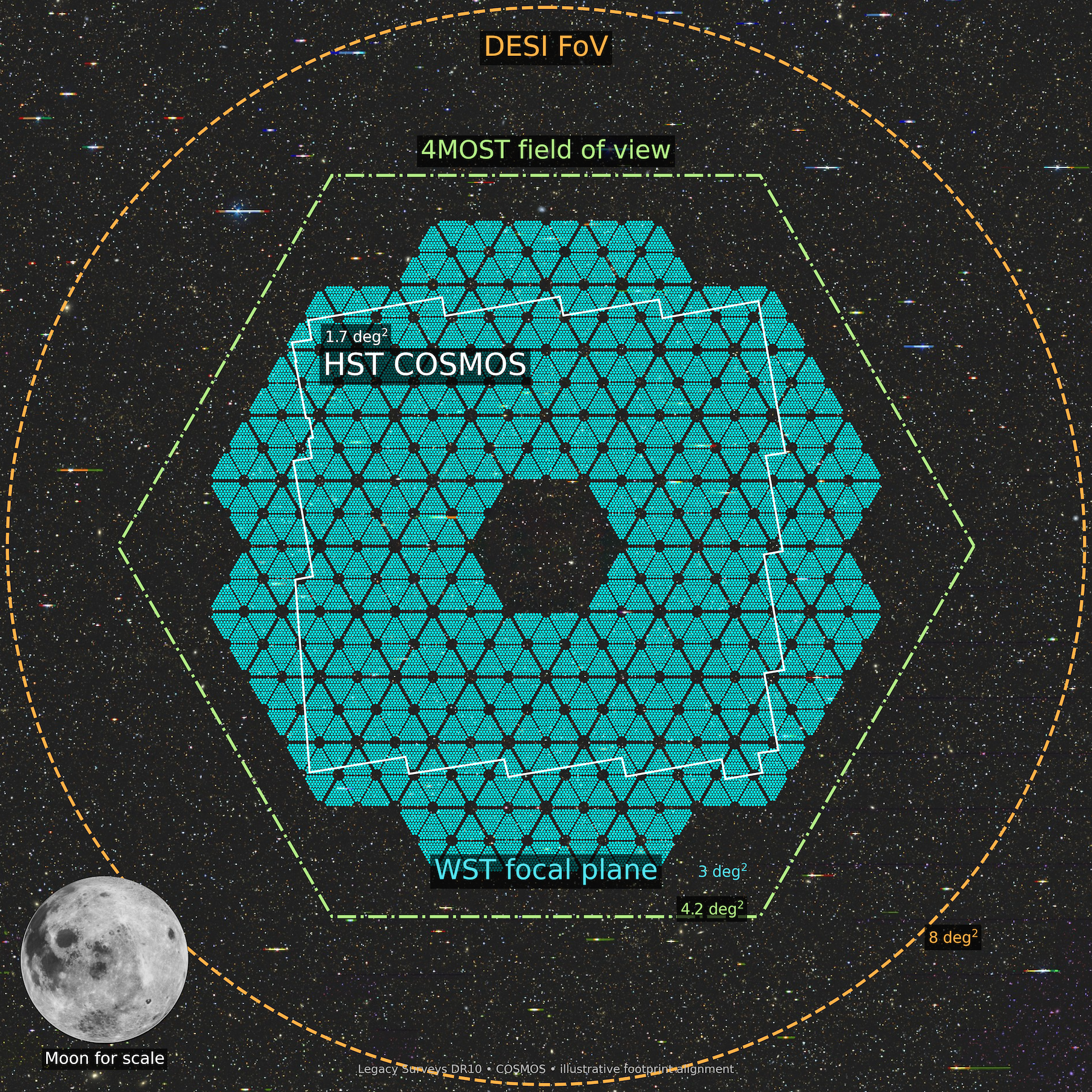} \\
    \includegraphics[width=\linewidth]{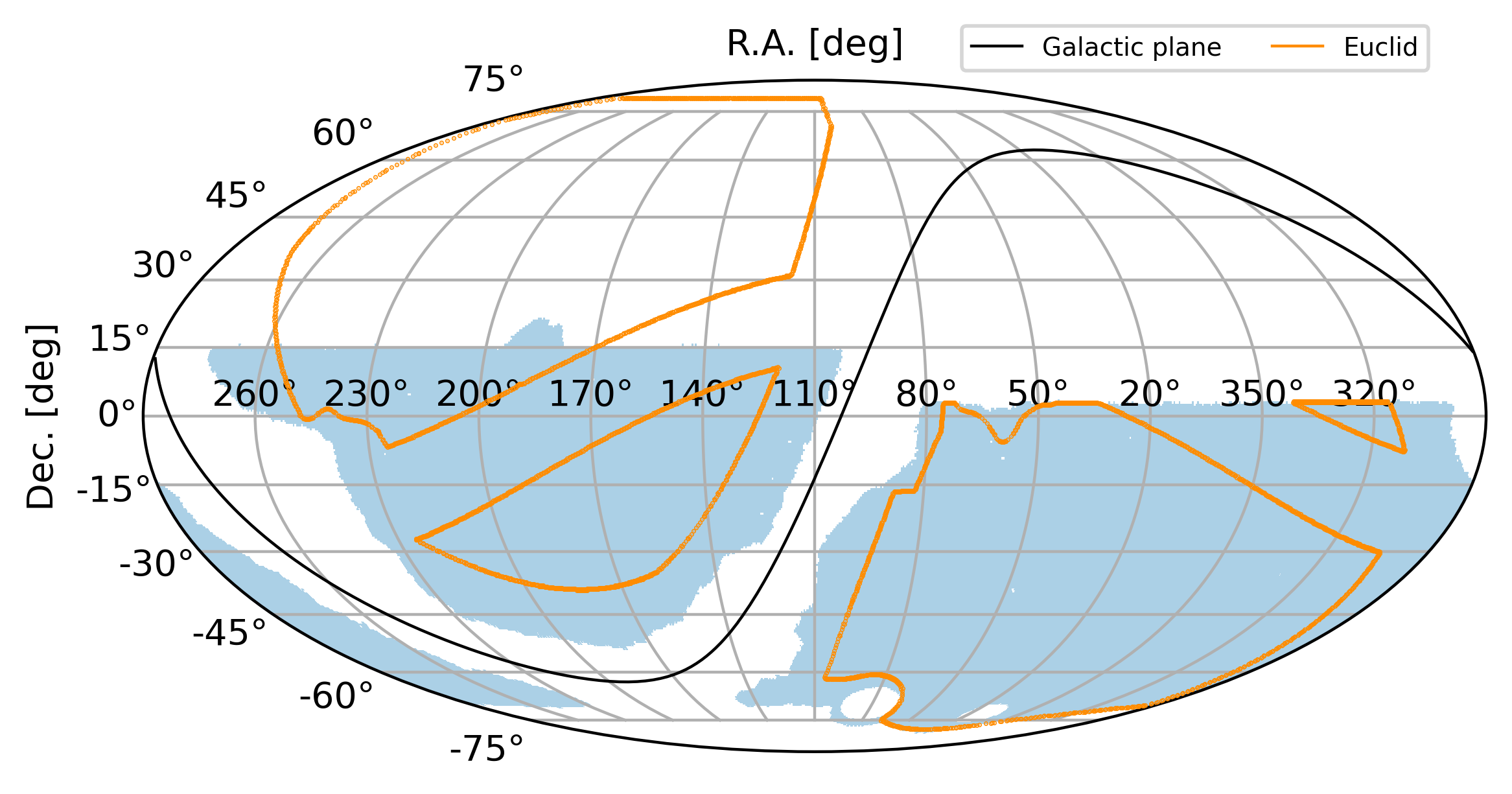} 
    \caption{Top panel: WST (lightblue fibers), 4MOST (green hexagon) and DESI (orange circle) fields of view (image from the Legacy Surveys viewer in the background). Bottom Panel: Baseline footprint of the WST cosmology survey, based on the Rubin LSST WFD after 10 years of operation, removing Magellanic clouds, along with the \textit{Euclid} survey footprint.}
    \label{fig:footprint}
\end{figure}

The WST is a 12-metre facility with a 3.1 deg$^2$ field of view (FoV) designed for wide-field spectroscopic surveys \citep{Bacon2024whitepaper}. This community-driven project will be proposed as the next ESO major facility in the context of the Expanding Horizons initiative (\url{https://next.eso.org/}) and, if selected, could enter operations in the 2040s from a site in Chile.
It is equipped with a highly multiplexed, fibre-fed, multi-object spectrograph (MOS), capable of simultaneously collecting spectra for up to 32,000 targets over 2.5 \sqdegrees (positioner FoV), operating in both low-resolution (MOS-LR, 30,000 fibres) and high-resolution (MOS-HR, 2,000 fibres) modes, together with a large 3$\times$3~arcminute integral field spectrograph (IFS)~\citep{Lee2026wsttechnical,Pernecker2026wsttechnical}. The top panel of \figreff{fig:footprint} shows the WST FoV projected on the sky over the COSMOS field\footnote{\url{https://science.nasa.gov/asset/webb/the-cosmos-webb-field-compared-with-the-hubble-acs-field/}}, where the unoccupied region at the centre of the focal plane corresponds to the location of the IFS.
WST is designed to encompass a very broad range of scientific areas from planetary and stellar formation to cosmology, where all the observations are shared over the focal plane similarly to the 4MOST design.
\\ 

In this work, we investigate WST's cosmological potential through the spatial clustering of several cosmological tracers—galaxies and quasars—as detailed in the following sections. For this purpose, the MOS-LR channel will be used extensively to obtain galaxy and quasar spectra, from which spectroscopic redshifts will be derived using continuum features together with emission and absorption lines. Owing to its larger collecting area relative to current Stage-IV spectroscopic facilities such as DESI, WST will reach fainter targets at fixed spectral signal-to-noise ratio (SNR). Assuming a reference spectroscopic survey with mirror diameter $D_{\textrm{base}}$ and exposure time $t_{\textrm{base}}$, the corresponding gain in limiting magnitude can be expressed as\footnote{This scaling assumes background-limited point-source detection at fixed exposure time. It neglects the effects of site quality, telescope image quality, fibre diameter, target size and detector noise, which most likely favour WST over DESI and 4MOST.}
\begin{align}
(\Delta m)_{\rm base\rightarrow WST} &= \Delta m_{\rm mirror} + \Delta m_{\rm exposure} 
\\ &=  2.5 \log_{10}\left(\frac{D_{\textrm{WST}}}{D_{\textrm{base}}}\right) + 2.5 \log_{10}\left(\sqrt \frac{t_{\textrm{WST}}}{t_{\textrm{base}}}\right),
\label{eq:scaling_mirror}
\end{align}
where $D_{\textrm{WST}}$ and $t_{\textrm{WST}}$ denote, respectively, the WST mirror diameter and the exposure time of the WST fibres. Comparing DESI ($D_{\textrm{DESI}} = 4$ and WST ($D_{\textrm{WST}}=12.5$) mirrors at fixed exposure time ($t_{\rm WST}=t_{\rm base}$), this scaling implies that WST can probe objects approximately $\Delta m_{\rm mirror} \sim 1.24$ magnitudes fainter than DESI for the same SNR. This simple estimate illustrates the potential of WST as a highly efficient facility for building extremely large spectroscopic redshift catalogues reaching very faint objects on comparatively short timescales.

\begin{figure}
    \centering
    \includegraphics[width=1\linewidth]{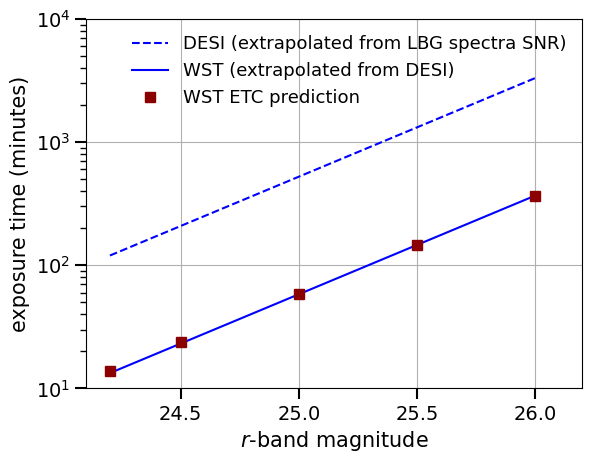}
    \caption{DESI and WST fibre required exposure time to reach a ${\rm SNR}=20$ in the $r$-band for an LBG (considered to have a flat spectrum) with magnitude $m$ in the $r$-band.}
    \label{fig:time_exp}
\end{figure}

We illustrate the gain from DESI to WST in terms of exposure time in \figreff{fig:time_exp} taking the example of the observation of $u$-dropout Lyman-Break Galaxies, high-redshift tracers at $2.5 < z < 3.5$ (see further details in \secreff{sec:target_selection}) considered to have a relatively flat continuum in the $r$-band at these redshifts. For a given LBG target with magnitude $m$ in the $r$-band, we represent the expected DESI exposure time that is required to achieve SNR=20 for the averaged flux in the $r$-band filter\footnote{This result is extrapolated from the $r=24.2$ LBG sample observed during 2 hours by DESI  \citep{RuhlmannKleider2024LBGCLAUDS} and using $t(m) = t_0 \times 10^{0.8m}$, see \appreff{app:exposuretime}. The corresponding DESI SNR/per-pixel is $\sim 0.5$ and WST SNR/per-pixel is $\sim 0.3$.}. For a $r=24.2$ target, 2 hours of DESI exposure are required, whereas only 15 minutes are required for WST when measuring it with the WST Exposure Time Calculator\footnote{\url{https://github.com/ferromatteo/pyetc_wst}}, also matching the DESI-to-WST rescaled exposure time-magnitude relation. For two hours of WST exposure, the same SNR can be reached for a $r\sim 25.4$ target.\\

The cosmological survey of WST is expected to map the three-dimensional distribution of galaxies and quasars across the observable Universe, enabling stringent tests of the laws of physics on cosmological scales. The priority is to have access to a large, deep, and homogeneous sample of galaxies and quasars over the sky available in Chile.  
As a baseline, we adopt the Rubin LSST Wide-Fast-Deep footprint (WFD; \citealt{LSST2009whitepaper}), shown in the bottom panel of \figreff{fig:footprint}, as representative of the area (18,000 \persqdegrees) accessible to most WST cosmology observations. Covering this footprint requires approximately 6,000 non-overlapping WST pointings. The exposure time of each target is based on the DESI strategy and performance. The current survey design considers the Bright, Grey, and Dark time programs (depending on the FLI: Fractional Lunar Illumination) with exposure times of 100, 600, and 1000~seconds ($\sim$15~minutes), respectively, considering nominal sky conditions~\citep{Mainieri2024wst}. ESO definition assumes FLI $> 0.7$ for bright time, $0.4<$ FLI $<0.7$ for grey time, and FLI $< 0.4$ for Dark time. These definitions differ slightly from those used by DESI, which defined Dark and Bright time based on survey speed (see Section 5.3 of~\citealt{schlafly2024}).  Default values we use throughout the paper are listed in \tabreff{tab:wst_observing_baseline}.
These exposure times will be revisited using WST ETC; however, from the first comparison with ETC (see \figreff{fig:time_exp}), the exposure time considered seems reasonable. In the following, we assumed these values when computing the survey time.  
\begin{table}
    \centering
    \begin{tabular}{c|c}
    WST observation setup & Values \\
    \hline
        Good observing conditions  & $80\%$ \\
        Observing hours/night & 9.5 hours \\
        Bright time/night & $33\%$\\
        Grey time/night & $15\%$\\
        Dark time/night & $52\%$\\
        Bright time exposure & 100 seconds\\
        Grey time exposure & 600 seconds\\
        Dark time exposure & 1000 seconds\\
    \end{tabular}
    \caption{Different default values for the WST observations we use throughout the paper.}
    \label{tab:wst_observing_baseline}
\end{table}

\subsection{Cosmological science with WST: Overview}

Stage-V cosmological surveys aim to provide new insights into the expansion history of the Universe, the formation of large-scale structure, and the statistical properties of the primordial density field. The cosmological science drivers of WST can be organised into several broad themes, each of which has already been extensively explored by previous and ongoing galaxy surveys.

\subsubsection{Fiducial science cases}

With the higher multiplexing capability of WST, we will obtain orders of magnitude more spectroscopic redshifts in the low-$z$ Universe ($z<2$). The resulting tracer densities will also be higher by an order of magnitude, reducing statistical noise. The higher density will further allow the small-scale clustering statistics to be measured more accurately. This, however, represents a challenge for the modelling, as fully exploiting the available information will require a more accurate treatment of non-linear physics. For the low-$z$ Universe, we limit ourselves to the standard BAO and redshift space distortion (RSD;~\citealt{1987kaiser}) science cases. A potential improvement would be to perform full-shape forecasts, which are becoming the new fiducial framework for extracting cosmological information from galaxy clustering. With the increased light-collecting capability of WST, cosmological surveys can be conducted at high redshift ($z>2$) by targeting fainter galaxies. Extending spectroscopic observations to higher redshifts offers new opportunities. First, it opens a new volume of the Universe containing additional cosmological information. The available cosmic volume at high redshift is also larger, which is particularly important for large-scale structure studies aiming to probe inflationary models. Moreover, the high-redshift Universe is more linear, allowing standard analyses to be extended to smaller scales. This is particularly valuable for constraining neutrino properties. In this subsection, we detail the different science cases considered in this work.

\subsubsection*{Baryon Acoustic Oscillations (BAO)} are imprinted in the large-scale distribution of galaxies, as relics of sound waves that propagated through the hot, ionised plasma of the early Universe. These features provide a standard ruler for measuring cosmic distances: extracting the characteristic BAO scale from the three-dimensional clustering of galaxies and quasars constrains the geometry and expansion history of the Universe, and in turn the properties of dark energy~\citep{Eisenstein2005, Levi2013desiwhitepaper}. Measurements of the BAO signal along and perpendicular to the line-of-sight yield two key geometric quantities—the distance-redshift Hubble $H(z_{\rm eff})r_d$ and the angular diameter distance $D_A(z_{\rm eff})/r_d$—evaluated at the effective redshift $z_{\rm eff}$ of the tracer sample, given the sound horizon $r_d$. We use the two Alcock-Paczynski (A-P) parameters,
\begin{equation}
    \alpha_\parallel=\frac{(H(z_{\rm eff})r_{\rm d})_{\rm fid}}{H(z_{\rm eff})r_d}\,\hspace{1cm} \alpha_\perp=\frac{D_{\rm A}(z_{\rm eff})/r_{\rm d}}{(D_{\rm A}(z_{\rm eff})/r_{\rm d})_{\rm fid}}.\label{eq:AP}
\end{equation}
Combined across multiple redshift bins and with suitable priors on other cosmological parameters, these measurements can be used to constrain the dark energy equation-of-state and probe, e.g., redshift evolution through the $w_0-w_a$ CPL parametrization~\citep{chevallier_accelerating_2001, Linder2003cpl} (see \appreff{app:forecast_details}).\\

\subsubsection*{Growth of structure and gravity tests} 
    
The growth of cosmic structure shapes both the spatial clustering of galaxies and their peculiar velocities. Redshift-space distortions (RSD;~\citealt{1987kaiser}) arise from peculiar motions of galaxies that shift their observed positions along the line of sight (los), creating apparent anisotropy in the clustering signal. RSD are commonly used to probe the growth rate of structure and the nature of gravity at cosmological scales by constraining the combination $f(z)\sigma_8(z)$, where $f(z)$ is the logarithmic growth rate, and $\sigma_8$ is the variance of the density field smoothed by a top-hat filter of 8 Mpc/$h$. Measuring $f$ directly tests general relativity on cosmological scales, as the growth-index parametrisation $f(z)\simeq \Omega_m(z)^\gamma$, with $\gamma = 0.55$ in the GR-$\Lambda$CDM model, can vary in alternative gravity models. In this work, we first consider the linear RSD (Kaiser) effect.
The Kaiser effect link the galaxy power spectrum $P_{\rm gg}$ to the matter power spectrum $P_{\rm m}$ by:
\begin{equation}
    P_{\rm gg}(k,\mu) = b_{\rm g}^2(z) (1+\beta\mu^2)^2 P_{\rm m}(k,z),\label{eq:linearrsd}
\end{equation}
where $b$ is the linear bias of the galaxy sample, $\beta =f/b_{\rm g}$, and $\mu=\boldsymbol{k}\cdot\boldsymbol{l}$ is the cosine of the angle between the wave-vector $\boldsymbol{k}$ and the line of sight $\boldsymbol{l}$.    
Recent advances in modelling the galaxy power spectrum using Effective Field Theory of Large-Scale Structure (\citealt{eftlss}) enable a perturbative treatment of mildly non-linear scales, allowing analyses to be extended to smaller scales and thereby increasing their constraining power \citep[e.g. the DESI DR1 full shape analysis, see][]{desi_DR1_fullshape}. 
To include the non-linear eﬀects and a more complex bias model, we use the velocileptors \citep{velociptor}, which takes the linear power spectrum and biases as inputs, and self-consistently calculates the non-linear redshift-space power spectrum, following the forecasting implementation of \cite{FISHLSS}.

WST’s spectral resolution, which is higher than DESI’s and comparable to that of 4MOST’s low-resolution spectrograph, could enable accurate stellar velocity-dispersion measurements for Fundamental Plane distance estimates, opening up an additional avenue for constraining cosmological parameters such as $\sigma_8$ and $f\sigma_8$ \citep[e.g.,][]{Dupuy, 2025arXiv251203226R}.  We do not explore this avenue in the present work, but note that it could provide a highly valuable and complementary probe at low redshift, where constraints from two-point clustering analyses are limited by the available cosmological volume.

\subsubsection*{Total neutrino mass $M_\nu$} 
Massive neutrinos leave a characteristic imprint on the growth of cosmic structure as they transition from relativistic to non-relativistic species, as well as on the Universe's geometry acting on the background energy density, accessible through BAO measurements. While relativistic, they contribute primarily to the radiation density; after becoming non-relativistic, they contribute to the matter density but retain significant thermal velocities, allowing them to free-stream out of overdense regions and suppress the growth of matter perturbations below their free-streaming scale (this is the effect we are focusing on in this work, see e.g. \citealt{LESGOURGUES_2006,Takada2006neutrinoinflation}). The strength of this suppression is governed by the neutrino fraction,
\begin{equation}
    f_\nu \equiv \frac{M_\nu}{93.14\,{\rm eV}\,h^2\Omega_{m}},
    \qquad M_\nu \equiv \sum_i m_{\nu_i},
    \label{eq:f_nu}
\end{equation}
and, on sufficiently small scales in the late-time linear regime, the relative suppression of the matter power spectrum can be approximated as $\Delta P_{m}/P_{m}\simeq -8f_\nu$, although its precise amplitude depends on scale, redshift, and cosmological parameters. This scale-dependent signature makes measurements of the full shape of the matter power spectrum particularly valuable for constraining $M_\nu$, beyond geometrical information alone (available from BAO measurements). Both approaches (BAO and power spectrum small-scale suppression) are essential to tighten constraints on the total neutrino mass. DESI DR2 BAO+CMB analysis \citep{Elbers_neutrino_DESIfullshape_2025} yields $\sum m_\nu < 0.0642\,\mathrm{eV}$, with a marginalized uncertainty of $\sigma(\sum m_\nu)=0.020\,\mathrm{eV}$, where the minimum value is set by oscillation experiments: $\sum m_\nu > 0.059\,\mathrm{eV}$ and $\sum m_\nu > 0.10\,\mathrm{eV}$ for the normal and inverted mass hierarchy, respectively. This precision favours the normal neutrino mass hierarchy over the inverted hierarchy, highlighting the potential of complementary large-scale structure probes to further constrain the absolute neutrino mass scale. High-redshift tracers are especially promising because their clustering can be modelled closer to the linear regime, reducing the impact of nonlinear gravitational evolution and baryonic feedback and enabling the use of smaller scales \citep{Takada2006neutrinoinflation}. \\

\subsubsection*{Primordial Non-Gaussianity (PNG)} 
Inflation remains the leading paradigm for the early Universe, and local PNG--quantified by the parameter $\fnl$--offers a key test of inflationary models. In the local PNG scenario, the primordial gravitational potential is conventionally parametrized as
\begin{equation}
    \Phi(x) = \phi_{\rm G}(x)
    + \fnl
    \left[
    \phi_{\rm G}^2(x)
    - \left\langle \phi_{\rm G}^2 \right\rangle
    \right],
    \label{eq:local_png}
\end{equation}
where $\phi_{\rm G}$ is the Gaussian primordial gravitational potential and $\fnl$ quantifies the amplitude of the primordial non-Gaussianity. The state-of-the-art single-field inflation model predicts a very small amplitude for local PNG in the primordial gravitational field, typically $\fnl \sim \mathcal{O}(\sim10^{-2})$. While the CMB has already provided strong constraints in favor of single-field inflation (with $\fnl = -0.9 \pm 5$, \citealt{Akrami2020}), classes of multi-field inflation models predict a potentially observable level of PNG, typically $\fnl \sim \mathcal{O}(1)$. This makes large-scale structure surveys the most promising avenue for future progress, as further improvements from CMB observations are fundamentally limited by cosmic variance.

Local-type PNG can leave a distinctive imprint on the large-scale clustering of galaxies through a scale-dependent correction to the galaxy bias \citep{Slosar2008,Dalal2008,Desjacques2010fNL}. This coupling between long- and short-wavelength potential fluctuations induces a scale-dependent modification to the linear galaxy bias, given by
\begin{equation}
    \Delta b(z,k) = b_\Phi \fnl
    \frac{3\Omega_{\rm m}H_0^2}
    {2k^2 T(k)D(z)},
    \label{eq:Deltab_fnl}
\end{equation}
where $b_\Phi$ describes the response of the tracer abundance to primordial potential fluctuations, $\Omega_{\rm m}$ is the present-day matter density parameter, $H_0$ is the Hubble constant, $T(k)$ is the matter transfer function, and $D(z)$ is the normalized linear growth factor. The characteristic $k^{-2}$ dependence of this correction on large scales enhances the sensitivity of galaxy clustering to local PNG at large scales. Consequently, high-redshift tracers covering large cosmological volumes are particularly valuable for probing these modes and obtaining competitive constraints. The tightest current constraint from the large-scale dependent bias, $\fnl = -3.6^{+9.0}_{-9.1}$, was obtained by \citet{Chaussidon2025fnldesiqsolrg} from large-scale power spectrum measurements of DESI quasars (QSOs, $0.8 < z < 3.1$) and luminous red galaxies (LRGs, $0.6 < z < 1.1$). Stage-V spectroscopic surveys such as the WST are expected to deliver independent and competitive constraints on $\fnl$ thanks to their larger footprint, higher number densities, and broader redshift coverage.

\subsubsection{Other cosmological science cases}

Our study is based on standard science cases, whereas future analyses may primarily focus on new cosmological probes. In particular, we currently limit ourselves to (template-based) power-spectrum analyses, while the bispectrum is already becoming a standard component of current DESI analyses~\citep{PS+BS}. Higher-order statistics are increasingly used to extract additional information from the non-Gaussianity of galaxy fields.
Applications include tests of parity violation~\citep{parity_violation}, constraints on inflationary models \citep{High_order_infl}, and measurements of gravitational redshifts~\citep{Gredshifts}. To extract non-Gaussian information, void science has matured into a precision tool for constraining the parameters of the standard cosmological model and its possible extensions \cite{PC_voids}. 
Galaxy clusters provide an independent cosmological probe through their abundance and scaling relations 
\citep[e.g.][]{Allen,Migkas}.
In addition, WST’s high spectroscopic target density, particularly for a magnitude-limited selection such as MagMax, could provide extensive spectroscopic coverage of clusters detected by eROSITA, Euclid, LSST and Sunyaev–Zel’dovich surveys such as ACT 
\citep[e.g.][]{Finoguenov,Capasso}. Spectroscopic follow-up of cluster member galaxies will enable stringent tests of general relativity on megaparsec scales through gravitational redshifts \citep{Tsaprazi2025gravitational_redshift_clusters_WST_white_paper,Tsaprazi2026gravitational_redshift_uncertainties_survey_requirements}. This probe is currently limited by the size of existing cluster samples and by the uncertainties in the determination of the cluster centres. Stage-V spectroscopic surveys such as WST will overcome these limitations by stacking spectroscopic data over unprecedentedly large samples of galaxy clusters.

At the power-spectrum level, there are also several interesting science cases beyond the standard analyses considered here. These include tests of General Relativity through measurements of the Doppler effect~\citep{Doppler}, constraints on modified gravity models~\citep{modifiedGravity}, and searches for alternative models of PNG \citep{High_order_infl}. WST forecasts will be performed in the future by the corresponding working groups, based on the tracer properties provided in this work.

By extending spectroscopic observations to fainter magnitudes, the WST low-z cosmology samples will probe a broader range of galaxy populations, including lower stellar masses at a given redshift, and potentially different host-halo mass distributions from those sampled by traditional DESI selections. For instance, WST will enable detailed investigations of dwarf-galaxy populations and their connection to the dark matter field~\citep{treiber_dwarf_2025, zhang_unexpected_2025} but also constrain how dense environments evolve with redshift to enable the construction of galaxy group catalogues, extending studies of central and satellite populations to fainter galaxies. WST would also be powerful for mapping cosmic filaments and the cosmic web itself to $z \sim 1$ over a very large area, which is directly relevant for missing-baryon studies with future X-ray missions such as newAthena~\citep{cruise_newathena_2024}. In particular, the connectivity of filaments (i.e the number of filaments feeding clusters’ cores) is a topological measure probing non-Gaussianity \citep{codis_connectivity_2018} and is regulated by the cold/warm/hot nature of dark matter~\citep{hyeonghan_weak-lensing_2024} (including massive neutrinos) and dark energy.

In addition, WST will allow in-depth tests of cosmic isotropy from galaxy number counts to accurately measure the cosmic dipole in redshift shells, and BAO/RSD measurements to be compared directly between different sky regions. This will provide a crucial assessment on the cosmic dipole as recent results reported tensions with the CMB-inferred kinematic dipole and with an isotropic expansion, from number-count dipoles from quasars and radio galaxies~\citep{secrest_challenge_2022, bashir_catwise2020_2025}, or directional variations of $H_0$ traced by galaxy cluster scaling relations that included spectroscopic velocity dispersions of galaxies within clusters ~\citep{pandya_examining_2024}, and other probes, e.g. local galaxies~\citep{boubel_testing_2025}, galaxy clusters~\citep{migkas2025galaxycluster} or SNIa~\citep{sah_anisotropy_2025, hu_testing_2024}. Particularly, a homogeneous magnitude-limited sample as the MagMax sample described in \secreff{ts:magmax} would be powerful for performing these tests. 

The WST combination of large collection area, high multiplexity and wide field of view opens up an alternative approach to derive cosmological constraints from the nearby Universe, particularly from the large galaxies in the Local Group that posed a major challenge to the previous generation of multi-object spectrographs on smaller telescopes. The first avenue is to estimate the local velocity field and to compare it with structures with the scale and mass of the Local Group from cosmological simulations. An early example is the work of \citet{Governato1997localgroup} who simulated a large sample of “local groups” in the CDM framework. WST can help to expand the velocity field probes to fainter magnitudes and wider angular distances. Another cosmological constraint is the content of the Local Group, in particular, the well-known missing satellites issue (e.g., \citealt{Klypin1999missingsattelites}). The impact of the WST here will be to trace the merging and accretion history of satellite galaxies in the halos of the nearby giant galaxies or even of the Milky Way by means of kinematic identification of streams (e.g., \citealt{Zhang2026sagittarius_stream}). WST will give access to less populated, sparser, and wider structures than the Sagittarius Stream. Measuring their masses is a direct constraint on the mass assembly history of the galaxies. Furthermore, the WST will allow us to estimate the abundances of stars originating in accreted galaxies or in nearby galaxies to depths and numbers that were not accessible before, leading to estimates of their contribution to the enrichment history of large galaxies (e.g. \citealt{Cerqui2025chemicalmilkyway}). WST will also support galaxy-evolution and astrophysical studies, such as the stellar and gas kinematics of galaxies in dense environments \citep{Raouf2021BGG} and, e.g., the interplay between AGN feedback and gas dynamics \citep{Raouf2024reviewAGN} affecting the galaxy-halo connection.

\section{Methodology of the paper} \label{sec:methodology_paper}
\begin{figure*}
    \centering
    \includegraphics[width=0.9\linewidth,
    trim=2 2 2 2,
    clip]{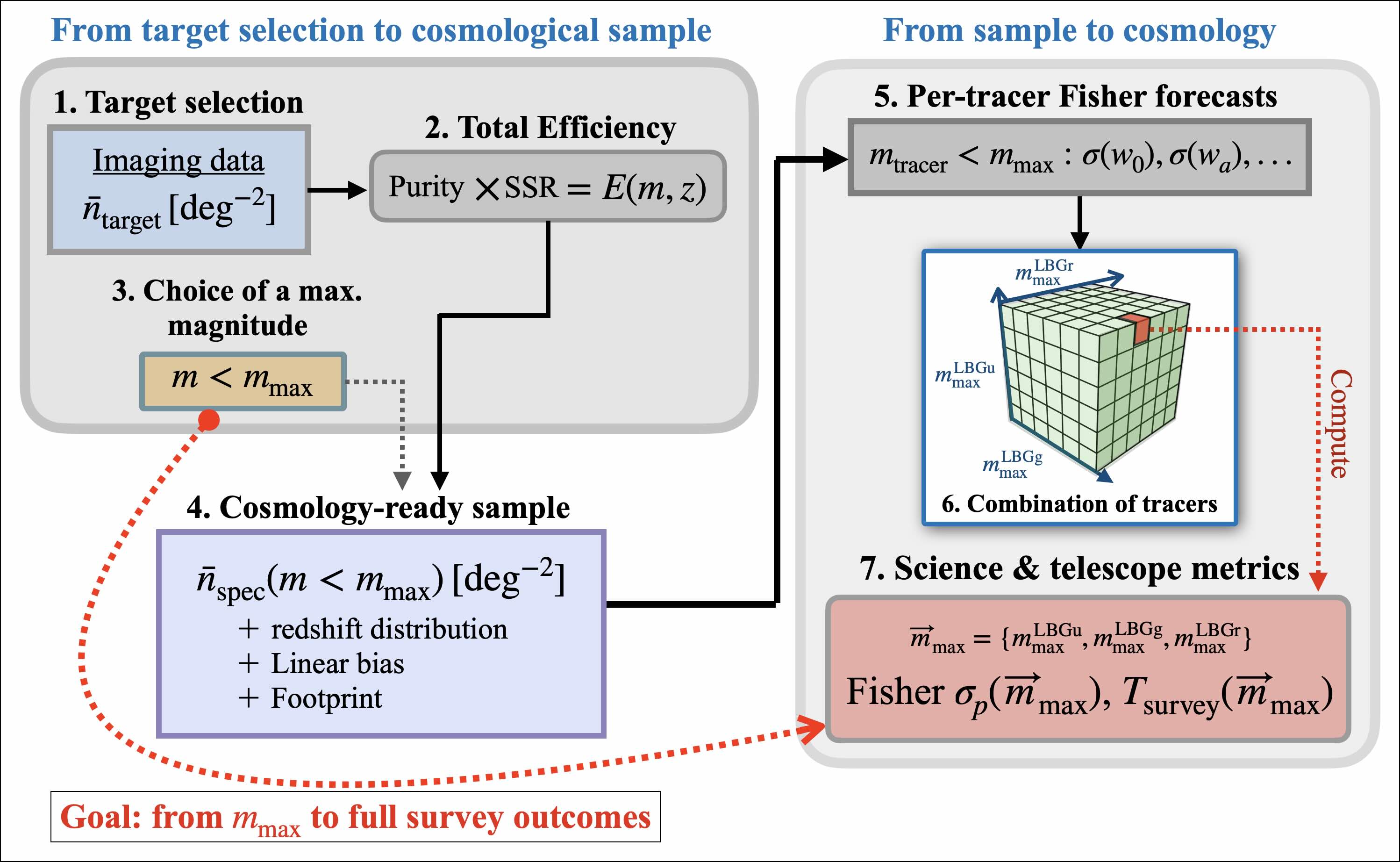}
    \caption{Overview of the pipeline designed in this paper, to go from target selection of tracers up to forecasts on cosmological parameters. }
    \label{fig:paper_logic}
\end{figure*}
In this paper, we aim to forecast the capabilities of WST to serve as an efficient cosmology experiment in the 2040s. To this end, we organise the work into a sequence of interconnected steps, summarised schematically in \figreff{fig:paper_logic}.

First, WST will target several classes of cosmological tracers, selected from Stage-IV photometric surveys, as presented in \secreff{sec:target_selection} (Step~1). Next, each targeted sample must be characterised in terms of its purity -- i.e., characterising the fraction of contaminants in the selection -- and the ease with which a reliable spectroscopic redshift can be obtained from its spectrum, as observed through the WST fibres. The combination of these two effects defines what we refer to as the ``total efficiency'' of a given tracer (Step~2; see \secreff{sec:modelling_wst_cosmo_survey}, and in particular \secreff{sec:wst_purity_specz_efficiency}). Defining a cosmological sample additionally requires specifying a maximum magnitude $m_{\rm max}$, since each tracer class is defined through a combination of colour cuts and a limiting magnitude in a given photometric band (Step~3). The magnitude limit will also define the density of the sample, as the number of objects exponentially increases with magnitude (\figreff{fig:recap_density_all}). Combining Steps~1, 2, and 3, we can then fully characterise a cosmology-ready sample through its spectroscopic redshift distribution, its linear bias, the resulting density of spectroscopically confirmed objects, and the survey footprint (Step~4). For a given choice of $m<m_{\rm max}$, this per-tracer information (Step~5) is used to compute both the survey time required to assign a fibre to every target (see \secreff{sec:survey_time_completeness}) and the corresponding cosmological forecasts (see \secreff{sec:forecast_methodology}). Per-tracer forecasts, obtained for different choices of $m_{\rm max}$, are presented in \secreff{sec:forecast}.

Beyond considering tracers individually, we can also combine several of them into a single, coherent survey design. In this case, a given survey design is defined by a specific set of tracers, each characterised by its own efficiency, redshift distribution $n(z)$, and other relevant properties at a chosen $m<m_{\rm max}$; varying $m_{\rm max}$ independently for each tracer defines a global hypercube of possible survey configurations (Step~6). For each point in this hypercube, we compute the corresponding scientific and telescope-time metrics (Step~7), which then allow us to identify the optimal multi-tracer survey design; we do this in particular for the LBG survey in \secreff{sec:forecast}.

\section{Target selection in the WST era}
\label{sec:target_selection}


\begin{table}
    \centering
    \resizebox{0.47\textwidth}{!}{%
    \begin{tabular}{lcccc}
        Tracers & Program &maximum mag. & $z_{\rm eff}$&$n_{\rm target}$ [\persqdegrees]\\
        \hline
        BG Bright & Bright &$r < 20.42$ & 0.27  &  2167\\
        BG Grey &Grey &$20.42< r < 21.07$& 0.41 & 1840 \\
        LRG & Grey&$z < 22.56$ & 0.91 & 1832\\
        ELG &Grey &$g< 25.06$ & 1.12 &  16,003\\
        \hdashline
        MagMax & Grey&$H < 21.4$  & 0.74 & 21,700 \\
        \hdashline
        QSO & Dark&$r < 24.24$&1.8&550\\
        LBG $u$-drop.& Dark&$r < 25$&3&5,100\\
        LBG $g$-drop.& Dark &$i < 25.5$&4&4,800\\
        LBG $r$-drop.& Dark&$z < 25.5$&5&1,000\\
    \end{tabular}}
    \caption{Summary information of the galaxy and QSO sample considered in this work. Maximum magnitudes, effective redshifts, and target densities are given for the baseline analysis setup.}
    \label{tab:summary_params_survey_design}
\end{table}

\begin{figure*}
    \centering
    \includegraphics[width=0.46\linewidth]{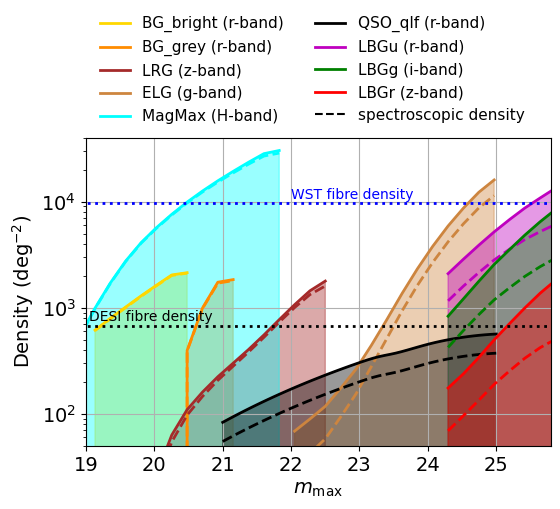}
    \includegraphics[width=0.5\linewidth]{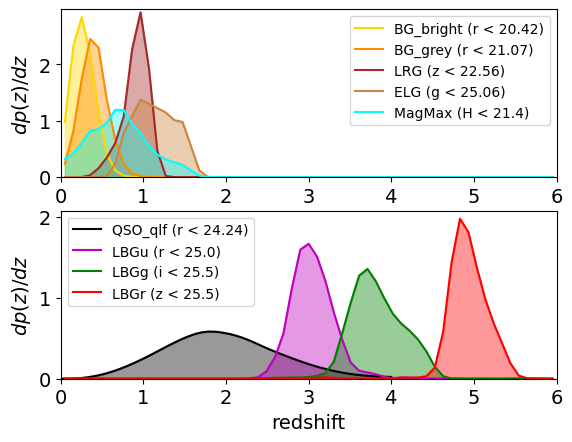}
    \caption{Left: Cumulative target densities and spectroscopically confirmed tracer densities for the different cosmological tracers, as a function of their maximum magnitude. Right: Corresponding redshift distribution of the spectroscopically confirmed tracers, for maximum magnitudes chosen to be in \tabreff{tab:summary_params_survey_design}. }
    \label{fig:recap_density_all}
\end{figure*}

Target selection is typically performed through a combination of colour cuts -- designed to (i) isolate a specific population within a given redshift range (ii) separate the galaxy or quasar sample from foreground stellar contaminants-- and a maximum magnitude cut $m<m_{\rm max}$. In this section, we define 3 types of target groups. First, a DESI-like sample, selecting Bright Galaxies (BG), Luminous Red Galaxies (LRG), and Emission Line Galaxies (ELG) at low redshift $z<1.6$ with magnitude limit rescaled using \eqreff{eq:scaling_mirror}. Alternatively, we define an $ H$-band magnitude-limited sample, called MagMax, up to redshift $z\sim 1.5$. Finally, we use a high-redshift $z>2$ sample using QSOs and LBGs. For convenience, we summarise in \figreff{fig:recap_density_all} the cumulative target density of each cosmological sample considered in this work as a function of the maximum limiting magnitude, together with the corresponding redshift distribution for a given choice of $m<m_{\rm max}$. 

\subsection{Photometric datasets in the WST era}
\label{sec:photometric dataset}
\subsubsection{Available wide-field imaging datasets in the 2040s}
\begin{figure}
    \centering
    \includegraphics[width=\linewidth]{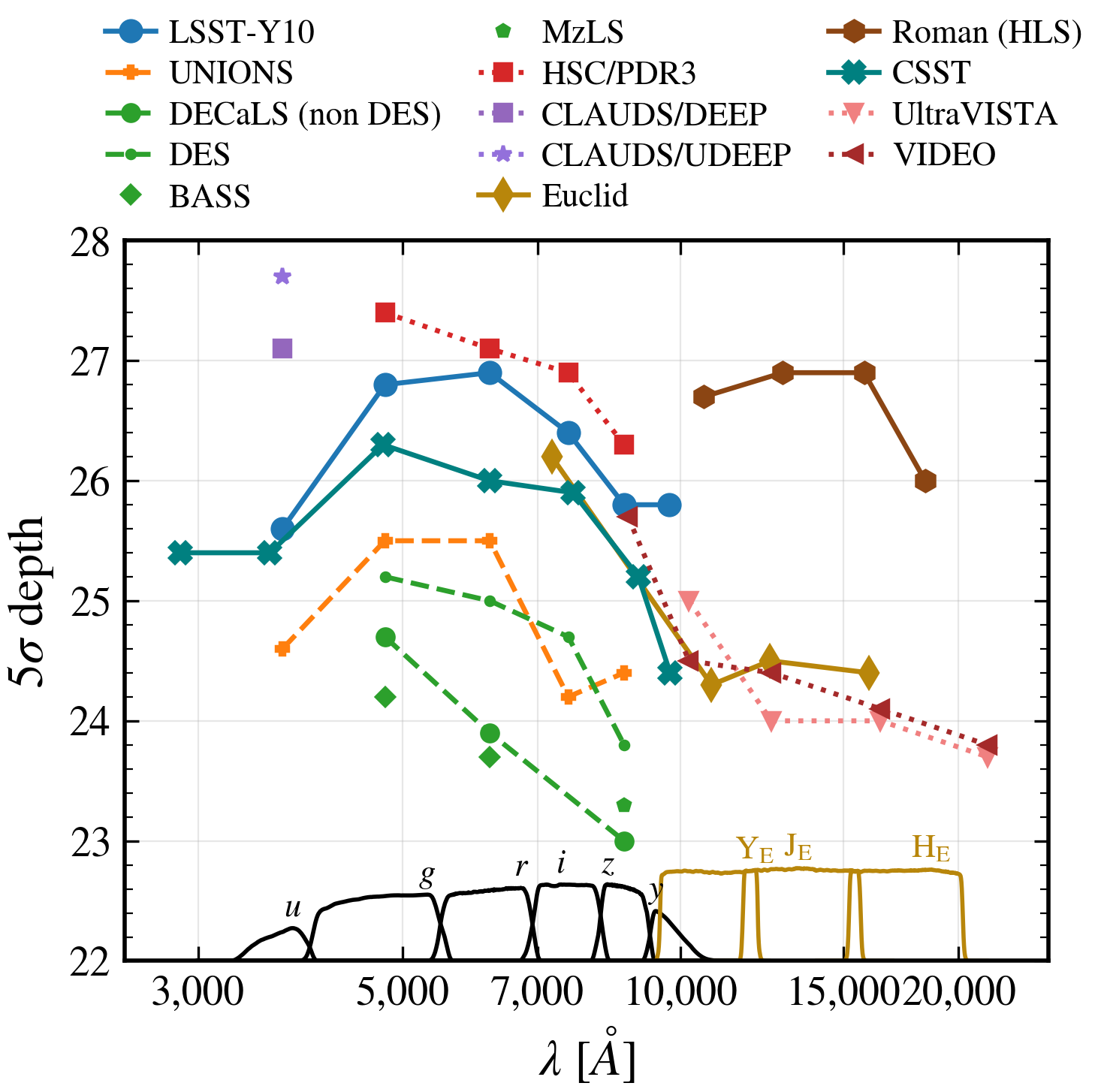}
    \caption{5$\sigma$ PSF depth versus $\lambda$ for some current and future photometric surveys, spanning the ground-based optical for wide surveys (UNIONS, DES, DECaLS, Rubin LSST, CSST) and deep fields (CLAUDS, HSC/PDR3). For optical-to-near-infrared: wide surveys such as \textit{Euclid}, \textit{Roman}, CSST, and deep fields from UltraVISTA. $ugrizy$ transmission curves are represented, along with the Euclid NISP $Y_\mathrm{E}$, $J_\mathrm{E}$, $H_\mathrm{E}$. 
        }
    \label{fig:imaging bands}
\end{figure}

Several Stage-IV photometric datasets are expected to be available by the time WST becomes operational, most notably
\begin{enumerate}
    \item the Rubin LSST~\citep{LSST2009whitepaper} will provide the $u,g,r,i,z$  and $y$ bands ($\sim$3000\AA \, to $\sim$11,000 \AA) respectively in the south hemisphere over 18,000 \sqdegrees. LSST began its main survey in June 2026. The Rubin LSST will allow wide coverage of $u$-band following the Ultraviolet Near-Infrared Optical Northern Survey (UNIONS;~\citealt{Gwyn2025unions}), which is actively observing the northern sky to cover a survey region of 6,250 deg$^2$.
    \item the~\textit{Euclid} satellite mission~\citep{Mellier2025euclid} is already on sky and has observed more than 6,000 deg$^2$ in August 2026, and will provide the $Y,J$ and $H$ bands on $\sim14,000$ deg$^2$. We make a clear distinction between the lower- and upper-case $y$ band in our whole analysis, as the $Y$ band in \textit{Euclid} is slightly different (from $\sim$ 9500\AA \, to $\sim$12,000 \AA) and is observed from space. \textit{Euclid} will also observe in the optical bands but with only one filter covering the same wavelength range as the $r,\, i$ and $z$ bands combined.
    \item Nancy Grace Roman Space Telescope~\citep{Akeson2019Romanwhitepaper} will provide 8 different bands that will cover the same wavelength range as the  $g,\, r,\, i,\, z,\, y,\, J,\, H$ and $K$s bands. The largest area covered by Roman is the High-Latitude Wide-Area Survey, which spans 5,100 deg$^2$.
    
    \item The Chinese Space Station Telescope (CSST;~\citealt{Gong2026csst}) providing $ugriz$ as well as the $YJK$ from space is expected for late 2026 or 2027, while Roman has already been successfully launched in August 2026.
\end{enumerate}  
These surveys will provide deep, multi-band imaging over large areas of the sky and large wavelength coverage, reaching depths well beyond those of current facilities. \figreff{fig:imaging bands} shows the 5$\sigma$ depth as a function of optical and near-infrared band for wide-area imaging surveys that are (i) currently available and (ii) expected to be accessible in the WST era, ranging from the Rubin LSST to \textit{Euclid}. 

\subsubsection{Mimicking Stage-IV photometric surveys}

Since we aim to evaluate the performance of WST for cosmological applications with the state-of-the-art photometric datasets expected in the WST era, we need to mimic these surveys with existing datasets. The first set is the DESI Legacy Surveys \citep{Dey_2019} data release 10.1 (LS DR10.1)~\footnote{DR11 (publicly released in August 2026) was not available at the time of the analysis}.

The core of LS DR10.1 is the imaging done with the Dark Energy Camera Legacy Survey (DECaLS), the Mayall $z$-band Legacy Survey (MzLS) and Beijing-Arizona Sky Survey (BASS, \citealt{bass_2017}) in the $g,r,i$ and $z$ bands.
In addition, we added DES imaging to the main data, as it uses the same camera as DECaLS in the same bands. 
Finally, complementary information from the photometry of the \textit{Gaia} mission~\citep{gaia2016} and the infrared data from the unWISE catalogue~\citep{Schlafly_2019} (respectively the $G_{\text{Gaia}}$ band and the set of $W1, W2, W3$ and $W4$ bands) were added to the targets after coordinate matching with all catalogues.
Altogether, they offer a wide set of multi-band imaging over a wide surface of the sky (more than 20,000 deg$^2$ since DR10.1)

The second set of data we use is deep photometric datasets in the optical and near-infrared covering relatively small fields of a few square degrees, such as in the COSMOS and XMM fields (see details here \citealt{Desprez2023CLAUDSHSCSSP}), largely used to conduct pilot clustering studies \citep{RuhlmannKleider2024LBGCLAUDS,Payerne2025lbg,choppin2026lbgcontrastivez}. In particular, the Canada-France-Hawaii Telescope Large Area U-band Deep Survey (CLAUDS;~\citealt{Sawicki2019CLAUDS}) and the Hyper Suprime-Cam Subaru Strategic Program (HSC-SSP;~(\citealt{Aihara2019HSC}) provide, respectively, $u$-band and $griz$-band coverage of this field significantly deeper than current Stage-III photometric surveys (see \figreff{fig:imaging bands}). At near-infrared wavelengths, COSMOS and XMM are also covered by the VISTA Deep Extragalactic Observations (VIDEO;~\citealt{jarvis2013video}) and UltraVISTA~\citep{McCracken2012ultravista}, which together provide near-infrared $Y$, $J$, $H$, and $K_s$-band imaging, similar to \textit{Euclid} photometric coverage. The COSMOS and XMM photometric catalogues considered in this work are constructed from source detections performed with \texttt{SExtractor}~\citep{sextractor_1996}. For each detected object, photometric redshifts are estimated using LePhare, a spectral energy distribution (SED) template-fitting code~\citep{lephare_2011}. We adopt the catalogue release providing photometric redshifts derived from the combination of 11 photometric bands ($ugriz$ plus near-infrared bands). A detailed description of the combined CLAUDS and HSC dataset over the COSMOS field is provided in~\cite{Desprez2023CLAUDSHSCSSP}. 

We note that the depth achieved in the five optical $ugriz$ bands by the CLAUDS+HSC imaging over COSMOS slightly exceeds the depth expected after ten years of Rubin LSST observations, the maximum to be expected for WST for now. As a result, the performance of the target selection presented in this work may be somewhat optimistic when compared to actual LSST-depth data. Alternatively, the COSMOS photometry could be artificially degraded to match the anticipated LSST Year-10 (Y10) depths, following the approach adopted in \citet{Hildebrandt2010phat,Payerne2025lbg,Payernelbgforecast}; we leave this refinement for future work.

\subsection{DESI-like low-redshift tracers}\label{ts:desi-like}

We first consider ``standard'' DESI-like low-redshift tracers: Bright Galaxies Sample (BGS), Luminous Red Galaxies (LRGs), and Emission Line Galaxies (ELGs). Each of these samples is primarily defined by a maximum magnitude cut in a specific photometric band, combined with specific colour cuts designed to (i) separate galaxies from stellar contaminants and (ii) isolate the target redshift range. To validate the target selection, DESI conducted a Survey Validation (SV) campaign to measure redshifts of the different tracers under different conditions and exposure times and with relaxed target selection cuts, to confirm that the main target selection samples were reaching the desired redshift success. For WST, we propose a low-$z$ sample that uses the same tracers as DESI with a higher magnitude cut. After the description of the DESI samples, we will present the extension for the WST low-$z$ target selection.

\subsubsection{Bright Galaxy Samples }

In DESI \citep{Hahn_2023}, two BGS subsamples were defined:
\begin{align}
(a) &\quad r < 19.5 \hspace{1.5cm}\text{(BGS-Bright)}, \\
(b)& \quad 19.5 < r < 20.175 \hspace{0.35cm}\text{+colour cut (BGS-Faint)}.
\label{eq:DESI_BG_cuts}
\end{align}

In addition to the magnitude cut, several colour cuts were added to target objects with high emission lines to maximise the redshift measurement success rate. Both samples are complemented with star-galaxy separation cuts, removing stars based on (i) \textit{Gaia} proper motion and parallax measurements and (ii) an additional star--galaxy separation cut, $G_{\rm Gaia}-r_{\rm raw} > 0.6$, where $r_{\rm raw}$ is the magnitude measured in the $r$ band not corrected for galactic extinction. Fibre-magnitude cuts\footnote{Defined as the magnitude of an object within a fibre of diameter 1.5$''$, under 1$''$ Gaussian seeing.} are further applied to exclude artefacts, following the approach adopted in \citealt{Hahn_2023} and \citealt{Verdier_2025}. Spatial masking is also applied to remove targets located near bright stars, large galaxies, and globular clusters\footnote{Spatial masking removes object detections near bright stars with ${\rm MAG}_{\rm VT} < 13$ (Tycho-2 sources) or $G_{\rm Gaia} < 16$ (\textit{Gaia} sources), using \texttt{BITMASK} 1 and 11 from the Legacy Survey catalogues (\url{https://www.legacysurvey.org/dr10/bitmasks/}), as well as sources located near large galaxies from the Siena Galaxy Atlas 2020 (SGA-2020) catalogue~\citep{moustakas2023} (\texttt{BITMASK} 12) and near globular clusters (\texttt{BITMASK} 13).}. 

\begin{figure}
    \centering
\includegraphics[width=1\linewidth]{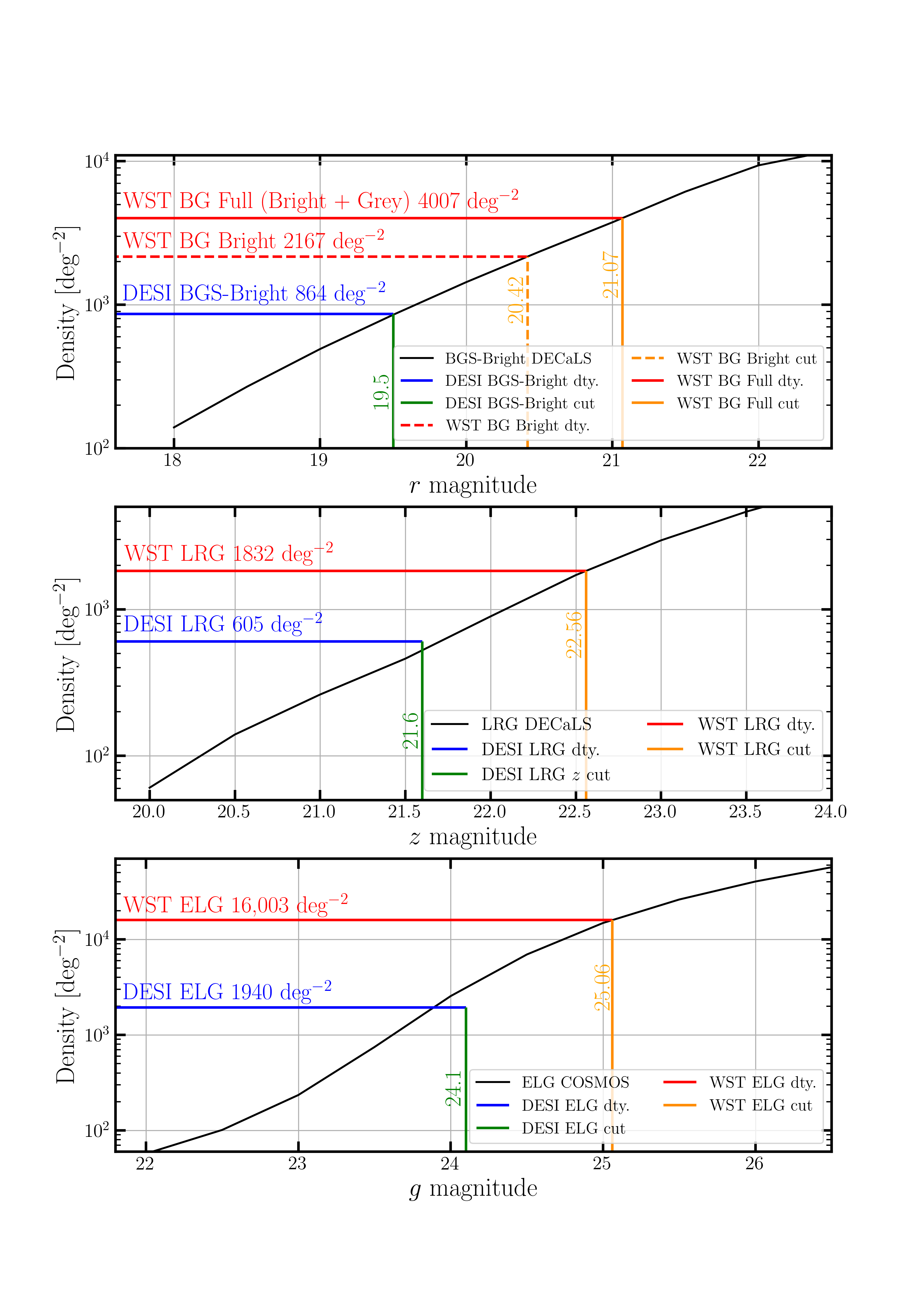}

    \vspace{-1.1cm}
    \caption{Total density versus maximum magnitude of the DESI and WST samples for each low-z DESI-like tracer. Top panel: DESI BGS-Bright, WST BG Bright, and full samples as a function of the $r$-band. WST BG Full is defined as the combination of the Bright and Grey samples. Middle panel: DESI and WST LRG target densities as a function of the $z$-band. Bottom panel: DESI and WST ELG target densities as a function of the $g$-band. DESI and WST magnitude limits for each sample are shown as green and orange vertical lines, respectively, and their densities as blue and red horizontal lines. The black curves correspond to the luminosity function of each sample in the DESI Legacy Surveys DR10.1 for BGs and LRGs and in the CLAUDS-HSC data for ELGs. The small mismatch in the LRG target density for the DESI cut arises because we used a subset of the Legacy Surveys files, which cover a smaller sky area than the full survey. For ELGs, it arises from the change in photometric dataset, from the Legacy Surveys to the deep photometry in the COSMOS field.
}
\label{fig:density_mag_DESILIKE_lowz}
\end{figure}

\figreff{fig:density_mag_DESILIKE_lowz} (upper panel) shows the cumulative target density of the BGS-Bright samples as a function of the maximum $r$-band magnitude cut. In a DESI-like configuration, the combined cumulative density of BGS-Bright and BGS-Faint reaches 1,400~\persqdegrees while BGS-Bright alone is 864~\persqdegrees. In the case of WST, we are only interested in extending the BGS-Bright sample. Spectra obtained with $\sim$180 seconds of exposure time were compared with a truth table made of the same objects observed during DESI SPV with cumulative co-adds of dark exposure combined with $\sim$300 seconds of exposure time on different nights in order to validate the redshift efficiency of the BGS-Bright sample~\citep{Hahn_2023}. The truth spectra thus have a minimum of 2000 seconds of exposure time. We validated the star--galaxy separation method by checking the object type using the obtained spectra.
From DESI SPV results, the DESI BGS-Bright sample is expected to reach $\sim99.5\%$ purity and $99\%$ completeness~\citep{Hahn_2023}.

\subsubsection{Luminous Red Galaxies}

Luminous Red Galaxies (LRGs) are massive, intrinsically bright galaxies dominated by old stellar populations and characterised by distinctly red colours \citep{baum_1955}. The pronounced $4,000\,$\AA\, break in their spectrum makes them straightforward to select photometrically and enables reliable redshift estimates from stellar absorption features (\citealt{Eisenstein_2001},\citealt{Zhou_2023}).
Owing to their higher number density at intermediate redshifts ($0.4 < z < 1.1$), LRGs have served as primary tracers of large-scale structure in past and current spectroscopic surveys, including SDSS and DESI Run 1. The LRG target selection strategy combines several colour cuts based on the optical $grz$ photometric bands with a limiting magnitude cut in the $z$-band. Using photometry from the DESI Legacy Surveys together with the infrared $W1$ band from unWISE, the DESI LRG selection is defined as follows:
\begin{align}
(i) &\quad z-W1>0.8 \times (r-z)-0.6 \\
(ii) &\quad (g-W1>2.9) \text{ or } (r-W_1>1.8) \\
(iii) &\quad [(r-W1)>1.8 \times(W1-17.14)\text{ and } \nonumber \\
&\quad (r-W1>W1-16.33)] \text{ or } (r-W1>3.3) \\
(iv) &\quad z_{\text{fiber}}<21.6
\label{eq:DESI_LRG_cut}
\end{align}
The first colour cut (i) exploits the 1.6 $\mu$m rest-frame stellar bump present in LRG spectra (\citealt{john_1988},\citealt{Sawicki_2002}) to separate galaxies from stellar contaminants. The second cut (ii) is designed to remove low-redshift interlopers while preserving high completeness for massive galaxies at $z > 0.4$. The third colour cut (iii), based on $r - W1$, is efficient at isolating the targeted redshift range (typically $0.6 < z < 1.3$), while the $W1$ magnitude limit selects the most luminous galaxies in $W1$ at any given redshift. Finally, a cut of $G_{\rm Gaia} > 18$ is applied to exclude bright \textit{Gaia} stars. As for the BGS selection, spatial masking is applied around bright stars, large galaxies, and globular clusters\footnote{Bright stars: \texttt{BITMASK} 1 and 11; large galaxies: \texttt{BITMASK} 12; globular clusters: \texttt{BITMASK} 13.}
The DESI selection imposes a subsequent fibre-magnitude cut ($iv$) of $ z_\mathrm {fibre} < 21.6$, yielding a total target density of 600 LRGs \persqdegrees\ (see \figreff{fig:density_mag_DESILIKE_lowz}).

To validate the target selection, a sample of "true" redshifts from deep co-added spectra was compared with the one obtained with an exposure time of $\sim$ 1000 seconds. Only 0.7$\%$ of the redshifts measured were considered as failures on a total of 15,379 spectra.


\subsubsection{Emission Line Galaxies}

Emission Line Galaxies (ELGs) are actively star-forming galaxies characterised by strong nebular emission lines, most notably the [OII] doublet ($3726$--$3729 \text{\AA}$), which is the strongest optical emission feature after the $H_\alpha$ Balmer line at $6562\,\text{\AA}$ \citep{Moustakas_2006}. The [OII] doublet has the advantage of being easily identifiable in optical spectra even at moderate resolution, and remains observable up to $z \simeq 1.6$, whereas the $H_\alpha$ line becomes inaccessible from ground-based observations above $z\simeq0.5$. ELGs reach their highest number density around $z\sim1.2$, coinciding with the peak of the cosmic star-formation rate density \citep{Madau_2014}.
In DESI, the main ELG target selection is made from two samples, LOP and VLO. The former aims to measure all ELGs in a redshift range $0.6 < z < 1.6$, while the latter gives more priority to targets in the redshift range $1.1<z<1.6$. In the following, we focus our results on the LOP sample. The main ELG-LOP selection strategy adopted in DESI ELG target selection \citep{Raichoor_2023} is defined by the following colour cuts:
\begin{align}
    (i)&\quad (r-z) < 0.15 \\
    (ii)&\quad (g-r) < 0.5 \times (r-z)+0.1 \\
    (iii)&\quad (g - r) < -1.2 \times (r - z) + 1.6 \\
    (iv)&\quad g_\mathrm{fibre} < 24.1
\label{eq:ELG_DESI_cut}
\end{align}
which together select ELGs within the redshift range $0.6 < z < 1.6$ and with a target density of $\sim$1940 \persqdegrees. For the DESI ELG sample, the criterion to select reliable redshifts is based on the [OII] flux measurement and the $\chi^2$ difference between the best and second-best-fit template. Applying this criterion leads to $\sim 70\%$ of reliable redshift measurements in the ELG LOP sample~\citep{Raichoor_2023}.

Although not used in this work (as we focus on target selection based on pre-existing photometric datasets), an independent view of the faint, line-dominated end of the ELG population is offered by WST integral-field spectroscopy; recent studies \citep{delMoralCastro2024eelgifs,GonzalezDiaz2025} have used the MUSE Hubble Ultra Deep Field Survey \citep{Bacon2023muse} to identify galaxies with strong line emission based on IFS. For example, \citet{delMoralCastro2024eelgifs} used the [O II] $\lambda\lambda$3726, 3729 emission lines (among others) to identify $z>0.5$ emission-line galaxies, where the $H\alpha$ line becomes inaccessible.

\subsubsection{WST low-redshift tracers}

In the following, we extrapolate the DESI selection, extending the sample to deeper magnitudes to define the WST low-$z$ samples following the mirror+exposure scaling relation in \eqreff{eq:scaling_mirror}. Besides the DESI-to-WST "miror" gain $\Delta m_{\rm mirror}=1.24$, we can compute from \tabreff{tab:wst_observing_baseline} the gain/deficit in magnitude arising from the difference in exposures from DESI to WST. The BG-Bright sample is expected to be observed in WST Bright Time, and the LRG and ELG samples in WST Grey Time. BGS are observed in DESI Bright Time, and LRG/ELG in DESI Dark Time (for DESI, 180 seconds and 1000 seconds are required for DESI Bright and DESI Dark Time, respectively ~\citealt{schlafly2024}).
We get
\begin{itemize}
    \item DESI Bright $\rightarrow$ WST Bright $\Delta m_{\rm exposure} = -0.32$
    \item DESI Bright $\rightarrow$ WST Grey $\Delta m_{\rm exposure} = -0.65$
    \item DESI Dark $\rightarrow$ WST Grey $\Delta m_{\rm exposure} = -0.28$
    \item DESI Dark $\rightarrow$ WST Dark $\Delta m_{\rm exposure} = 0$
\end{itemize}
Combined with $\Delta m_{\rm mirror}$, we define three samples using the same target selection strategy as DESI, increasing the magnitude limit of each sample:
\begin{itemize}
    \item WST BG-Bright: $r < 20.42$
    \item WST LRG: $z_{\text{fiber}} < 22.56$
    \item WST ELG: $g_{\text{fiber}} < 25.06$
\end{itemize}
The target densities reached for WST BG-Bright, LRG, and ELG samples are, respectively, $\sim$2,100 \persqdegrees, $\sim$1,800 \persqdegrees and $\sim$16,000 \persqdegrees. 

In addition, the WST BG-Bright sample is extended into a fainter sample, called WST BG-Grey, which is designed to be observed during WST Grey Time. Switching from DESI Bright Time to WST Grey Time in \eqreff{eq:scaling_mirror} allows an additional gain in magnitude of $\Delta m_{\rm exposure} = 0.93$ compared to the WST BG-Bright sample. The selection of the BG-Grey sample is similar to the WST BG-Bright sample but with a magnitude range:
\begin{itemize}
    \item WST BG-Grey: $20.42 < r < 21.07$,
\end{itemize}
such as the minimum WST BG-Grey $r$-cut corresponds to the maximum $r$-cut of the WST BG-Bright. The WST BG-Grey sample reaches a target density of 4,200 \persqdegrees. The WST BG-Full (defined as the combination of the BG-Bright and BG-Grey samples) leads to a target density of 7,100 \persqdegrees. All WST sample target densities and magnitude limits are shown in \figreff{fig:density_mag_DESILIKE_lowz} together with the DESI samples.

Regarding LRGs, beyond simply increasing the limiting magnitude, an alternative approach to increasing the LRG target density would be to shift the third colour cut in \eqreff{eq:DESI_LRG_cut}, involving $z-W1$, toward bluer colours, as investigated in \citealt{Schlegel2022desi2}. Conversely, one could instead reduce the target density while keeping the same limiting depth, as was done for the 4MOST-CRS target selection \citep{Verdier_2025,Bandi_2026}. We do not explore these scenarios here, as we rely on the increase in the maximum magnitude for later use. Regarding ELGs, since the DESI Legacy Surveys photometry ($grz$, from DECaLS) is too shallow to reach the extended magnitude limit for WST (the 5$\sigma$ point-source depth in the $g$-band is $24.5$; see \figreff{fig:imaging bands}), the selected sample would be severely affected by photometric incompleteness (see e.g. Appendix E of~\citealt{Payernelbgforecast}); we rely on deep photometric datasets available over the COSMOS field (see \secreff{sec:photometric dataset}).

The WST BG Bright is designed to be observed during Bright Time with 100s exposures, and WST LRG and ELG during Grey Time with 600s exposures in nominal conditions, while in DESI, LRG and ELG samples are observed during Dark Time. As already mentioned in \secreff{sec:WST overview}, the definition of observational time between DESI and WST (ESO) is different, and comparing predictions from the WST ETC showed that the same SNR ratio is obtained for WST samples during ESO Grey Time compared to the DESI samples in (DESI) dark time. Switching these samples in Grey time allows keeping ESO dark time only for the observation of high-$z$ tracers.

A summary plot in the left panel of \figreff{fig:recap_density_all} shows the cumulative target densities as a function of the limiting magnitude for the BG-Bright, BG-Grey, LRG, and ELG samples, and, for the baseline selections described above, their respective photometric redshift distributions in the right panel of the same figure. 

\subsection{MagMax: maximum-magnitude-limited samples} \label{ts:magmax}

Beyond the tracer class-based approach described above, we explore an alternative and more general strategy for target selection: defining a magnitude-limit sample, with minimal reliance on colour information. We refer to these samples as MagMax. In DESI, the BGS-Bright sample already has a design similar to a MagMax sample, as well as the WST BG-Full extension, i.e., being an $r$-band magnitude-limited sample with a star-galaxy separation cut.
The motivation for this approach is twofold. First, from a survey-design perspective, magnitude-limited selection offers a simpler, more robust, and more easily reproducible strategy than multi-colour tracer selection, since it depends on a single photometric quantity rather than on a combination of colour cuts whose efficiency and completeness can be sensitive to the depth and homogeneity of the underlying photometry. Second, from a cosmological perspective, a magnitude-limited sample maximises the number density of galaxies at a given depth, and therefore constitutes a natural benchmark against which the performance of colour-selected tracers (BGS, LRG, ELG) can be compared and understood. One advantage of such samples is that imaging systematics are simpler to control, and the modelling of galaxy bias, redshift evolution, or magnification bias is easier to perform. In addition, these samples can be useful to test the cosmological principle, i.e., the cosmic dipole, using galaxy counts, which need a homogeneous sample with no colour selection~\citep{secrest_challenge_2022}.   
Several cosmological experiments already used a MagMax sample; however, they often combined it with a photometric-redshift cut, as in the DES MagLim sample~\citep{weaverdyck2026} and the 4MOST-WAVES experiment~\citep{waves_2019}. In our work, we try to avoid any photometric redshift-dependent selection as it introduces additional complexity~\citep{weaverdyck2026}.

However, the stellar contamination of the MagMax sample needs to be minimised, as stars can be a source of noise in our data if they are misclassified and can lower the efficiency of the cosmological experiment, as each observation of a star could have been used for a galaxy instead\footnote{The stars observed in the MagMax sample are very valuable for studying the Milky Way’s accretion history and halo substructure as we are observing faint stars and can be complementary to the dedicated Milky Way survey within WST.}.
The former strategy used in SDSS, DESI, and 4MOST-CRS uses the unWISE $W1$ band together with the \textit{Gaia} catalogue to remove stellar contaminants, as adopted for the BGS, LRG, and ELG target selections described above. However, when targeting a magnitude-limited sample, the aim is to push significantly deeper in a single optical band, and stellar separation techniques need to be revisited.
In addition, infrared datasets such as unWISE (providing $W1$ photometry) will remain considerably shallower than the future multi-band $ugriz$ catalogues delivered by the Rubin LSST~\citep{rubin_2026}, and will not be substantially improved in terms of depth by SphereX~\citep{Melnick_galaxy_2024} either, motivating, for instance, the use of forced photometry to associate infrared fluxes with optically detected LSST sources (that we do not consider in this work for the MagMax sample). Similarly, the \textit{Gaia} catalogue will not reach the depths required for robust star--galaxy separation in the MagMax sample. To address this limitation, we used \textit{Euclid} photometry that will provide infrared bands $Y$, $J$, and $H$. However, the maximum wavelength accessible in $H$ is $\sim 2 \, \mu$m, while $ W1$ and $W2$ cover a higher wavelength range, $ 2.5 < \lambda < 5.5 \, \mu$m, preventing us from performing similar star--galaxy separation. Instead, we explore in this work an alternative approach to perform star--galaxy separation based on optical datasets, using LSST-, CSST-, and IR \textit{Euclid}-like photometry. The COSMOS deep field provides photometry for the $u$, $g$, $r$, $z$, $i$, $y$, $Y$, $J$, and $H$ bands, closely mimicking the photometric depth and coverage expected from the combination of LSST and \textit{Euclid}. 

\begin{figure}
    \centering
\includegraphics[width=1\linewidth]{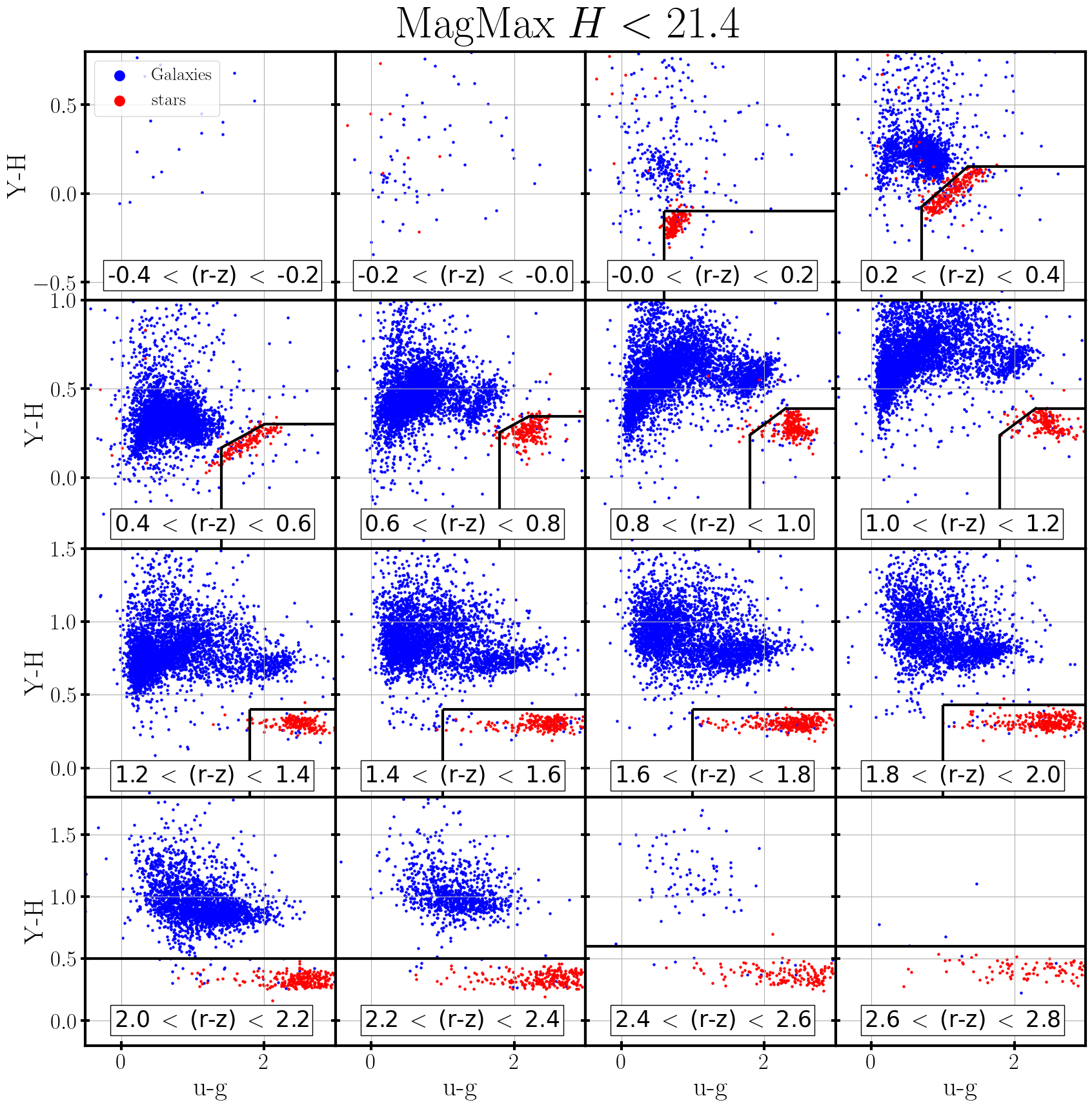}
    \caption{Three-dimensional colour-space diagram used to separate galaxies (blue) from stars (red), based on the $g$, $r$, $z$, $Y$, and $H$ photometric bands~\citep[][in prep.]{Verdier_stargal}. We define an object as a star if the column \texttt{OBJ$\_$TYPE} equals 2 in the CLAUDS-HSC catalogue. Each panel shows projected $(r-z)$ bins of length 0.2 from $-0.4$ to $2.8$. The black line indicates the adopted star--galaxy separation cuts. The sample shown here corresponds to the $H < 21.4$ MagMax sample.}
    \label{fig:StarGal}
\end{figure}

In this work, we consider a MagMax sample based on an $ H$-band magnitude limit. The star--galaxy separation is designed in a three-dimensional colour-space selection, based on $u-g$ versus $Y-H$ colours in bins of $r-z$. Details on the separation performance and strategy are described in~\citet[][]{Verdier_stargal} (in prep.). As illustrated in \figreff{fig:StarGal}, this approach identifies a star--galaxy separation boundary that reaches a purity of approximately 99$\%$ for galaxies and 95$\%$ for stars. This 3D selection introduces one colour cut per $(r-z)$ bin to remove the stellar locus from the sample. This contradicts our previous statement that we do not want to rely on colour-space cuts in the MagMax sample. 
However, in this case, the colour cuts separate two different regions (stars and galaxies) of the colour space and affect the galaxy sample only slightly (reaching 99\% purity). The 3D colour-space diagram can be seen as a feature map where galaxies and stars do not populate the same area. The galactic extinction is not taken into account in the MagMax sample as we only worked in the COSMOS field. However, the shift in the ($r-z$) colour due to galactic extinction is of the order of the size of the bin used in~\figreff{fig:StarGal}. When the LSST and \textit{Euclid} data are available for target selection, we will need to be very careful with extinction correction.
Using this refined star--galaxy separation, we define the $ H$-band MagMax sample with a single baseline magnitude cut of
\begin{equation}
    (i) \, H<21.4,
\end{equation} 
to reach a target density of $\sim$22,000 deg$^{-2}$, roughly corresponding to the sum of the baseline WST BG-Bright, WST BG-Grey, LRG and ELG (i.e. full DESI-like WST Bright+Grey times), has been set for the WST telescope design (10,000 fibres/\persqdegrees), where we can reach better than 90\% completeness with 4 overlapping tiles (see \figreff{fig:completeness_model_no_priority})
. The cumulative target density as a function of the magnitude limit of the $H$ MagMax sample is shown in the left panel of \figreff{fig:recap_density_all}, with the corresponding redshift distribution shown in the right panel of the same figure.

\subsection{Tracers of the high redshift Universe}

\subsubsection{Quasars}

\begin{figure}
    \centering
    \includegraphics[width=\linewidth]{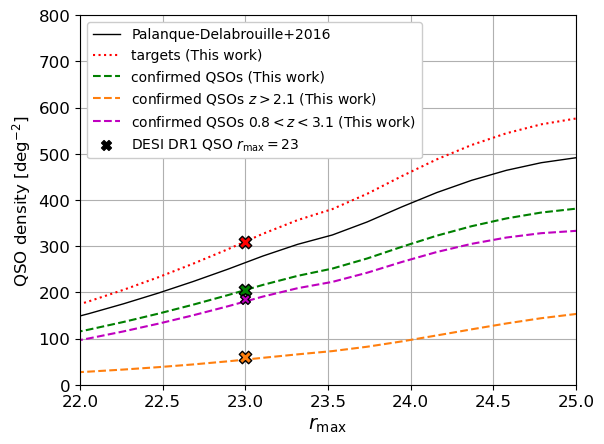}
    \caption{Models for the angular density of QSOs (either target density of spectroscopically confirmed QSOs). DESI DR1 QSO results are shown for $r_{\rm max}=23$.}
    \label{fig:qso_density_mag}
\end{figure}

Quasars (quasi-stellar objects, QSOs) are of the most luminous objects in the Universe, powered by accretion of matter onto the supermassive black holes that reside at the centres of active galaxies (see \citep{kellermann_discovery_2014, peterson_introduction_1997} for an introduction to QSOs and a historical review). Their spectra are characterised by strong, broad emission lines due to thermal broadening. Their high intrinsic luminosity makes them visible out to very high redshift, and therefore there are efficient to trace the large-scale structures of the matter field over an large cosmic volume through clustering measurement \citep{myers_clustering_2007, zarrouk_preliminary_2021, neveux_completed_2020, desi2025baodr2} (as for BGS, LRGs, ELGs) or via their Lyman-$\alpha$ forest\footnote{At high redshift (i.e. higher than $z\>2.1$) that traces the line-of-sight neutral hydrogen in the intergalactic medium, which can be used to probe the distribution of matter}~\citep{du_mas_des_bourboux_completed_2020, collaboration_desi_2025, collaboration_desi_2026DESI}. 
Most quasars are spatially unresolved and therefore appear as point sources, making them difficult to separate from stars. Quasars have a characteristic blue continuum and classical UV excess~\citep{ross_sdss-iii_2013, richards_spectroscopic_2002}. In addition, the hot dust torus surrounding a quasar's accretion disc produces a strong infrared excess (notably in $W_1-W_2$ from WISE) that stars do not show. Successful selection of a highly complete and pure QSO sample is originally based on their UV ~\citep{ross_sdss-iii_2013, richards_spectroscopic_2002} or on the detection of their near-infrared excess, as done in eBOSS~\citep{myers_sdss-iv_2015} or DESI~\citep{chaussidon2023qsots}. Selection based on both aspects has also been performed using optical and IR photometry~\citep{eltvedt_vst_2023, chehade_2qdes_2016} and is currently used for 4MOST-CRS~\citep{Bandi_2026}. However, extending QSO selection to deeper magnitudes is not straightforward, as it introduces significant galaxy contamination. For instance, going to $r<24.5$ yields a target density of more than $2,000$ targets/\persqdegrees, which is inconsistent with predictions from the QSO luminosity function (QLF;~\citealt{Palanque2016QSOQLF, li_predicting_2026}) that predict $\sim 400$ true QSO/\persqdegrees for $r<26.9$. Therefore, QSO selection needs to be redefined for future Stage-V cosmological surveys, which requires dedicated work beyond the scope of this paper. In this work, we want a model for the QSO target density as a function of the $r$-band magnitude $n_{\rm target}(r<r_{\rm max})$. Rather than applying a specific QSO target selection to COSMOS, we adopt a model-based approach. We start from the DESI QSO selection~\citet{chaussidon2023qsots} based on a Random Forest algorithm, which uses $r$-band magnitudes between $16.5<r<23$. This selection includes approximately 310 targets~\persqdegrees, of which 205 \persqdegrees are confirmed as genuine QSOs. The density as the sum of confirmed QSOs $n_{\rm true}^{\rm sel.}(r<23)$ and contaminants $n_{\rm contaminants}(r<23)$ can be express as:
\begin{align}
   n_{\rm target}(r<23) = c\times n_{\rm true}(r<23) + (1-p) \times n_{\rm target}(r<23)
\end{align}
where $c$ define the completeness and $p$ the purity of the selection. The selection function equation $\Phi =c/p$ allows us to compute the target density $n_{\rm target}(r<23)$ as a function of the density of true QSOs $n_{\rm true}(r<23)$. We now have two unknowns, $c$ and $p$. For $c$, we use
\begin{equation}
    c = \frac{n_{\rm true}^{\rm DESI\ sel.}(r<23)}{n_{\rm true}^{\rm QLF}(r<23)} = 0.77
\end{equation}
where $n_{\rm true}^{\rm QLF}(r<23)=264$ \persqdegrees is obtained using the Quasar Luminosity Function (QLF) of~\citet{Palanque2016QSOQLF}. Then, $p$ is given by
\begin{equation}
   p= \frac{n_{\rm true}^{\rm DESI\; sel.}(r<23)}{n_{\rm target}^{\rm DESI}(r<23)} = 0.67,
\end{equation}
that finally leads to $\Phi=1.15$.
We use this relation to compute the target density of QSOs as a function of any other given maximum magnitude illustrated in \figreff{fig:qso_density_mag} up to $r<25$. Let us note that this ratio of $\Phi=1.16$ is a zeroth-order parametrisation, as it may change with magnitude $\Phi(r_{\rm max})$, as well as the depth of the imaging and the selection method. We will keep using it but raise a warning here. The cumulative target density of the QSO sample versus the $r$-band and corresponding redshift distribution are shown in \figreff{fig:recap_density_all}. As a WST baseline setup, we choose to use the DESI $r$-mag maximum cut, complemented by the WST-to-DESI "mirror" gain of about 1.24.

\subsubsection{Lyman Break Galaxies}
\label{sec:ts_lbg}

Lyman Break Galaxies (LBGs, \citealt{Steidel1996LBG}) are young, actively star-forming galaxies. They have historically served as central tracers for studying galaxy formation and evolution at high redshift \citep{Steidel1996LBG,Steidel1999LBG,Giavalisco2004LBG,Reddy2008LBG,Hildebrandt2009lbg,Harikane2023lbg}, and have more recently emerged as powerful cosmological probes \citep{WilsonWhite2019dropout,payerne2026lbgcmblensingunions}. Large $U$-band coverage such as provided by the Rubin LSST will be essential to select LBGs \citep{Crenshaw2025lbgsellsst}, which makes them promising tracers for exploring the $z>2$ Universe with high-density samples.

The rest-frame spectra of LBGs exhibit a strong decrement in flux blueward of the Lyman limit at $912\,\text{\AA}$, together with a broader Lyman break feature spanning from $912\,\text{\AA}$ to $1216\,\text{\AA}$, produced by absorption and resonant scattering of neutral hydrogen along the line of sight through the intergalactic medium. Redward of $1216\,\text{\AA}$, the spectrum is well described by a flat continuum with different emission and absorption features.
LBG dropout selection (see, e.g., \citealt{RuhlmannKleider2024LBGCLAUDS}) relies on identifying this flux decrement in a given \textit{dropout} photometric band (either $u$, $g$, or $r$) relative to its neighbouring, redder band ($g$, $r$, or $i$, respectively). The $u$-, $g$-, and $r$-dropout selections together span a broad redshift range, from $z\sim2.5$ to $z\sim5.5$. For the $u$-, $g$-, and $r$-dropout samples, respectively, the limiting-magnitude band is taken to be $r$, $i$, and $z$.

For the $u$-dropout selection, with $r<r_{\rm max}$, we adopt the selection criteria of \citet{RuhlmannKleider2024LBGCLAUDS}:
\begin{align}
    &(i)\ u - g > 0.3, \\
    &(ii)\ -0.5 < g - r < 1, \\
    &(iii)\ [u - g > 2.2 \times (g - r) + 0.32] \\
    &\ \cup \ [u - g > 0.9 \ \cap \ u - g > 1.6 \times (g - r) + 0.75]
\end{align}
The first cut enforces the dropout criterion, while the additional diagonal cut in $u$-$g$-$r$ colour space is designed to avoid contamination from the stellar locus. For the $g$-dropout sample, we impose $i<i_{\rm max}$ together with the colour cuts
\begin{align}
&(i)\ g - r > 1, \\
&(ii)\ -1.5 < r - i < 1, \\
&(iii)\ g - r > 1.5\times(r - i) + 0.8.
\end{align}
For the $r$-dropout sample, we impose $z < z_{\rm max}$, together with
\begin{align}
&(i)\ 1.7 < r - i < 5, \\
&(ii)\ r - i > 1.6\times (i - z) + 1.2, \\
&(iii)\ -0.5 < i - z < 1.5.
\end{align}
The cumulative target densities of the three LBG samples as a function of the $r$-, $i$-, and $z$-band limiting magnitude are shown in the left panel of \figreff{fig:recap_density_all}, with the corresponding redshift distributions, for a representative choice of magnitude cut, shown in the right panel of the same figure. For LBGs, we have no strong requirement for their baseline setup scaled from DESI (as we did for the other tracers). Based on the quite stable overall efficiency of the LBG selection, we consider the reasonable cuts $r<25$, $i<25.5$ and $z<25.5$ for the LBG $u$-, $g$- and $r$-dropout samples. We will see in \ref{sec:forecast_Lbg_mag} how modifying these magnitude cuts impact survey time and cosmological constraints. 

\subsubsection{Lyman-$\alpha$ Emitters}

Although Lyman-$\alpha$ Emitters (LAEs) are not considered in this paper, we nonetheless emphasise their importance for exploring the high-redshift Universe. LAEs are young, low-mass, actively star-forming galaxies at high redshift, typically selected from narrow- or intermediate-band imaging surveys as galaxies exhibiting a large rest-frame equivalent width in Ly$\alpha$ emission~\citep{Ouchi2020lae, layana-astudillo_new_2026, Ebina2026laeclustering, Ebina2025laeclustering, Steidel1996LBG}. LAEs are closely related to LBGs, but significantly less massive, and can in principle be regarded as a subset of the broader LBG population; however, given the differences in their selection functions and in the magnitude and stellar-mass ranges they probe, LAEs and LBGs are generally treated as distinct tracer classes. The Intermediate-Band Imaging Survey (IBIS, Dey et al. in prep.) employs a set of medium-band filters on the Dark Energy Camera (DECam, \citealt{honscheid_dark_2008}) at the Blanco Telescope in Chile. The IBIS medium-band filter set enables the selection of LAEs for clustering studies~\citep{Ebina2025laeclustering, Ebina2026laeclustering}, paving the way for Stage-V spectroscopic surveys and their precursors, such as DESI-II~\citep{Schlegel2022desi2}. A comparable narrow-band approach could similarly be pursued with the Vera C. Rubin Observatory following the completion of its ten-year LSST program. A proof-of-concept study~\citep{Herrera2025lbg} has shown that Lyman-$\alpha$ forest analysis can be extended to higher-redshift tracers like LBGs and LAEs, which are fainter than QSOs, making the measurement of small-scale clustering extremely promising for Stage-V spectroscopic surveys.

\section{Modelling the WST Cosmology survey}
\label{sec:modelling_wst_cosmo_survey}

In this section, we investigate how these different samples can be combined and jointly optimised to maximise the scientific return of the survey, subject to constraints on total telescope time, spectroscopic efficiency, and related observational factors. Designing an optimal survey strategy requires striking a careful balance between scientific outcomes and practical operational considerations, such as how observing time is allocated and how efficiently fibres are exposed. In the following, the observing time is considered to be dedicated only to the WST cosmological surveys, i.e., all the WST fibres are used for cosmology targets. Therefore, as the WST survey design includes targets with other science cases, the observing time should depend on the fraction of fibres allocated to cosmology. A complete assessment of the total efficiency of a survey would require performing the full target selection and running an end-to-end pipeline, including fibre assignment, observation specifics, and spectroscopic redshift reconstruction. We do not attempt this here; instead, we adopt a simplified parametrisation that captures these effects as an overall modification of the target redshift-magnitude distribution. We adopt a generic formalism in which, for a given tracer, the density of spectroscopically confirmed objects (i.e. tractable and useful for cosmological analysis) is given by
\begin{align}
\label{eq:cumulative}
    n_{\rm spec}(m< m_{\rm max})&= \int\limits_{-\infty}^{m_{\rm max}}\int\limits_{0} ^{+\infty}{ \ E(m,z)\, \phi^{\rm TS}(m,z) \ \rm d}m\, { \rm d}z 
\end{align}
where $\phi^{\rm TS}(m,z)$ is the target redshift-magnitude distribution and $E_(m,z)$ is an arbitrary, tracer-dependent efficiency function that captures the efficiency of a target to pass all the requirements and fall into the spectroscopic sample used for cosmological analysis. By construction, $E(m,z) \in [0,1]$ and encompasses any effects that prevent a target from being used for cosmological analysis.

In the left part of \figreff{fig:paper_logic}, we show the workflow we designed to go from target-selection outputs to a cosmology "ready-to-use" sample (surface density of spectroscopically confirmed tracers, spectroscopic redshift distribution, and tracer bias). Varying $m_{\rm max}$ for each sample allows us to explore how constraining power on some cosmological parameters of interest improves as the sample size increases; see the sections below. 

\subsection{Purity and spectroscopic redshift success rate}

\label{sec:wst_purity_specz_efficiency}

Once a target sample has been defined (characterised by $n_{\rm target}(z_{\rm phot})$, as shown in \figreff{fig:recap_density_all}), the path toward a certified spectroscopic sample suitable for cosmological analysis is not yet complete. Two distinct effects must still be accounted for: the purity $\mathcal{P}$ of the target sample and the spectroscopic redshift success rate denoted as SSR. 

The purity $\mathcal{P}$ quantifies the fraction of objects in your target catalogue that are successfully identified as the intended class of object at the correct redshift range, namely
\begin{equation}
    \mathcal{P} = \mathrm{Proba}(\mathrm{target\in class})
\end{equation}
Each sample is generally not composed exclusively of the intended tracer population, i.e., target samples can be contaminated by interlopers such as stars, galaxies in the QSO samples, or lower-redshift galaxies (like ELGs in LBGs samples). The purity depends on the quality of the imaging data and the flexibility of the adopted colour cuts~\citep{Verdier_2025}. 
The purity ($\mathcal{P}$) quantifies the fraction of objects in your target catalogue that are successfully identified as the intended class of object at the correct redshift range. Depending on the target class, different interlopers will degrade the sample purity and therefore reduce the final sample usable for the cosmological analysis.

In addition, the cosmological sample, used for cosmological analysis, is further reduced by the spectroscopic redshift success rate. The SSR, given by
\begin{equation}
    \mathrm{SSR}=\mathrm{Proba}(\mathrm{good}\, z_{\rm spec}|\mathrm{target\in class})
\end{equation}
describes the probability for objects that have the correct target class but where the spectroscopic pipeline fails to provide a correct redshift measurement. The SSR depends on (i) the intrinsic properties of the tracer itself, in particular the number of relevant spectral features falling within the spectrograph's wavelength range that can be used to derive a reliable spectroscopic redshift, (ii) the instrumental and survey characteristics, such as the mirror size, spectral resolution, and exposure time, and (iii) the specific algorithm used to derive the spectroscopic redshift. For instance, at low signal-to-noise flux ratio, the redshift success rate typically increases approximately linearly with SNR before saturating at a plateau \citep{Krolewski2025desi}. We define the spectroscopic efficiency $E(m,z) =\mathcal{P}\times \mathrm{SSR}$ as the ratio of spectroscopically confirmed targets belonging to the desired tracer type/class to the target density (a combination of purity and SSR).

Assuming a spectroscopic efficiency comparable to the one achieved by DESI, we adopt, for simplicity, a constant spectroscopic redshift reconstruction efficiency across redshift and magnitude for the DESI-like tracers BGS, LRG, ELG, but also for MagMax and for the QSO samples. Namely, for BGs we use $E_{\rm BG}(m,z) = 0.99$ for $z < 1$, and $0$ otherwise; for LRGs we use $E_{\rm LRG}(m,z) = 0.99$ between $0.4 < z < 1.1$, 0 otherwise; and for ELGs, we use $E_{\rm ELG}(m,z) = 0.73$ for $0.6 < z < 1.6$, reflecting the redshift range over which the [OII] doublet is observable~\citep{DESI_2024_II}. For the MagMax sample, we adopt the same efficiency as the BGs, $E_{\rm MagMax}(m,z) = 0.99$. Finally, for QSOs, we consider the overall efficiency obtained at $r<23$ in~\citet{chaussidon2023qsots}, such as $E_{\rm QSO}(m,z) = 0.66$ for all redshifts and magnitudes. We represent in \figreff{fig:qso_density_mag} the QSO spectroscopic density model for different redshift intervals, along with DESI DR1 QSO measured values. LBGs constitute a new tracer population in this context and are more challenging to measure. Therefore, we adopt a more elaborate redshift efficiency model. First, we define the purity $\mathcal{P}$ to be zero at $z<2$, 1 otherwise, since most candidates at lower redshifts are QSO and ELG contaminants. 

\begin{figure}
    \centering
    \includegraphics[width=\linewidth]{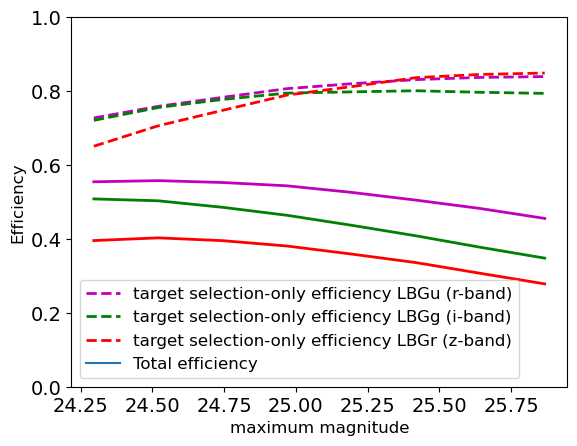}
    \caption{
    Efficiency of the LBG $u$-, $g$-, and $r$-dropout selections, either target-selection efficiency or total efficiency as a function of the corresponding limiting-band magnitude. }
    \label{fig:lbg_efficiency}
\end{figure}

\figreff{fig:lbg_efficiency} shows the purity defined as $n_{z_{\rm phot}\in[z_1, z_2]}/n_{\rm target}$ (hereafter referred to as the target selection purity) as a function of the maximum magnitude of the $u$-, $g$-, and $r$-dropout LBG selections, evaluated using $ugriz$ CLAUDS+HSC/PDR3 broadband photometry over the COSMOS field. $[z_1, z_2]$ denotes the redshift range of the photometric redshift distribution targeted by each LBG dropout selection, with $z_1=\{2,3,4\}$ and $z_2=\{4,5,6\}$ for the $u$-, $g$-, and $r$-dropout samples, respectively. For all three samples, the selection efficiency increases with the magnitude limit and asymptotes to a plateau of approximately $80$--$85\%$, indicating that the large majority of selected targets lie within the intended high-redshift range and are selected at high magnitude in their detection bands (these are faint galaxies). The dominant low-redshift interlopers in the LBG selection are ELGs, whose Balmer break can be mistaken for a Lyman break, and QSOs at intermediate redshift~\citep{choppin2026lbgcontrastivez,RuhlmannKleider2024LBGCLAUDS}.

\begin{figure}
    \centering
    \includegraphics[width=\linewidth]{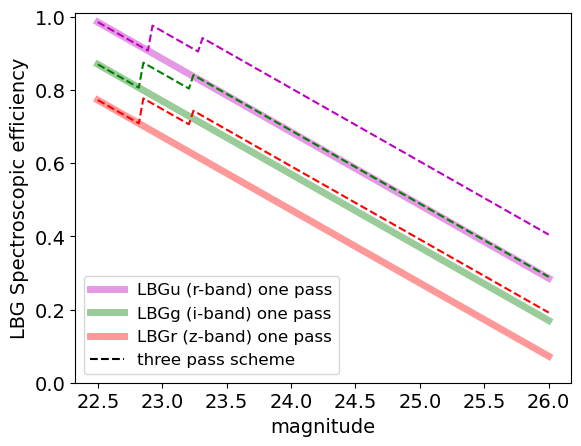}
    \caption{Spectroscopic redshift efficiency for the three LBG samples, assuming either a single observing pass or 3 passes.}
    \label{fig:lbg_ssr}
\end{figure}

For the LBG $\mathrm{SSR}$, we follow the methodology developed in~\citet{Doumerg2023forecast}, rescaling the LBG spectroscopic redshift efficiency measured for the $u$-dropout sample in~\citet{RuhlmannKleider2024LBGCLAUDS} to the performance expected for WST. Assuming that faint LBG spectra are sky-noise dominated, and that the spectroscopic efficiency depends primarily on (i) the spectrum SNR, which improves from DESI to WST as derived above, and (ii) the specific redshift-fitting algorithm used in DESI\footnote{This algorithm is a convolutional neural network (CNN) based on QuasarNET~\citep{Busca2018quasarnet} -- which is the algorithm used for QSOs in DESI as part of the spectroscopic classification pipeline -- applied to DESI spectra to perform LBG classification and redshift regression. It is trained to identify 14 absorption lines and 2 emission lines, outputting a confidence level (CL) for each line. The 16 CLs are ranked, and a spectrum is classified as an LBG if the fifth-highest CL exceeds a given threshold. The CNN redshift is taken from the highest CL. This is then refined using RedRock (RR)~\citep{Guy2023RROCK}, which uses the CNN redshift as a prior and LBG-specific templates built from 840 visually inspected spectra from the DESI pilot survey of XMM-LSS.}. We obtain the rather simple rescaling relation
\begin{equation}
    \mathrm{SSR}_{\rm WST}^{\rm udrop}(m) = \mathrm{SSR}^{\rm udrop}_{\rm DESI}(m - \Delta m_{\rm DESI\rightarrow WST}),
\end{equation}
where $\Delta m_{\rm DESI\rightarrow WST}=1.23\ \mathrm{(mirror)} -1.13\ \mathrm{(time)}$\footnote{The exposure times considered here are 2 hours for DESI and 15 minutes for WST Dark Time.}
and $\mathrm{SSR}^{\rm udrop}_{\rm DESI}(m)=-0.2(m-23.5) + 0.75$~\citep{RuhlmannKleider2024LBGCLAUDS}, with $m$ denoting the $r$-band magnitude. We note that in~\citet{Doumerg2023forecast}, the corresponding rescaling was instead performed using the efficiency SSR$^{\rm udrop}_{\rm MSE}(m)=-0.18m + 4.8$, measured by injecting LBG template spectra into the MSE Exposure Time Calculator \citep{Percival2019MaunaKea}. This relation applies specifically to the $u$-dropout LBG sample. To extend this to the $g$- and $r$-dropout samples (denoted generically as X-dropout, with detection bands $i$ and $z$, respectively), we adopt a similar argument based on the decrease in spectral SNR due to distance dimming, under the assumption that the intrinsic physical properties of LBGs remain fixed over the redshift range considered. Specifically, we assume that the portion of the LBG spectrum sampled in the $i$-band for a $g$-dropout galaxy corresponds to the same spectral region sampled in the $r$-band for a $u$-dropout galaxy, such that only the SNR degradation from volume/redshift dilution needs to be accounted for. This yields
\begin{equation}
    \mathrm{SSR}_{\rm WST}^{\rm Xdrop}(m) = \mathrm{SSR}_{\rm WST}^{\rm udrop}(m - \Delta m_{\rm udrop \rightarrow Xdrop}),
\end{equation}
where
\begin{equation}
    \Delta m_{\rm udrop \rightarrow Xdrop} = 2.5\log_{10}\left(\frac{(1+z_u)D^2(z_u)}{(1+z_X)D^2(z_X)}\right),
\end{equation}
and $m$ denotes the $i$-band and $z$-band magnitude for the $g$- and $r$-dropout selections, respectively, and $z_u$ ($z_X$) denotes for the mean redshift of the $u$-dropout selection (respectively of the $X$-dropout selection).

\figreff{fig:lbg_ssr} shows the resulting spectroscopic efficiency models for the $u$-, $g$-, and $r$-dropout LBG samples as a function of their respective limiting-band magnitude. We additionally show the corresponding efficiency obtained under a three-pass observing scenario, inspired by \citet{Doumerg2023forecast}, in which the overall spectroscopic efficiency is increased by shifting the efficiency curve by $\Delta m(n_{\rm pass}) = 2.5\log_{10}(\sqrt{n_{\rm pass}})$. In this scenario, the sample is divided into three magnitude intervals, with the two brighter bins required to reach a minimum efficiency $p_{\rm min}$ using one, two, or three exposures, respectively. This strategy boosts spectroscopic efficiency specifically at the faint end of the LBG sample. The total density of spectroscopically confirmed LBGs, combining the target-selection and spectroscopic efficiencies through \eqreff{eq:cumulative}, is shown in  \figreff{fig:lbg_efficiency}. 

\begin{figure}
    \centering
    \includegraphics[width=\linewidth]{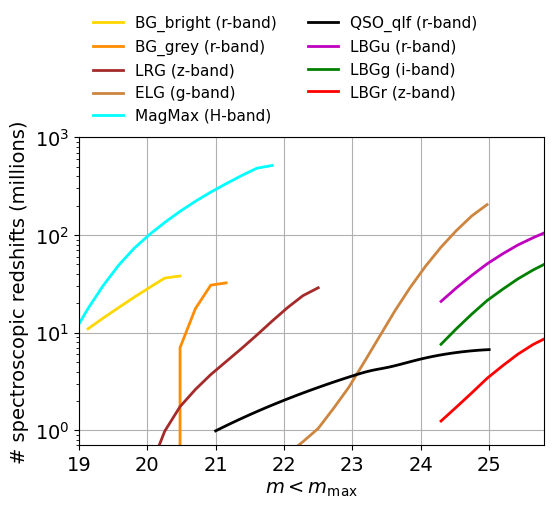}
    \caption{Number of spectroscopic redshifts per tracer. This number is evaluated accounting for the target density, the survey volume, and the total efficiency (purity and SSR). }
    \label{fig:number_of_spectroscopic}
\end{figure}

Finally, accounting for the purity and spectroscopic success rate for each tracer, \figreff{fig:number_of_spectroscopic} displays the number of spectroscopic redshifts that can be securely measured with WST observations across the different target classes. 

\subsection{Survey time and target completeness}
\label{sec:survey_time_completeness}

The telescope time allocation will define the overall completeness of the target catalogue, i.e., some targets might not be observed due to the lack of observing time or due to fibre assignment competition. 
The total survey fibre time, $T_{\rm fibre}$, required for a given set of tracers selected up to a maximum magnitude $m_{\rm max}^{\rm tracer}$, is given by~\citep{Doumerg2023forecast}
\begin{equation}
    T_{\rm fibre}=\frac{4\pi \,f_{\rm sky}}{N_{\rm fib}}  \sum_{{\rm tracer}=i}\int\limits_{-\infty}^{m_{\rm max}^{{\rm tracer},i}}\int\limits_{0}^{+\infty}{\rm d}z\, {\rm d}m \; t_{\rm exp}^{{\rm tracer},i}(m)\;\phi^{{\rm TS},i}(m,z),
\end{equation}
where $t^{\rm exp}_{\rm tracer}(m)$ is the exposure time required for a given tracer\footnote{We note that, in general, $t^{\rm exp}_{\rm tracer}(m)=\langle n_{\rm pass}(m)\rangle\times t_0$, where $t_0$ is the unitary exposure time, e.g., 1000~seconds.} at magnitude $m$ (with $m$ corresponding to the relevant limiting-magnitude band for that tracer), and $\phi^{\rm tracer}(m)$ is the corresponding per-tracer surface density per unit magnitude.

Because each fibre reaches its own sky area, a spectrograph cannot observe tightly packed targets simultaneously, leading to fibre incompleteness~\citep{pinon_mitigation_2025}. Consequently, completing a survey at a fixed target density requires multiple passes over the same FoV. The required number of passes scales directly with target density, as denser samples demand more observation epochs to achieve full completeness. 

\begin{figure}
    \centering
    \includegraphics[width=\linewidth]{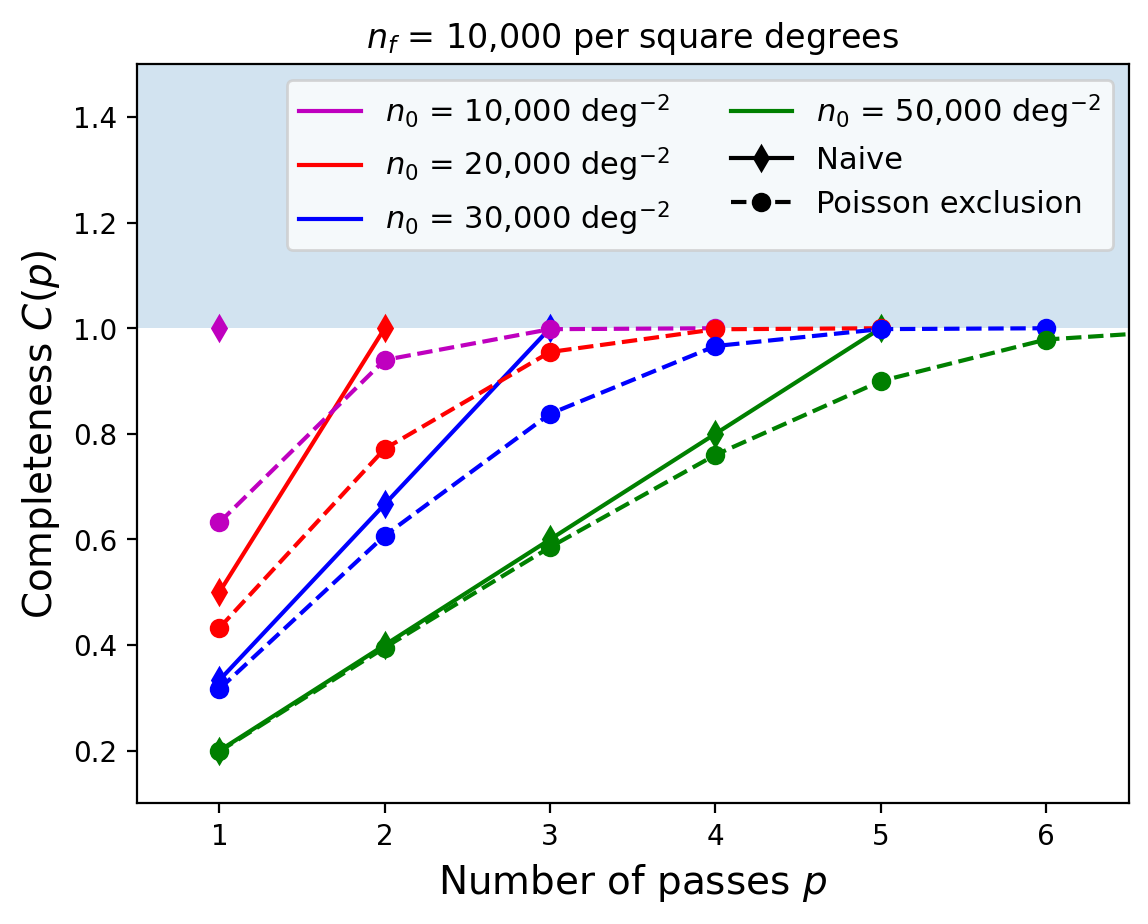}
    \caption{Fiber completeness as a function of the number of passes, for different values of the target density between 10,000 and 50,000 targets/\persqdegrees. The initial number of fibres considered is $n_f=10,000$/\persqdegrees based on the WST design. The lines with diamond markers represent the naive case, where all fibres are filled with a target, and the dashed lines with circles show the results using the Poisson exclusion model. }
\label{fig:completeness_model_no_priority}
\end{figure}

Following Poisson statistics, we adopt a fibre-completeness model that gives the number of targets assigned with a fibre, $n_{\rm assigned}(p)$, as a function of the number of passes ($p$) for a given target density $n_f$:
\begin{equation}
    n_{\rm assigned}^{(p=1)} = n_{f}\left(1 - \exp\Big[-\frac{n_{\rm pointings}}{n_f}\Big]\right).
\end{equation}
After $p+1$ passes, the number of remaining unobserved target, $n_{\rm not\, assigned}^{(p+1)}$, follows the recursive relation
\begin{align}
n_{\rm assigned}^{(p+1)} &= n_f \left( 1 - \exp\Big[-\frac{n_{\rm not\, assigned}^{(p)}}{n_f}\Big] \right).
\end{align}
From the above equations, we can evaluate the completeness $C(p|n)$ of a given field after $p$ passes as
\begin{align}
    C(p|n_{\rm pointings}) = 1 - \frac{n_{\rm not\, assigned}^{(p)}}{n_{\rm pointings}}.
\end{align}
\figreff{fig:completeness_model_no_priority} shows an example of the evolution of the completeness as a function of the number of passes, compared to a simplistic linear case with no collision, where all fibres are assigned a target at each pass with a completeness given by $C(p) = 1-p/p_{\rm max}$, where the maximum number of passes is the ratio of the number of target pointings $n_{\rm pointings}$ divided by the number of fibres $n_f$. The survey time after $p$ passes is given by\footnote{We ignore here that fields will always need a bit of overlap to tile the sky. We also consider no overhead (time spent on each exposure that is not spent collecting science light. It is added on top of the exposure time itself (slewing the telescope to the next field and settling, reading out the detectors, etc.)}
\begin{equation}
    T(p) = p\times t_{\rm exp}\frac{\Omega}{\mathrm{FoV}}
\end{equation}
The ``total'' survey time $T_{\rm pass}$ is obtained by considering a completeness threshold $C > C_{\rm min}$\footnote{to be chosen to be 95$\%$ (almost all the targets are observed)} such as $ T_{\rm pass} = T(p^*)$ where $
p^\star\equiv \{C(p^\star)= C_{\rm min}\}$. This Poisson approximation provides a simple estimate of the number of passes required to reach a specified target completeness; more detailed fibre-assignment simulations, including the geometrical constraints of real positioners, can yield different completeness values (e.g. \citealt{Saunders2014Poissoncompleteness}). We have in general $T_{\rm pass} > T_{\rm fibre}$, since $T_{\rm fibre}$ optimistically considers that all fibres are always filled with a target, regardless of Poisson exclusion. 
Finally, the total calendar survey time, $T_{\rm calendar}$, is related to $T_{\rm pass}$ through
\begin{equation*}
    f_{\rm obs, \ nights}\times f_{\rm good,\, obs,\, conditions}\times T_{\rm calendar} = T_{\rm pass},
\end{equation*}
where $ f_{\rm obs, \ nights}$ and $f_{\rm good,\, obs,\, conditions}$ account, respectively, for the fraction of observable nights and the fraction of nights with favourable observing conditions. 

\subsection{Methodology for WST cosmological forecasts}
\label{sec:forecast_methodology}

\begin{figure}
    \centering
    \includegraphics[width=\linewidth]{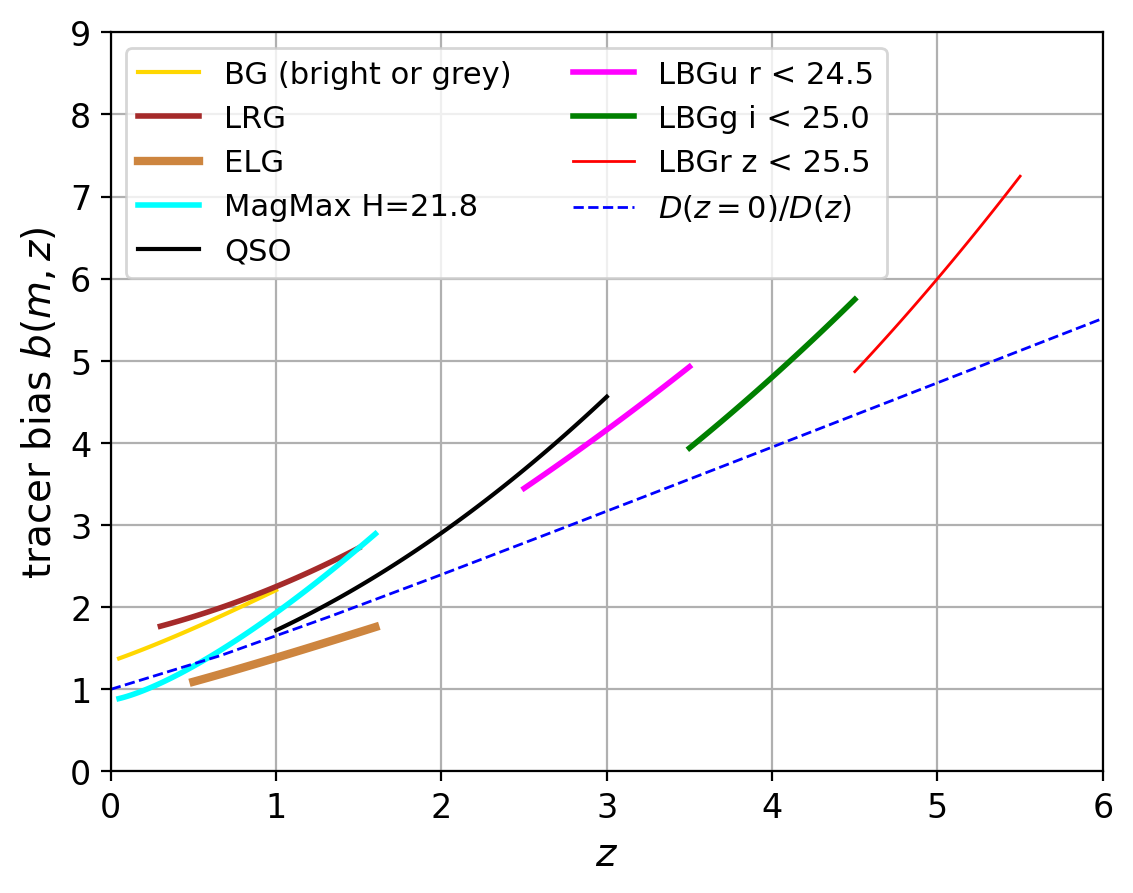}
    \caption{Linear galaxy/QSO bias as a function of redshift for the different tracers we consider in this work.}
    \label{fig:summary_bias}
\end{figure}

\paragraph*{Fisher matrix:} Among the various low- and high-redshift tracers considered in this work, our primary cosmological probe is the redshift-space galaxy power spectrum, $P_{\rm gg}(k,z)$, from which cosmological information is extracted. To assess the constraining power of the different analysis configurations, we compute the Fisher matrix element $F_{\alpha\beta}$~\citep{FISHER} for a set of cosmological parameters $\{\theta_\alpha\}_{1\leq \alpha \leq N}$, given by
\begin{align}
    F_{\alpha\beta} = \frac{V_{\rm sur}}{4\pi^2} \int &{ \rm d}k\, k^2 \times \\\int\limits_{-1}^{1} &\frac{{ \rm d}\mu}{2}
\left(\frac{P_{\rm gg}(k,\mu)}{P_{\rm gg}(k,\mu) +1/\bar{n}(z)}\right)^2 \frac{\partial\ln P_{\rm gg}}{\partial \theta_\alpha}\, \frac{\partial\ln P_{\rm gg}}{\partial \theta_\beta}.
\end{align}
The corresponding parameter variance matrix is bounded from below by the Cramér--Rao inequality \citep{CRRau} $\mathrm{\sigma}(\theta_\alpha) \geq [F^{-1}]^{1/2}_{\alpha\alpha}$. On linear scales, the galaxy power spectrum is related to the underlying matter power spectrum through the linear galaxy bias. This Fisher-matrix formalism provides a direct link between the cosmological observables and the achievable precision on a given set of cosmological parameters, allowing us to explore the constraining power on cosmology of the different tracer samples, spanning a wide range in redshift, number density, and bias. We consider $k_{\min}$ to correspond to the maximum length scale, while $k_{\max}$ depends on the accuracy of the theoretical model on non-linear scales \citep{Doumerg2023forecast}. We take by default
\begin{equation}
    k_{\min} = \frac{2\pi}{V^{1/3}}
    \; [h\,\mathrm{Mpc}^{-1}],
    \qquad
    k_{\max} = 0.1
    \; h\,\mathrm{Mpc}^{-1}.
\end{equation}
and
\begin{equation}
V_{\mathrm{sur}}(z,\Delta z,f_{\rm sky})
=
4\pi f_{\rm sky}
\int_{z-\Delta z/2}^{z+\Delta z/2}
\frac{c\,\chi^2(z')}{H(z')}\,\mathrm{d}z'
\end{equation}
\paragraph*{Galaxy bias:} For the BG samples, we adopt $b_{\rm BG}(m,z) = 1.34/D(z)$ \citep{DESIreport2016}, for LRGs we use $b_{\rm LRG}(m,z) = 0.209\times (1 + z)^2 + 1.415$ \citep{Chaussidon2025fnldesiqsolrg}, for ELGs, we use $b_{\rm ELG}(z) = 0.84/D(z)$ \citep{Mostek2013elgbias}\footnote{Let us note that a slightly smaller ELG bias of $0.76/D(z)$ was found in \citet{Rocher2023hod}.}, for MagMax, we use $b_{\rm MagMax}(m,z) 
= 1.1 + 0.25\,(27.8-m_H)\,z^{0.9}.$ based on clustering analysis of $H$-selected samples in the \textit{Euclid} Flagship simulation \citep{Castander2025simulation}. Details on the bias estimation for the MagMax sample are provided in \appreff{app:magbias}, for QSOs we use $b_{\rm QSO}(z) = 0.237 \times(1 + z)^2 + 0.771$ \citep{Chaussidon2025fnldesiqsolrg}, for the $u$-, $g$-, and $r$-dropout LBG samples we use the redshift-magnitude dependent prescription of \citet{WilsonWhite2019dropout}, giving $b_{\rm LBG}(<m,z)$ (see their Eq. (2.7)). \figreff{fig:summary_bias} summarises the bias models that we use in this paper as a function of redshift. Let's note that the MagMax and LBG bias models are also magnitude-dependent, so we specify the chosen magnitude value.

\paragraph*{The BAO case:} We consider two forecasting approaches. The first follows the standard approach of \cite{seo_improved_2007}. In this method, the BAO signal is assumed to be exponentially damped, with fixed parameters, and the AP parameters (\eqreff{eq:AP}) are allowed to vary, effectively shifting the BAO template (cf. \appreff{app:forecast_details}). A scale-independent galaxy bias is assumed, as well as linear redshift-space distortions. To account approximately for the effects of reconstruction, the damping parameters are reduced by a factor of 0.5. The main advantage of this approach is its computational efficiency.
A more realistic method is implemented in \cite{FISHLSS}, which accounts for non-linear effects and explicitly models the reconstruction procedure. The corresponding Fisher matrix can be computed using the \texttt{FishLSS} package. The authors found that the resulting uncertainties on the AP parameters agree with those from the standard approach at the $10-15\%$ level (we also test this in \appreff{app:forecast_details}), which is sufficient for qualitative forecasting. Since the standard approach is significantly faster, we use it as our default method for running multiple forecasts and exploring different magnitude cuts and tracer fractions. For the final quantitative forecasts of the ($w_0-w_a$) constraints, however, we will use the more realistic approach implemented in \cite{FISHLSS}. Details on the $\{F_{\alpha_\parallel,\,\alpha_\perp}(z_i)\}_i \rightarrow F_{w_0,\,w_a}$ procedure are reported in \appreff{app:forecast_details}

\paragraph*{The growth case:} We also consider two approaches. The first is a fixed-template method, in which the matter power spectrum at ($z=0$) is fixed, while ($f\sigma_8$) and ($b\sigma_8$) are allowed to vary according to \eqreff{eq:linearrsd} (se \appreff{app:forecast_details_growth})\footnote{A natural extension of this work would be to forecast $f\sigma_{12}$ instead of $f\sigma_8$, where $\sigma_{12}$ is the rms matter fluctuation in spheres of radius $12\,{\rm Mpc}$, rather than $8\,h^{-1}{\rm Mpc}$ \citep{Sanchez2020sigma12}. This choice has several advantages. First, $\sigma_8$ is defined on a scale that depends on $h$, so for different values of $h$ it probes different physical scales. Defining the smoothing scale in ${\rm Mpc}$ removes this dependence on $h$. Second, since $h$ is no longer a hidden nuisance parameter in the definition, the constraints on $f\sigma_{12}$ are less sensitive to the fiducial value of $h$ and to the degeneracy between $h$ and the amplitude of fluctuations. This makes comparisons between tracers, surveys and cosmological models more direct, which is particularly relevant for full-shape analyses where $h$ is not fixed.}. This approach is also computationally very efficient. However, it does not capture the full constraining power of a full-shape analysis extending into the mildly non-linear regime. We therefore also consider the forecasting outputs from \texttt{FishLSS}, which enables us to forecast ($f\sigma_8(z_i)$) using the full shape of the power spectrum, with non-linear effects modelled up to one-loop order. This approach is computationally more expensive, and we therefore only apply it to the optimised WST samples, while using the faster fixed-template procedure for the qualitative forecasts.

\paragraph*{The local-type PNG case:} We adopt the modified universality relation $b_\Phi = 2\,(b(z) - p_\Phi)\,\delta_c$ \citep{Slosar2008}, where $\delta_c$ is the critical density for spherical collapse, $T(k)$ is the matter transfer function, and $D(z)$ is the linear growth factor. The parameter $p_\Phi$ depends on the physical mechanism by which a given tracer population traces the underlying dark matter halo distribution, and is therefore expected to vary between tracer types~\citep{barreira_galaxy_2020}. We adopt $\fnl=0$ as our fiducial value. Since $p_\Phi$ remains poorly characterized for most of the tracers considered here, we assume $p_\Phi=1$ for all high-redshift tracers as a working assumption, except for QSO, where we use 1.6 \citep{Slosar2008}. For the local PNG analysis exclusively, we adopt the same procedure as for RSD, by fixing the template power spectrum and only Fisher-analysing $\fnl$ and the linear bias. So the Fisher matrix is a $2\times 2$ matrix. One important systematic effect, that is not studied here, is the presence of spatial variations in survey depth and Galactic dust—that can induce spurious clustering signal at large scales, the latter being mitigated by removing spurious correlation between imaging features and local density. It was found to be an important budget in the variance of the power spectrum at large scales for photometrically selected tracers \citep{Chaussidon2025fnldesiqsolrg,payerne2026lbgcmblensingunions}. 

\paragraph*{The neutrino mass case:} For this parameter, we rely on a linear Fisher-matrix forecast built directly from the power spectrum. Although this treatment is admittedly simpler than the MCMC-based analyses with non-linear power spectrum corrections typically used in the most recent neutrino-mass studies, our purpose here is not to obtain the tightest possible absolute bound on $M_\nu$, but rather to compare how the achievable precision responds to changes in the observational configuration of the survey. This comparative goal is further supported by the fact that even the most advanced neutrino-mass pipelines currently in use do not converge on a single, consistent modelling approach. We therefore expect any systematic offset between our simplified treatment and a more sophisticated $M_\nu$ modelling to affect all of the survey configurations considered in a similar way, leaving our relative, configuration-to-configuration conclusions essentially unaffected. Our fiducial cosmological model consists of the six standard $\Lambda$CDM parameters supplemented by the total neutrino mass, $\{p_i\}=\{M_\nu,H_0,\omega_c,\omega_b,n_s,A_s\}$. The Fisher matrix is then marginalized over all cosmological parameters (along with the tracer bias) other than $M_\nu$, adopting the Planck priors summarized in \tabreff{tab:prior_fiducial}. The limiting factor in the constraint on $M_\nu$ is the prior used for the amplitude of primordial fluctuations $A_s$ \citep{Byle2021neutrino}. When taken from CMB data, it is intrinsically linked to the precision on the optical depth $\tau$\footnote{As CMB does not constrain $A_s$ directly, but the combination $A_s e^{-2\tau}$.}. Although we consider the \textit{Planck} prior on $\tau$ from \citet{Planck2020cosmofull}, let us note that any improvement on $\tau$ (for instance from future large-scale CMB polarisation or 21 cm measurements, see e.g., \citealt{Liu2016tau}) translates directly into a tighter prior on $A_s$. This matters for the neutrino mass because the suppression of the small-scale power spectrum by massive neutrinos is measured relative to the overall amplitude of the spectrum.

\section{Application to the WST Cosmology survey}
\label{sec:forecast}

\subsection{Completeness and duration of the different WST Cosmology surveys}

\begin{figure*}
    \centering
    \includegraphics[width=0.47\linewidth]{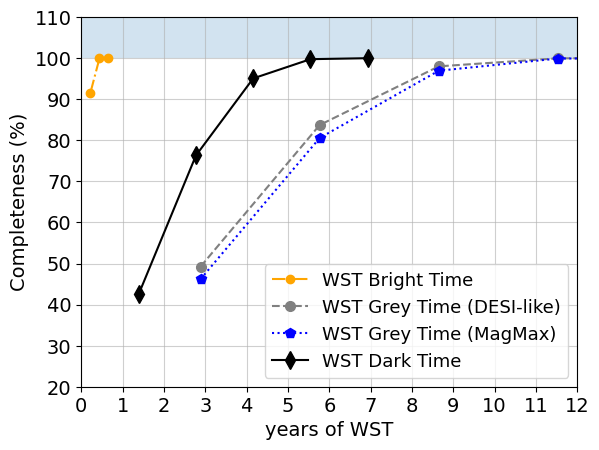}
    \includegraphics[width=0.47\linewidth]{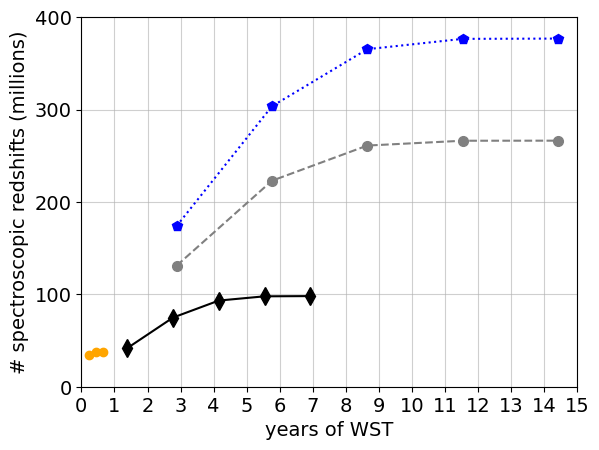}
    \caption{Left: Completeness of the different WST sub-surveys as a function of elapsed time since the start of the WST survey. Two configurations are shown for the Grey Time survey, corresponding to either the DESI-like tracers or the MagMax sample. Each marker corresponds to one more full coverage of the WST footprint in the corresponding WST Bright/Grey/Dark Time. Right: Number of spectroscopic redshifts associated with each WST Bright/Grey/Dark Time since the start of the WST survey.}
    \label{fig:completeness_wst}
\end{figure*}

\begin{figure}
    \centering
    \includegraphics[width=1\linewidth]{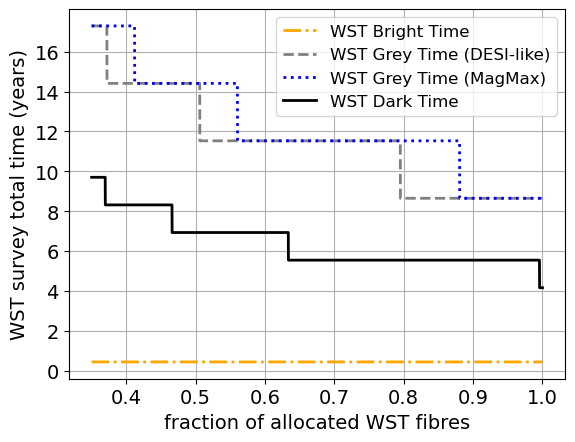}
    \caption{WST Cosmology Bright, Grey and Dark Time total duration as a function of the fraction of allocated fibres for the cosmology program.}
    \label{fig:wst_fibre_allocation}
\end{figure}

We define three distinct WST sub-surveys, corresponding to the three main observing-condition categories: Bright Time (composed of the BG-Bright sample), Grey Time (composed of either the DESI-like samples: BG-Faint, LRG, and ELG samples, or the MagMax sample) and Dark Time (composed of the QSO sample together with the three LBG samples). We adopt the baseline limiting magnitudes listed in \tabreff{tab:summary_params_survey_design}; we apply the formalism presented in \secreff{sec:survey_time_completeness} to compute, for each sub-survey, the number of passes required per field of view to assign a fibre to every target, and the corresponding total sub-survey time (using default values in \tabreff{tab:wst_observing_baseline}). The left panel \figreff{fig:completeness_wst} shows the resulting completeness of the WST Bright, Grey, and Dark Time surveys as a function of elapsed survey time. We find that the Bright Time survey reaches $>95\%$ completeness within less than a year. The Grey Time survey (either DESI-like or MagMax) targetting a total of $\sim 22,000$ galaxies, requires approximately nine years for completion. The Dark Time survey is completed in approximately five years. The right panel of \figreff{fig:completeness_wst} displays the number of spectroscopic redshifts over time, achieving $\rm 40$ million redshifts of BG-Bright, $\sim 400$ million MagMax redshifts (in Grey Time) or $\sim 250$ million DESI-like low-redshift (BG-Grey, LRG, ELG in Grey Time), and $\sim 100$ million redshifts of intermediate- and high-redshift tracers (combination of QSOs and LBGs). The WST will allocate fibres to different scientific programs, not only large-scale structure surveys for cosmological applications.  The \figreff{fig:wst_fibre_allocation} shows the total time of the WST Cosmology Bright, Grey and Dark surveys\footnote{Computed as the time necessary to reach a completeness higher than $95\%$.} as a function of the fraction of WST fibres allocated to cosmology. Intuitively, if half the fibres are allocated for cosmology, WST sub-survey times (except the Bright Time) are multiplied approximately by a factor of two; the Grey Time WST cosmology survey with MagMax is completed in less than 13 years, with DESI-like tracers in 9 years, and the Dark Time in 8 years. 

\subsection{ Cosmology with DESI-like tracers at low and intermediate redshifts}

At redshifts $z<1.6$, BG, LRG, and ELG samples have been the main tracers of low-redshift large-scale structure, through RSD and BAO measurements from BOSS/eBOSS~\citep{Alam2021sdsseboss} and DESI~\citep{desi2025baodr2}. Alternatively, the low-redshift Universe can be probed with magnitude-limited galaxy samples, labelled as MagMax. At intermediate redshifts, QSOs extend clustering measurements to $z=2.1$, and to even higher redshifts through QSO Ly$\alpha$ forests.
We first explore how improved telescope capabilities and access to fainter galaxies improve constraints on the growth of structure and dark energy using DESI-like tracers. We then investigate the potential of the MagMax sample as an alternative probe of cosmology.

\subsubsection{Impact of telescope capability: from DESI to WST}

\subsubsection*{Growth of structures via RSD}
\begin{figure*}
    \centering
    \includegraphics[width=0.45\linewidth]{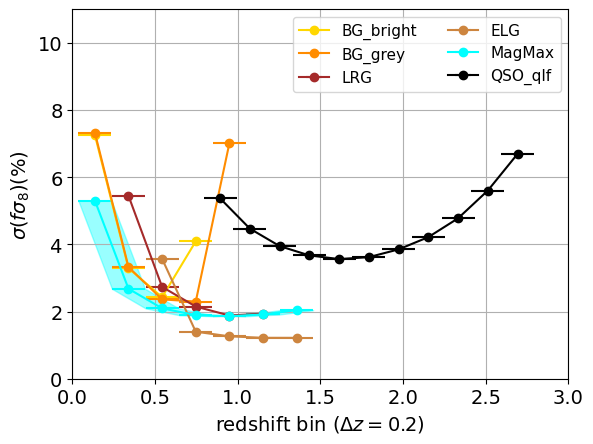}
    \includegraphics[width=0.45\linewidth]{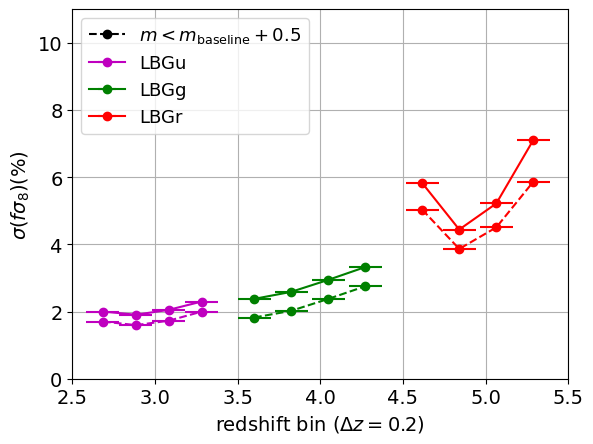}
    \caption{Left: Baseline uncertainty on the RSD parameter $f\sigma_8$ for the DESI-like tracers (BG, LRG, ELG, and QSO) and the MagMax sample per redshift bin of $\Delta z=0.2$, for the baseline maximum magnitude cuts listed in \tabreff{tab:summary_params_survey_design}. Right: Same for the LBG samples. The points connected by dashed lines are for samples with fainter magnitude limits.}
    \label{fig:rsd_per_bin}
\end{figure*}
\begin{figure*}
    \centering
     \includegraphics[width=0.45\linewidth]{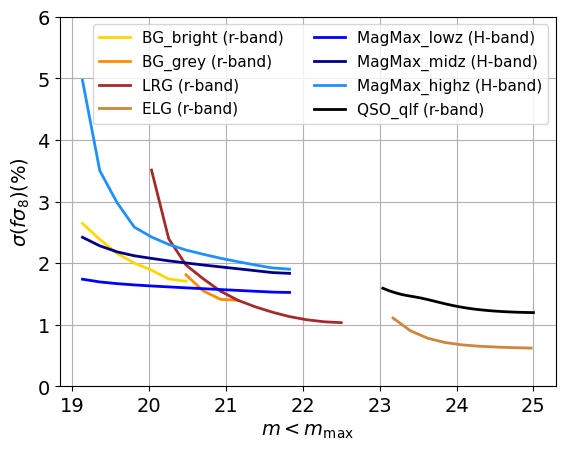}
     \includegraphics[width=0.45\linewidth]{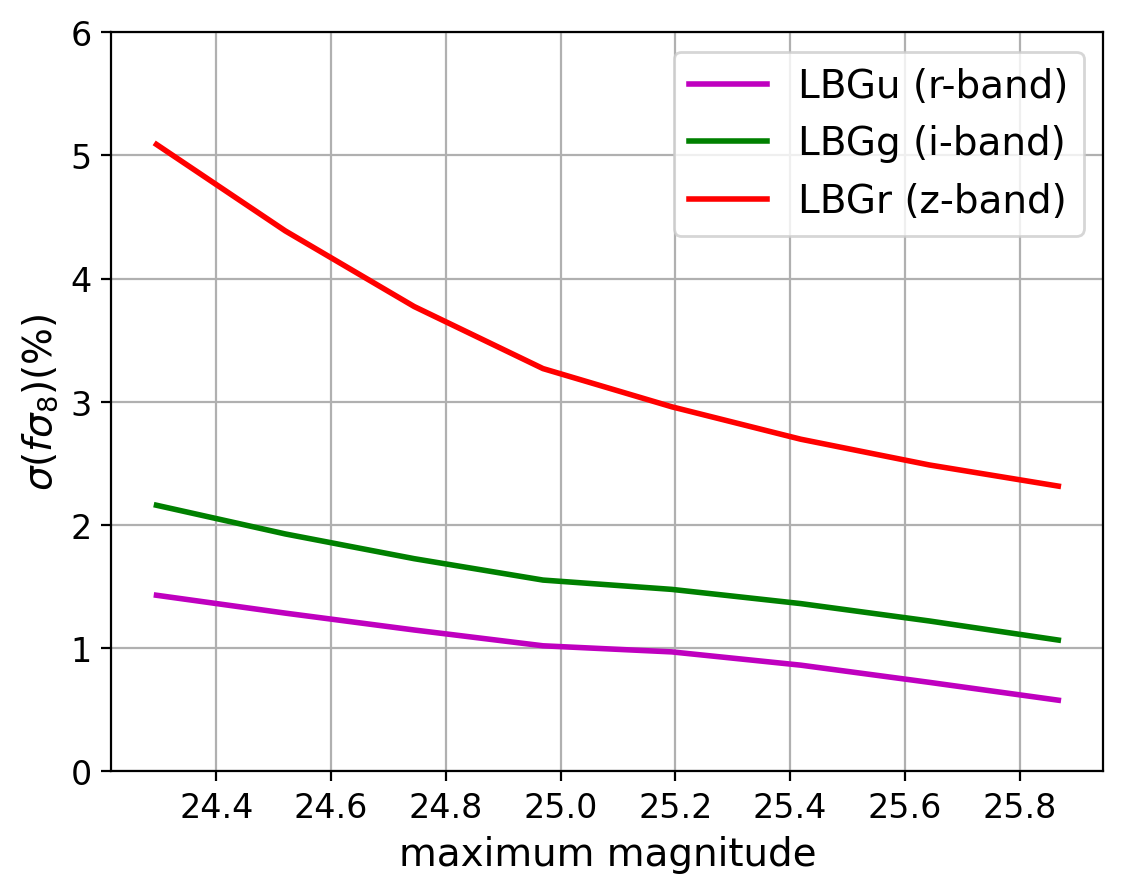}
    \caption{Left: Combined error on the RSD parameter $f\sigma_8$ at the effective redshift of each tracer (see \tabreff{tab:summary_params_survey_design}), for the DESI-like tracers and the MagMax sample. The MagMax sample is split into three redshift bins to isolate its redshift evolution. Right: same for LBG samples. }
    \label{fig:RSD_params_BG_ELG_LRG}
\end{figure*}

The precision on $f(z_{\rm eff})\sigma_8(z_{\rm eff})$, marginalized over $b(z_{\rm eff})\sigma_8(z_{\rm eff})$ (see \appreff{app:forecast_details_growth}), is shown in the left panel of \figreff{fig:rsd_per_bin} for redshift bins of $\Delta z=0.2$. The constraints are combined over the full redshift range of each DESI-like tracer and shown as a function of the limiting magnitude in the left panel of \figreff{fig:RSD_params_BG_ELG_LRG}. The full set of WST DESI-like tracers achieves better than $10\%$ precision on $f(z_{\rm eff})\sigma_8(z_{\rm eff})$ over the full redshift range probed.
As shown in \figreff{fig:RSD_params_BG_ELG_LRG}, going to fainter magnitudes provides only limited gains in the growth constraints. The constraints saturate for all tracers except QSOs. This indicates that, for these standard tracers and within the linear regime, increasing the number density provides little additional information once the measurements approach the cosmic-variance limit. The DESI DR1 ShapeFit-only normalized $f\sigma_8$ measurements \citep{Adame2025fullshapeRSD} yield the following relative uncertainties: BGS, $20\%$ over $0.1 < z < 0.4$; LRG1, $13\%$ over $0.4 < z < 0.6$; LRG2, $12\%$ over $0.6 < z < 0.8$; LRG3, $10.5\%$ over $0.8 < z < 1.1$; ELG2, $9.5\%$ over $1.1 < z < 1.6$; and QSO, $12\%$ over $0.8 < z < 2.1$. Overall, the DESI DR1 relative uncertainties range from $9.5\%$ to $20\%$, with ELG2 providing the smallest relative uncertainty and BGS the largest.

\subsubsection*{Baryon Acoustic Oscillations and Dark Energy}
\begin{figure*}
    \centering
    \includegraphics[width=0.45\linewidth]{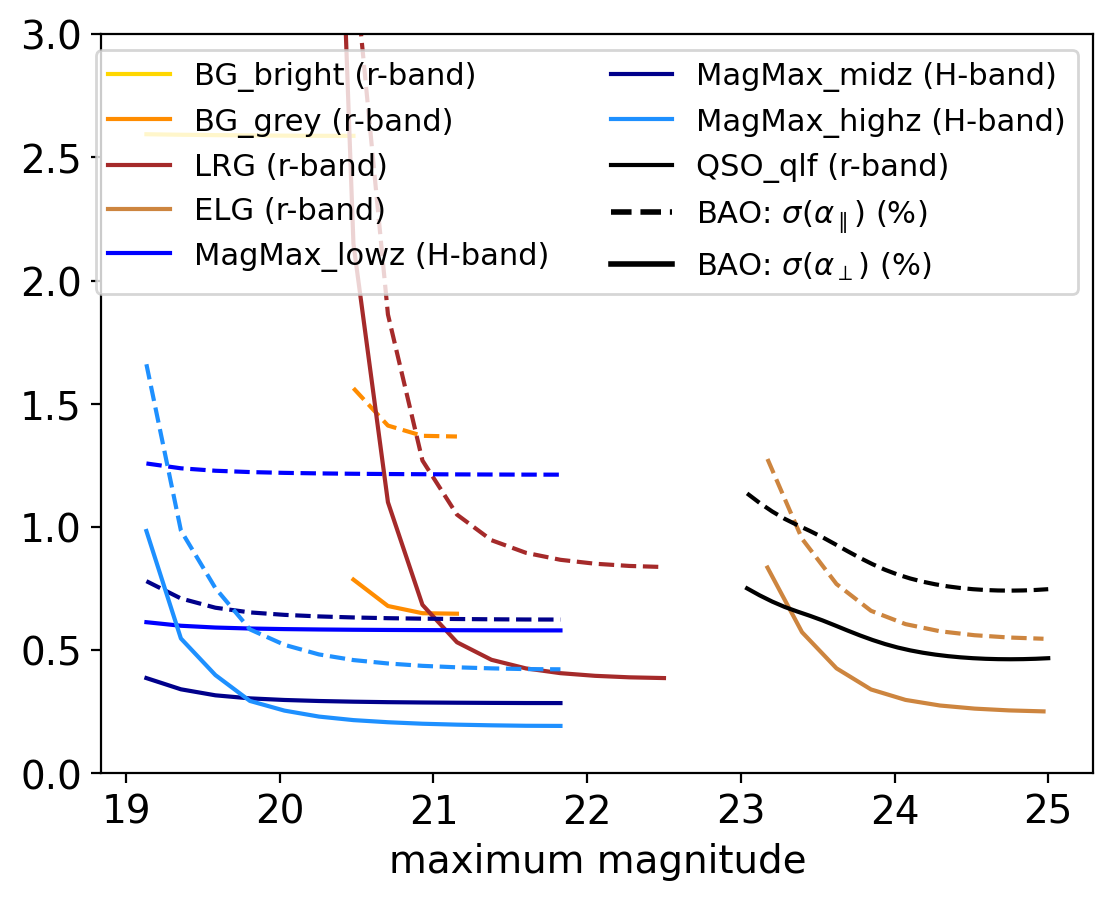}
    \includegraphics[width=0.45\linewidth]{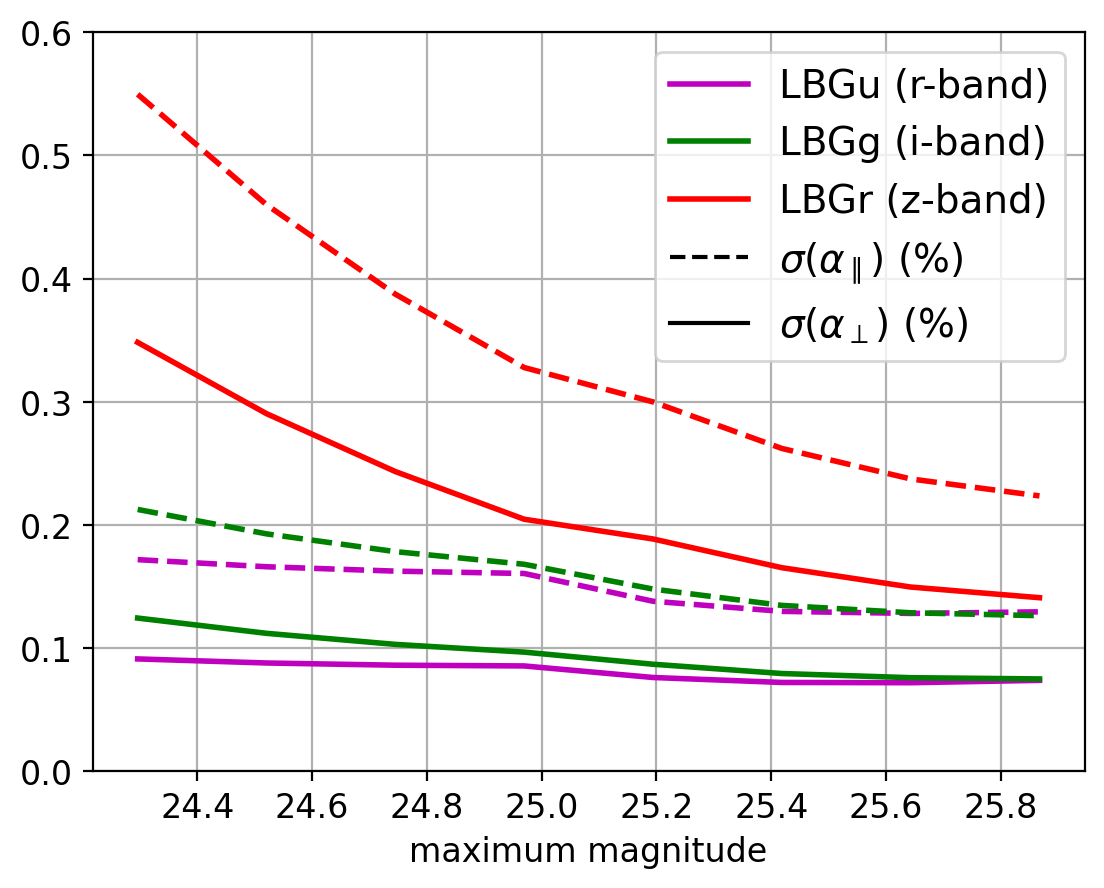}
    \caption{Left: BAO constraints ($D_A$ and $H$) from (i) DESI-like tracers in their specific redshift range (BG, LRG, ELG, QSO)  and (ii) alternative MagMax sample as a function of their respective maximum magnitude. Right: BAO constraints ($D_A$ and $H$) for the 3 LBG samples as a function of their respective maximum magnitude.}
    \label{fig:BAO_params_combined}
\end{figure*}

\begin{figure*}
    \centering
    \includegraphics[width=0.9\linewidth]{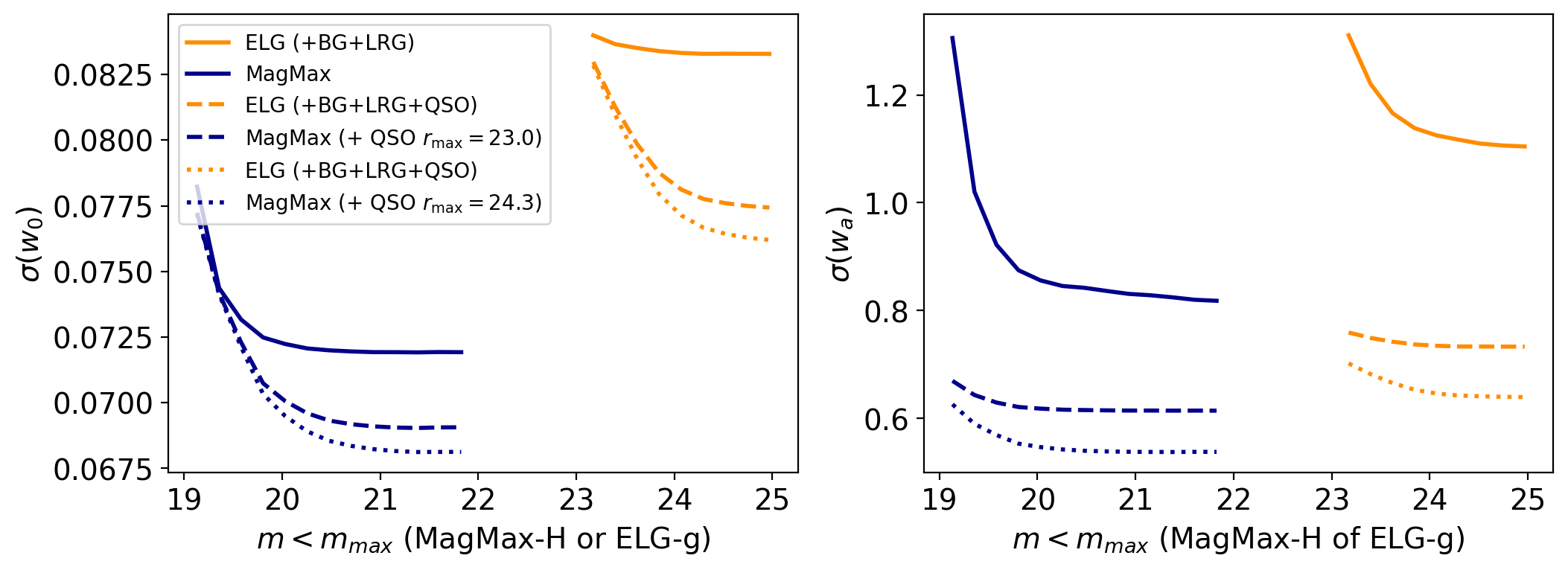}
    \caption{Forecasts on $w_0w_a$ cosmology from the combination of DESI-like tracers, or MagMax.}
    \label{fig:BAOwowa_params_BG_ELG_LRG}
\end{figure*}

The power spectrum of cosmological tracers can be used to extract the BAO features of the matter density field. For each DESI-like sample (BG, LRG, ELG, and QSO), the left panel of \figreff{fig:BAO_params_combined} shows the relative uncertainty 
DESI measurements are not fully cosmic-variance limited. Increasing the DESI-like tracers' number density by extending the selection to fainter magnitudes therefore improves the BAO constraints.

The BAO scale can be used to constrain the properties of dark energy (see details in \appreff{app:forecast_details}), in particular the evolving dark-energy equation-of-state parameters $w_0$ and $w_a$, which are among the primary cosmological targets of Stage-V surveys, given the recent hints of dark-energy evolution reported by DESI~\citep{desi2025baodr2}. To this end, we combine all DESI-like tracers by assigning them adjacent, non-overlapping redshift intervals centred on their effective redshifts $z_{\rm eff}$. We consider $0<z< 0.4$ for the BG-bright sample, $0.4 < z < 0.7$ for the BG-faint sample, $0.7 <z < 1.1$ for the LRG sample, and $1.1 < z<1.6$ for the ELG sample.

This allows us to extract the BAO feature from separate cosmological volumes and sum their respective $w_0-w_a-\Omega_m-H_0$ Fisher matrices (see \appreff{app:forecast_details}).
In \figreff{fig:BAOwowa_params_BG_ELG_LRG}, we show the marginalised uncertainties on $w_0$ and $w_a$, marginalising over $\Omega_m$\footnote{We use an uninformative uniform prior on $\Omega_m$.}, when combining all DESI-like tracers, with and without QSOs, as a function of the maximum ELG magnitude. We vary the ELG magnitude since this is the sample for which the constraints do not saturate.
For the maximum ELG magnitude corresponding to the DESI-to-WST baseline, we obtain $\sigma(w_0)=0.0825$ ($0.0775$) and $\sigma(w_a)=1.1$ ($0.6$) when combining the low-redshift tracers alone (when adding QSOs).

\subsubsection{WST: DESI-like tracers vs MagMax}

A central question for future Stage-V facilities is whether the DESI-like colour-selection strategy remains optimal as the instrument's depth and multiplexing increase substantially, or whether a simpler selection can achieve comparable or better performance. We address this question by comparing the cosmological constraining power of the DESI-like tracers, rescaled to the WST depth, with that of the MagMax sample. The latter requires no tracer-specific colour cuts, relying only on star-separation cuts; see \secreff{ts:desi-like} and \secreff{ts:magmax}.

Adopting the target-selection strategy described for the MagMax sample, the left panel of \figreff{fig:rsd_per_bin} shows the per-redshift-bin constraints on $f(z_{\rm eff})\sigma_8(z_{\rm eff})$ for a MagMax target density approximately matching the combined density of the BG, LRG, and ELG samples. The corresponding constraints combined over the redshift range of each sample are shown in \figreff{fig:RSD_params_BG_ELG_LRG}. We split the MagMax sample into three redshift bins (MagMax$\_$lowz: $0 < z < 0.7$, MagMax$\_$midz: $0.7< z <  1.1$, and MagMax$\_$highz: $1.1< z <  1.6$), chosen such that their effective redshifts approximately match those of the BG, LRG, and ELG samples.
At low redshift ($z\lesssim0.5$), MagMax provides significantly tighter constraints than BG and LRG, with improvements of up to $30\%$, due to a lower galaxy bias value which boosts the RSD contrast term $\beta=f/b$. At higher redshift, however, the intrinsically bright MagMax selection results in a higher galaxy bias, reducing the relative amplitude of the RSD contrast. Consequently, despite having comparable target densities, MagMax provides weaker constraints than ELG at $z\sim1$. The ELG sample, with its lower galaxy bias, is forecast to provide constraints on $f\sigma_8$ that are approximately $30\%$ tighter than those from MagMax in this redshift range. It is worth noting that we neglect the magnitude dependence of the ELG, lRG, and BG galaxy biases in these forecasts. Accounting for this dependence could further improve the constraints obtained with the DESI-like tracers for WST.
In \figreff{fig:rsd_fullshape}, we repeat the analysis using the full-shape approach and show the relative constraints on $\sigma_8$ as a function of redshift. The two results are in excellent agreement; the MagMax sample provides tighter constraints than the DESI-like tracers at $z<1$, while the DESI-like tracers yield tighter constraints at $1<z<1.6$.

We repeat the same exercise for the BAO parameters, shown in the left panel of \figreff{fig:BAO_params_combined} as a function of the limiting magnitude of the three MagMax sub-samples. We then compare the corresponding $w_0-w_a$ constraints in \figreff{fig:BAOwowa_params_BG_ELG_LRG}, both with and without QSOs. We obtain $\sigma(w_0)=0.0725$ ($0.0675$) and $\sigma(w_a)=0.8$ ($0.5$) when combining the low-redshift MagMax samples alone (when adding QSO). Overall, MagMax yields approximately $10\%$ tighter $w_0-w_a$ constraints than the DESI-like tracer combination.
On the left panel of \figreff{fig:wowa_money}, we report the constraints on $\alpha_\parallel$ as a function of redshift for the full WST redshift range, using MagMax at low-$z$, compared with the cosmic variance (CV) limit. We find that the measurements are fully CV-limited up to $z=1.6$. In the range $1.6<z<2.5$, we only consider BAO measurements from QSO clustering, resulting in substantially weaker constraints. At higher redshifts, the constraints from LBG clustering approach the CV limit, although a gap remains between the two. In the range $2<z<2.5$, the inclusion of the Lyman-$\alpha$ forest BAO measurements from QSOs would further improve the constraints.
On the right panel of \figreff{fig:wowa_money}, we show the corresponding $1\sigma$ contours in the $w_0$-$w_a$ plane. Both the low-$z$ and high-$z$ probes contribute significantly to the constraints on DDE. In particular, including the high-$z$ measurements improves the uncertainty on $w_a$ by a factor of $\sim2$, and that on $w_0$ by a factor of $\sim1.3$, highlighting the gain with high-$z$ observations on DDE. 

The key result is that the MagMax sample, despite requiring no dedicated colour selection, can achieve constraining power comparable to the combined DESI-like tracers when its limiting magnitude is chosen to reach a similar total target density. This shows that, within our forecasting assumptions, a purely magnitude-limited sample can reproduce the RSD and BAO performance of a multi-tracer colour selection at low redshift.
This result has practical implications for survey design. MagMax requires no tracer-specific colour cuts, reducing the complexity of the target-selection strategy and its calibration. It also replaces three independently tuned tracer selections with a single, density-driven selection parameter. This makes the approach simpler and more transparent, while retaining comparable cosmological constraining power in our forecasts. In particular, it suggests that a magnitude-limited strategy could provide a viable alternative to conventional multi-tracer colour selections for future Stage-V surveys.

\subsection{Cosmology with high-redshift tracers}

As for the low- and intermediate-redshift tracers, LBGs provide a way to probe the growth of cosmic structure and the expansion history of the Universe in the largely unexplored regime $z>2$ using galaxies as tracers of the density field.  While current surveys such as DESI already probe $z>2$ through quasars and the Ly$\alpha$ forest, LBGs offer a complementary galaxy-based tracer at these redshifts. 

Stage-V spectroscopic surveys offer the opportunity to target the large population of faint LBGs at high redshift. WST could obtain millions of LBG redshifts, enabling their use as a high-redshift cosmological probe. Beyond measurements of dark energy, their large-scale clustering could also provide sensitivity to primordial physics and inflation.

\subsubsection{Impact of LBG maximum magnitude}
\label{sec:forecast_Lbg_mag}
We note that the mean LBG bias decreases when magnitude increases, since fainter, lower-mass, and less strongly biased galaxies make up a growing fraction of the sample as the magnitude limit is pushed deeper. This trend partially offsets the reduction in shot noise obtained by including more targets, thereby degrading the net constraining power (although the precise behaviour depends on the specific LBG sample considered, since it differs between the $u$-, $g$-, and $r$-dropout selections). To mitigate this effect, we split each LBG sample into two magnitude bins at $m=25$, and adopt a multi-tracer forecast approach for $m_{\rm max}>25$, following the methodology described in~\citet{Doumerg2023forecast} and ~\citet{Doppler}. 

\subsubsection*{Growth of structures} As for the low- and intermediate redshift tracers, we show in the right panel of \figreff{fig:rsd_per_bin} the LBG constraints per redshift bin on $f(z_{\rm eff})\sigma_8(z_{\rm eff})$, and the combined constraints per LBG sample in the right panel of \figreff{fig:RSD_params_BG_ELG_LRG} as a function of the corresponding limiting magnitude. Contrary to low-$z$ tracers, we are not saturating for any of the LBG samples, showing that despite WST capabilities, we are still in a shot noise regime. Typically, pushing to magnitude $\sim 25.5$ we get a $50\%$ improvement compared to the 24.5 magnitude limit (sparse densities.

\begin{figure*}
    \centering
    \includegraphics[width=0.95\linewidth]{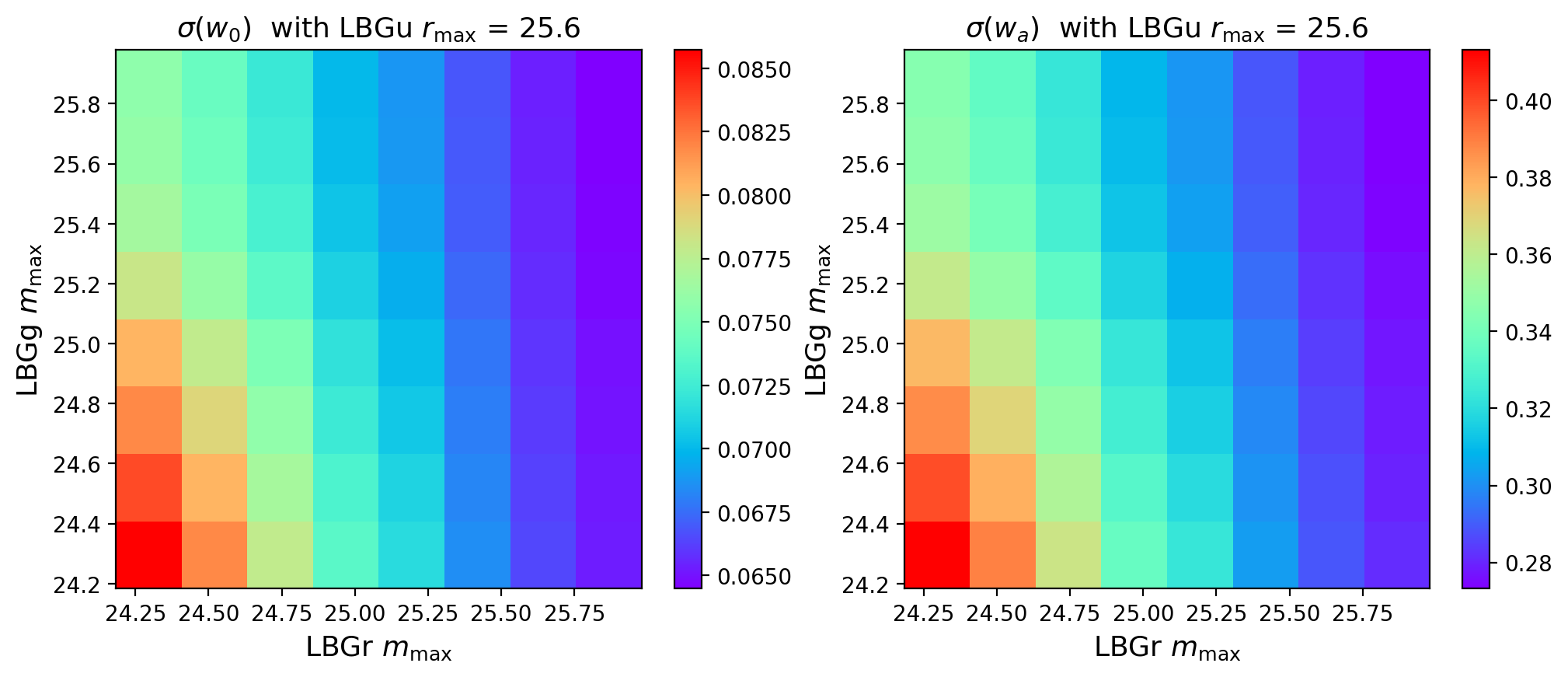}
    \caption{Dark energy $w_0w_a$ constraints from the combination of the three LBG samples. Left: $w_0$ constraints as a function of the maximum magnitude of the $g$- and $r$-dropout LBG sample, for fixed maximum magnitude of the $u$-dropout LBG sample. Right: same as left panel but for $w_a$.}
    \label{fig:wowa_lbg}
\end{figure*}

\begin{figure*}
    \centering
    \includegraphics[width=0.95\linewidth]{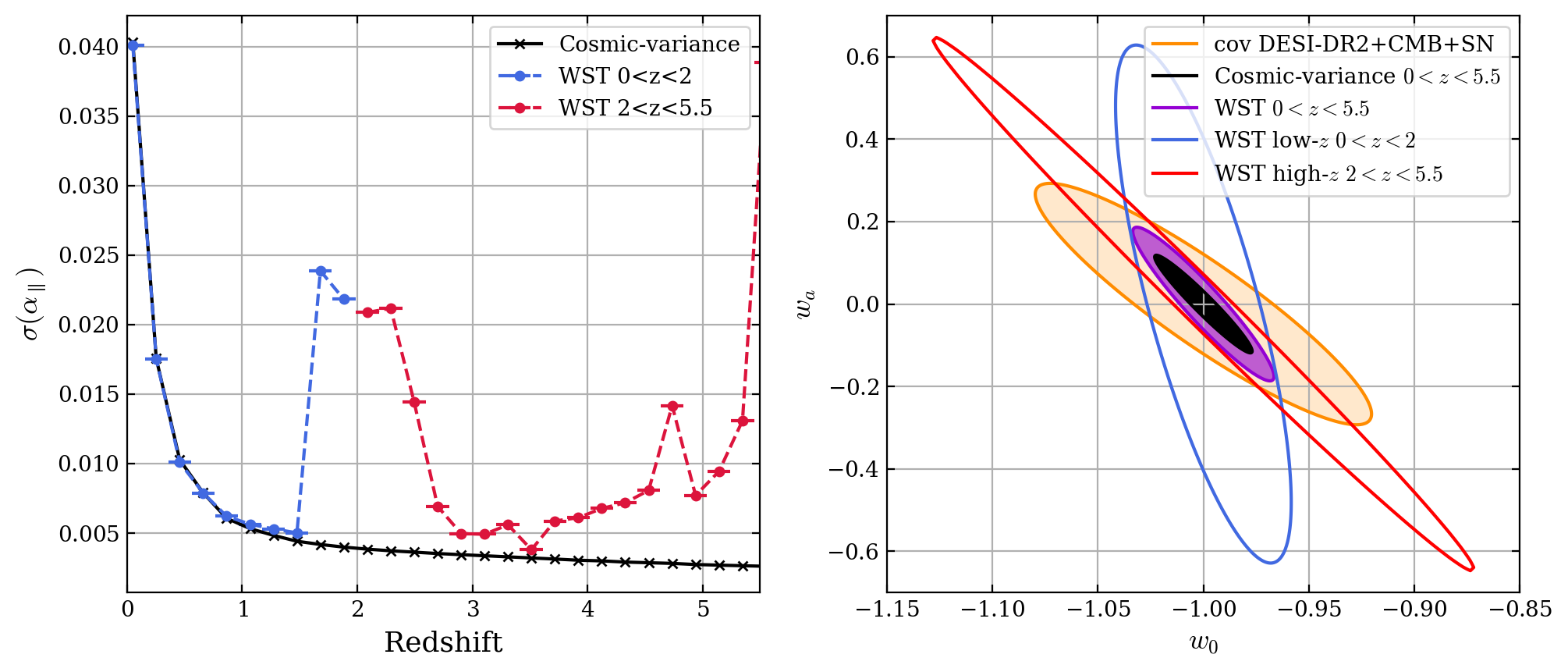}
    \caption{Left: Forecasts on $\alpha_\parallel$ over the full WST redshift range, compared with the cosmic variance limit. Forecasts are evaluated in $\Delta z=0.1$ bins, assuming a survey area of $18,000$ deg$^2$. Right: 1$\sigma$ contours of $w_0$-$w_a$  for a DDE model, centred on the $\Lambda$CDM fiducial point $(w_0,w_a)=(-1,0)$. The orange contour shows the equivalent constraints from the DESI DR2 BAO+CMB+SN measurements (same covariance, but the mean is re-centred on $\Lambda$). The blue and red contours show the constraints from the WST low-$z$ ($z<2$) and high-$z$ ($z>2$) surveys, respectively, while the purple contour corresponds to the full WST redshift range [let us note that $r_d$ is considered perfectly known here]. The black contour shows the cosmic variance limit. The WST DESI-like and MagMax samples yield nearly identical constraints at low-$z$, as the measurement is cosmic-variance limited for $z<1.6$. }
    \label{fig:wowa_money}
\end{figure*}

\subsubsection*{BAO and Dark Energy}
The right panel of \figreff{fig:BAO_params_combined} shows the per-LBG sample constraints on the 2 BAO parameters. The \figreff{fig:wowa_lbg} shows the $w_0$ and $w_a$ precision after combining the 3 LBG samples. The LBG $u$-dropout sample is expected to reach a higher target density than the $g$- and $r$-dropout samples~\citep{Crenshaw2025lbgsellsst}. However, the $g$- and $r$-dropout samples probe higher redshifts and are therefore better suited to anchoring the high-redshift pivot of any redshift-dependent dark energy parametrization. As shown in \figreff{fig:wowa_money}, we find a high degree of complementarity between low- and high-redshift BAO measurements in constraining the $w_0$-$w_a$ parameter space. The constraint on $w_0$ is driven primarily by the low-redshift measurements, which are already close to the cosmic-variance limit, as expected since $w(z=0)=w_0$. At high redshift, the BAO measurements exhibit a strong anti-correlation between $w_0$ and $w_a$, with, for example, $w(z=3)=w_0+0.75w_a$. Combining this high-redshift information with the precise low-redshift constraint on $w_0$ therefore breaks the degeneracy between the two parameters, leading to tight constraints on both $w_0$ and $w_a$. This degeneracy breaking allows the WST measurements to achieve substantially tighter constraints than the combination of current DESI BAO, CMB, and SN data.

\begin{figure*}
    \centering
    \includegraphics[width=0.47\linewidth]{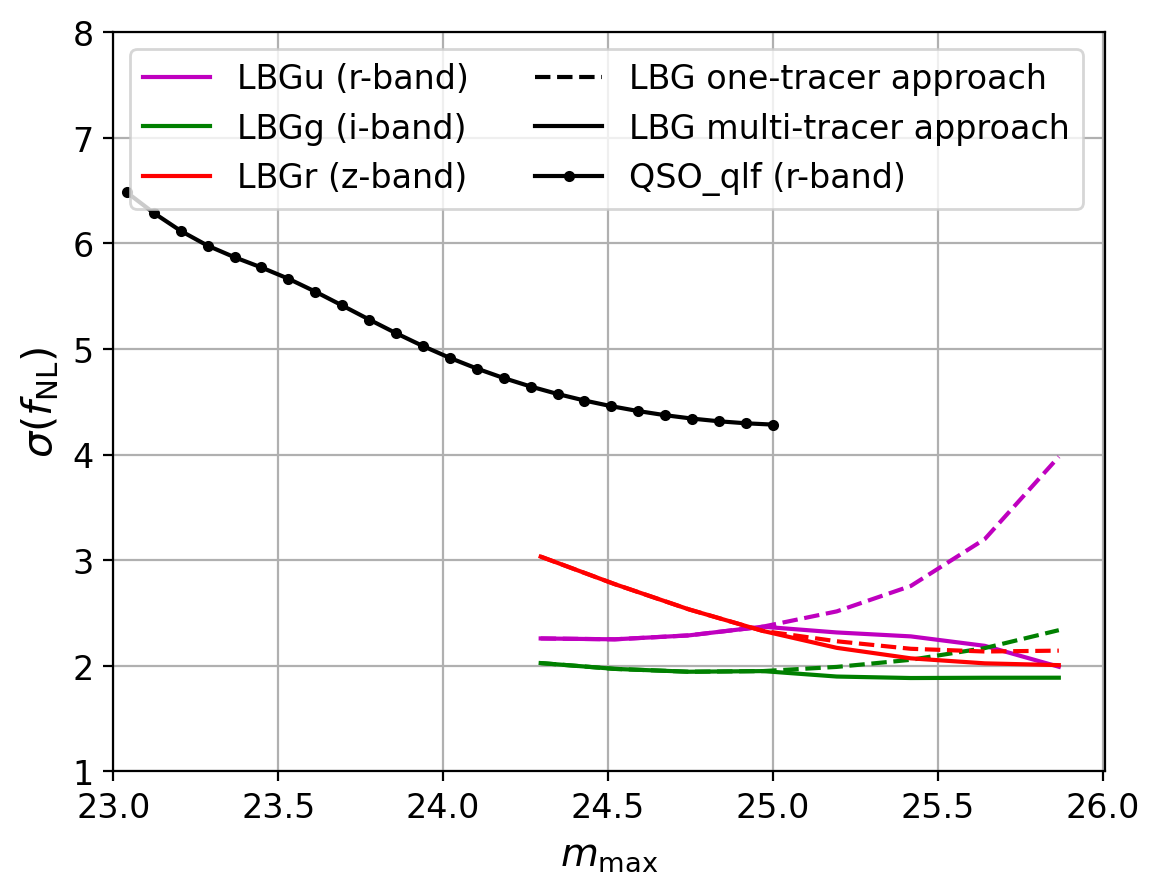}
    \includegraphics[width=0.47\linewidth]{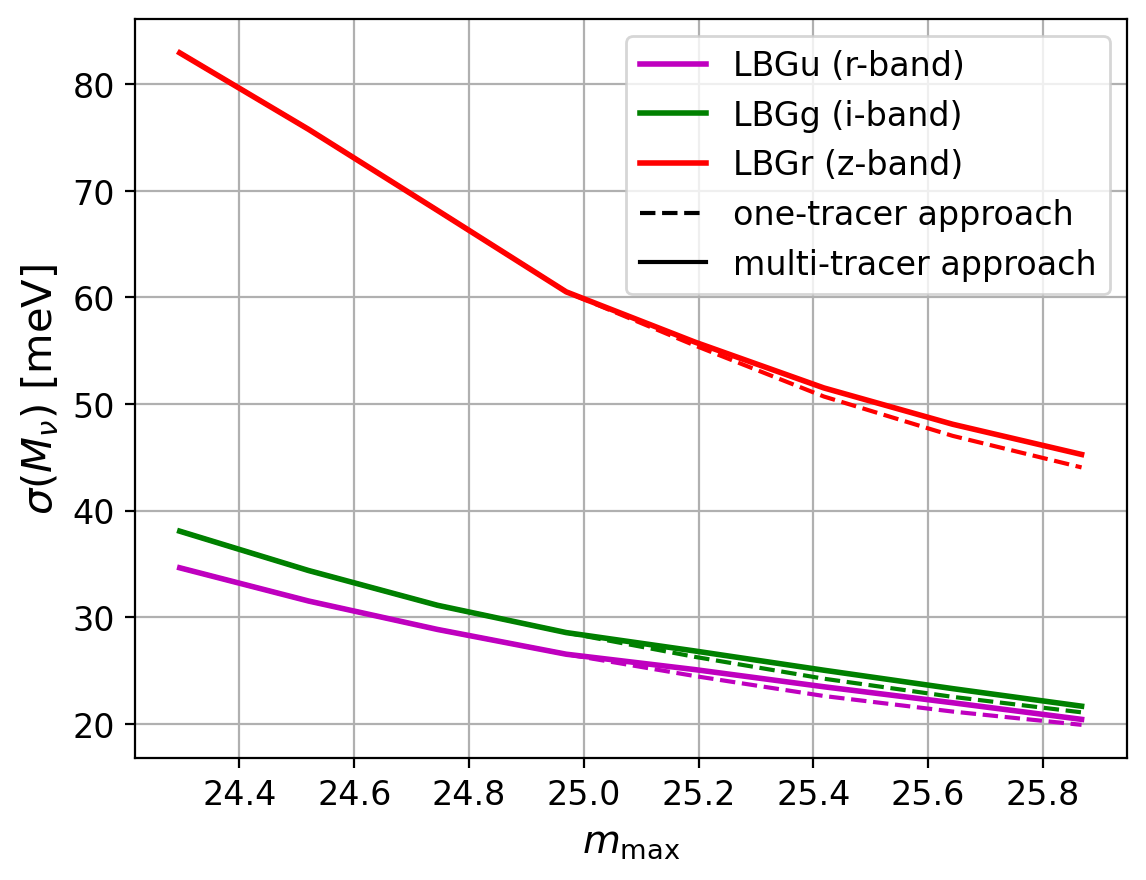}
    \caption{Left: local Primordial Non-Gaussianity constraints ($\fnl$) as a function of the LBG sample maximum magnitudes. The one-tracer approach and multi-tracer approach are displayed together. Right: Same for fot the total neutrino mass $M_\nu$.}
    \label{fig:fnl_mnu_mag_lbg}
\end{figure*}

\subsubsection*{Local primordial non-Gaussianities $\fnl$}
High-redshift tracers offer a particularly valuable probe of local primordial non-Gaussianity, which requires large cosmological volumes but not high number densities to achieve competitive precision. These conditions can naturally be met by LBG samples at $z>2$. The left panel of \figreff{fig:fnl_mnu_mag_lbg} shows the forecasted precision on $\fnl$ for the three LBG samples. We compare a single-tracer analysis with the multi-tracer approach introduced above. For the latter, we split each LBG sample into two magnitude-selected tracers at $m=25$, where $m$ denotes the $r$-, $i$-, or $z$-band magnitude for the respective dropout sample.

This split improves the constraints on $\fnl$ by exploiting the different galaxy biases of the two tracers and partially cancelling cosmic variance. In contrast, in the single-tracer case, increasing the number density eventually leads to cosmic-variance-limited measurements, while the effective bias can decrease as fainter galaxies are included. The resulting trade-off between number density and bias can therefore limit, or even degrade, the $\fnl$ constraints.
The three LBG samples exhibit different regimes. The $u$ and $g$-dropout samples reach the cosmic-variance limit relatively quickly, so increasing their depth provides little additional information. The $r$-dropout sample remains the only LBG sample that is significantly shot-noise limited, and therefore continues to benefit from increasing the depth of the selection. The three LBG samples contribute similarly to the constraints, with an uncertainty $\sigma(\fnl)\sim 2$. Combining the three, we get an effective constraint $\sigma(\fnl)\sim 1$.

\subsubsection*{Total neutrino mass $M_\nu$}

The precision on the total neutrino mass, $M_\nu$, obtained from the three LBG samples is shown in the right panel of \figreff{fig:fnl_mnu_mag_lbg}. Interestingly, none of the tracers reaches the cosmic-variance limit, indicating that the increased depth and number density enabled by WST can be fully exploited for this measurement. The $r$-dropout sample has the lowest number density and is therefore more strongly limited by shot noise, yielding constraints better than a factor of two weaker than those from the $u$- and $g$-dropout samples. The higher galaxy bias of the $g$-dropout sample compared to that of the $ u$-sample increases the clustering amplitude and partially compensates for its lower effective number density, resulting in comparable constraints between the $u-$ and $g-$ samples.

\subsubsection{Optimising the WST high-redshift LBG survey}
\begin{figure*}
    \centering
    \includegraphics[width=0.97\linewidth]{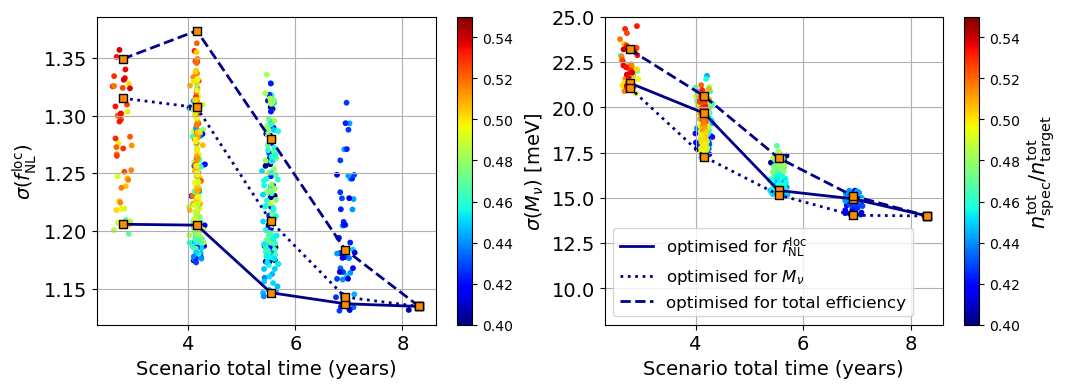}
    \caption{Left: Forecasts on $\fnl$ and $M_\nu$ for the different possible LBG survey designs, corresponding to different total survey times. Colour indicates the total spectroscopic efficiency of the tracer combination. We add a small scatter on the $x$-axis for readability.}
\label{fig:LBG_ugr_fnl_time}
\end{figure*}
For the high-redshift WST survey, LBG samples can be optimised for science returns. They are not fully studied for now; they are still in progress as promising tracers for future large-scale surveys. For the LBG sample, we address the interplay between maximum magnitudes (defining the LBG samples), the total survey time required to get all possible spectroscopic redshifts, and the return in terms of precision on cosmological parameters. For each set of $\vec{m}_{\rm max}^{\rm LBG}=\{m_{\rm max,i}\}_{\rm i \in [LBGu,\ LBGg,\ LBGr]}$, we compute the corresponding total errors on $p=\{\fnl, M_\nu\}$, denoted as $\sigma_{\rm tot}(p|\vec{m}_{\rm max}^{\rm LBG})$, as well as the total survey time and the total spectroscopic efficiency (see \secreff{fig:paper_logic}) defined as
\begin{equation}
    \mathrm{Eff}(\vec{m}_{\rm max}^{\rm LBG})=\frac{\sum\limits_{\rm i \in [LBGu,\ LBGg,\ LBGr]} n_{\rm spec}(m< m_{\rm max,i})}{\sum\limits_{\rm i \in [LBGu,\ LBGg,\ LBGr]} n_{\rm target}(m< m_{\rm max,i})}.
\end{equation}
From these three-dimensional matrices, we evaluate the pipeline shown in \secreff{fig:paper_logic} to obtain the total forecasted error on $\fnl$ and $M_\nu$, for every $\vec{m}_{\rm max}^{\rm LBG}$, and report result in \figreff{fig:LBG_ugr_fnl_time} as a function of the corresponding total survey time (in years). The total survey time takes discrete rather than continuous values, since it is computed in units of instrument passes $T_{\rm survey} = n_{\rm pass/FoV}\times \frac{\Omega}{\mathrm{FoV}}\times t_{\rm pass}$, see \secreff{sec:survey_time_completeness}. Each matrix element (i.e. choice of limiting magnitudes) is colour-coded according to the total spectroscopic efficiency. The resulting forecasts are shown in \figreff{fig:LBG_ugr_fnl_time}, displaying the different choices, the selected total forecasted error on $M_\nu$ and $\fnl$.  We find that, $\sigma(\fnl)$ cannot be significantly improved beyond the range $1.3$–$1.1$, whereas $\sigma(M_\nu)$ improves from $25$ to $15$~meV. This indicates that local PNG constraints already saturate their achievable precision at comparatively short survey times (i.e., shallow limiting magnitudes), whereas constraints on $M_\nu$ continue to improve significantly between 2- and 7-year survey durations. 

\begin{figure*}
    \centering
    \includegraphics[width=0.95\linewidth]{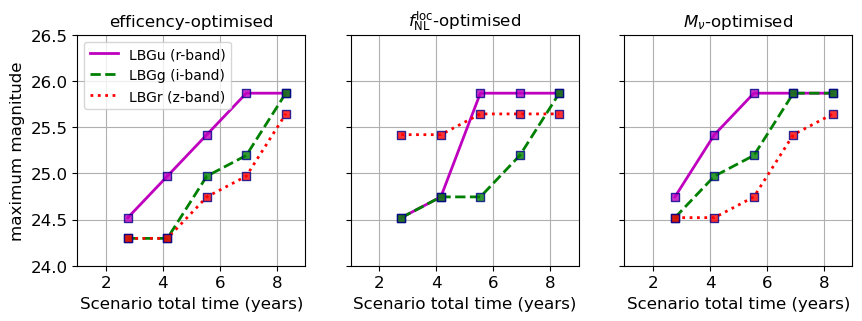}
    \includegraphics[width=0.95\linewidth]{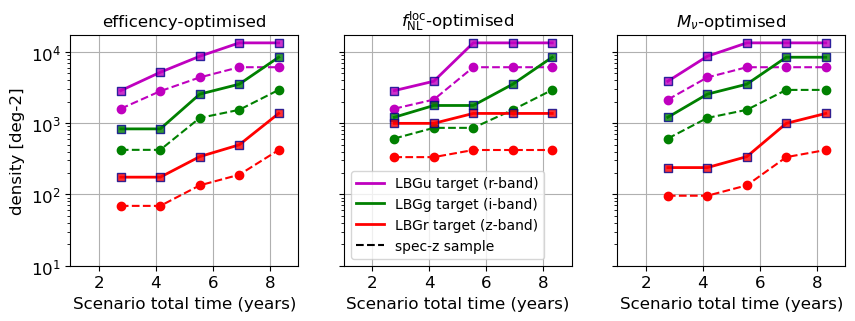}
    \caption{Top: Maximum magnitudes required in the different LBG detection bands for the three different scenarios explored in this section. Bottom: Same as top panel, but for target and spectroscopic densities.}
    \label{fig:LBG_maxmag_band}
\end{figure*}

Among all possible choices $\vec{m}_{\rm max}^{\rm LBG}$, we select, among groups of survey designed sorted by total survey time, $\vec{m}_{\rm max}^{\rm LBG*}$ corresponding to (i) $\vec{m}_{\rm max}^{\rm LBG*}\equiv \max\, \mathrm{Eff}(\vec{m}_{\rm max}^{\rm LBG})$, i.e. maximizing the spectroscopic efficiency (ii) $\vec{m}_{\rm max}^{\rm LBG**}\equiv \min\, \sigma_{\rm tot}(p|\vec{m}_{\rm max}^{\rm LBG})/\mathrm{Eff}(\vec{m}_{\rm max}^{\rm LBG})$, i.e. minimizing the total error on parameters, which are $\fnl$ and $M_\nu$, but also accounting for total efficiency constraints. \figreff{fig:LBG_maxmag_band} displays the maximal magnitudes of LBG samples for the three optimisation schemes.  We also display the resulting density of spectroscopic objects to give a better sense of the properties of the observational samples. We see that the $\fnl$ optimal strategy relies on observing the three tracers, including a large fraction of $r-$ dropout. The cosmic variance regime is reached quickly, which explains why there is not much improvement after several years. On the other hand, to get good constraints on $M_\nu$ in the first years, we rely solely on $u-$ and $g-$ dropout. $r-$ only starts to matter after 4 years. 

\section{Discussions on limits of our study and upcoming work}
\label{sec:discussion_limitation}
The framework developed in this work connects the design of the target selection (\secreff{sec:target_selection}), the efficiency of having measurements suitable for cosmological analysis (\secreff{sec:wst_purity_specz_efficiency}), and observing time to the cosmological performance of a WST-like survey (\secreff{sec:modelling_wst_cosmo_survey}). Our results identify MagMax as a promising alternative to DESI-like selections at low redshift, and illustrate how the preferred high-redshift observing strategy impacts the cosmological results. These conclusions nevertheless rely on specific assumptions of the tracer populations and survey performance. The forecasts should therefore be interpreted as a basis for comparing and refining survey configurations, with their absolute precision and preferred magnitude limits subject to further observational validation.\\

\noindent\textit{Photometric datasets:}
A first limitation concerns the photometric datasets used to characterise the samples. Our estimates of target densities and redshift distributions rely on deep imaging over the small COSMOS field, which cannot fully capture the sample variance and spatial variations in observing conditions expected across a wide-area survey. Although these data provide a useful approximation to the photometry available at the time of WST, they do not capture the statistical and observational complexities of a wide-area survey that can modify both the density and the purity of the selected samples, particularly near the magnitude and colour-cut boundaries. The star--galaxy separation considered for the $H$ band MagMax sample requires further validation over a wider range of imaging conditions and stellar populations. The \textit{Euclid} $H$ band extends to longer wavelengths than the ground-based UltraVISTA $H$ band used here, whose coverage is restricted by atmospheric absorption: approximately $1.5$--$2.0,\mu\mathrm{m}$ compared with $1.5$--$1.8,\mu\mathrm{m}$. This difference may improve the separation between stars and galaxies when infrared colours are combined with $(u-g)$, as suggested by the use of the $K_{\mathrm{s}}$ band in \citet[][]{Verdier_stargal} (in prep.). Therefore, the selection boundaries and resulting purity will need to be reassessed using the \textit{Euclid} photometry.\\

\noindent\textit{Spectroscopic efficiency model:}
A second limitation of this work is the assumption of the spectroscopic efficiency, which can affect the forecasted performance. For DESI-like tracers, this efficiency is based on current DESI results at the DESI limiting magnitude and we consider that this efficiency will not be affected when going to fainter magnitudes. Although this provides an observationally motivated reference model, its validity at fainter magnitudes remains to be established. For the MagMax sample, since \textit{Euclid} NIR data are not yet publicly available over a sufficiently large area, we cannot currently derive any quantity from observations, and we assume a constant efficiency within the adopted redshift ranges similar to that of the BGS-Grey. However, it is not guaranteed that the redshift efficiency will be constant for the full magnitude range at fixed exposure times and needs to be validated. The 4MOST-WAVES survey \citep{waves_2019}, which is designed to target an $r$-band magnitude-limited sample, will provide an initial opportunity to assess redshift recovery for this type of selection. 
At high redshift, our LBG spectroscopic model extrapolates DESI measurements for $u$-dropouts to WST and extends them to $g$- and $r$-dropouts through SNR scaling. This assumes that the relevant intrinsic spectral properties remain sufficiently similar across the redshift range considered. Changes in spectral features, instrumental throughput, and sky background may affect redshift recovery. Dedicated spectral simulations using the WST ETC, together with spectroscopic measurements, will be needed to validate these efficiencies. Because efficiency determines both the tracer density and the observing time, such improvements may change the results of \secreff{sec:forecast_Lbg_mag}. In addition, the redshift distributions considered in this work are based on photometric-redshift estimates and can also be improved by using other calibration techniques like clustering-redshift where overlapping spectroscopic data are available~\citep{WZ_euclid}.\\

\noindent\textit{Duration of WST programs (Bright, Grey and Dark Times):}
Our observing-time estimates likewise rely on an approximate treatment of survey completeness. The Poisson-based fibre-assignment model captures the reduced efficiency of successive passes as the remaining target population becomes sparse, but does not reproduce the full geometry of the positioners or the spatial clustering of targets. More realistic survey simulations, incorporating target priorities, field overlaps, and observing conditions, will be required to refine the relation between magnitude limits, completeness, and survey duration.\\

\noindent\textit{Galaxy bias model:}
The choice of the galaxy bias models also introduces a further uncertainty in the forecast performance. For the DESI-like tracers, we adopt relations calibrated using DESI samples and extrapolate them to fainter magnitude limits without explicitly modelling their luminosity dependence. Fainter selections may probe different galaxy populations and therefore need not follow the same bias evolution. For MagMax, we account for magnitude and redshift dependence using the \textit{Euclid} Flagship catalogue. This prediction remains dependent on the galaxy population model implemented in the simulation. Existing wide-area imaging can help constrain the bias of most DESI-like selections through angular clustering measurements and can be a future direction to improve our work. However, for ELG and MagMax selection, LSST-depth $g$-band and \textit{Euclid} NIR imaging are required to enable a corresponding validation of the bias model.\\

\noindent\textit{Cosmological forecasts:} Finally, the cosmological forecasts presented in this work are restricted to the linear regime. Consequently, the additional information provided when increasing the galaxy density at smaller scales, where non-linear clustering becomes important, is not captured here. Exploiting smaller scales will strengthen the scientific return of denser datasets, but it requires a more complex treatment of non-linear scales and their associated modelling uncertainties, which we leave to dedicated future studies. The primordial non-Gaussianity forecasts also depend on the assumed relation between linear galaxy bias and the response to primordial fluctuations, including the adopted value $p_{\Phi}=\{1, 1.6\}$ for high-redshift tracers. Calibrating this response and controlling large-scale observational systematics will be necessary to assess the achievable precision on $f_{\rm NL}^{\rm loc}$. We also mention that this work focused on the constraints that can be obtained from lower-order clustering, and ongoing WST follow-up papers will explore how higher-order clustering and other independent analyses will further strengthen the constraints.

\section{Conclusions}
\label{sec:conclusions}

Stage-V spectroscopic surveys will open a discovery space in cosmology, extending precision large-scale-structure measurements to redshifts and volumes far beyond the reach of Stage-IV facilities such as DESI. Realising this potential, however, requires target selection strategies that are explicitly designed around the capabilities of the next generation of instruments, rather than a straightforward extrapolation of existing Stage-III/IV recipes. In this paper, we have developed and characterised such strategies in the specific context of the Wide Field Spectroscopic Telescope (WST), combining instrumental design, photometric target selection, spectroscopic efficiency modelling, and cosmological forecasting into a single, self-consistent framework.

We began by detailing the WST facility itself: its 12-metre aperture, wide field of view, and highly multiplexed spectrograph, which together allow WST to reach magnitudes $\Delta m \sim 1.24$ deeper than DESI at fixed exposure time. Building on this instrumental gain, we revisited the target selection strategies of the classic low-redshift tracers used by SDSS, BOSS/eBOSS, DESI, and 4MOST — Bright Galaxies, Luminous Red Galaxies, and Emission Line Galaxies — rescaling their limiting magnitudes to exploit the full depth accessible to WST. Beyond these established tracers, we introduced and characterised new selection strategies more specifically tailored to the WST era. First, we proposed the MagMax sample, a purely magnitude-limited selection requiring no tracer-specific colour cuts, and showed that a three-dimensional colour-colour star–galaxy separation, exploiting deep optical and near-infrared photometry, can achieve stellar contamination levels comparable to classic colour-selected samples even at the faint magnitudes targeted by WST. Second, we extended target selection to $z>2$ through Lyman-Break Galaxy dropout samples, characterising their target density, purity, and spectroscopic efficiency using deep multi-band photometry over the COSMOS field, and building an explicit rescaling of DESI-measured LBG spectroscopic efficiencies to the depth and instrumental characteristics of WST.

A central result of this work is that the MagMax sample constitutes a genuine alternative to the standard suite of DESI-like colour-selected tracers at low redshift. When tuned to match the combined target density of the BG, LRG, and ELG samples, MagMax recovers comparable precision on the growth-rate parameter $f\sigma_8$ and on the BAO distance parameters $D_A$ and $H$, and consequently on the dark energy equation-of-state parameters $w_0$ and $w_a$ — all while requiring only a single tuning parameter (its limiting magnitude) rather than the several independently calibrated colour cuts needed for each classic tracer. Given the sensitivity of colour-selected samples to the depth and homogeneity of the underlying photometry, this simpler and more robust alternative is particularly attractive for a survey whose target selection must remain reliable across a range of photometric datasets and depths not yet fully characterised today.

At higher redshift, we showed that LBG samples provide access to the growth of structure and expansion history at $z>2$, complementing and extending the low-redshift BAO and RSD measurements. We further demonstrated that these same high-redshift, large-volume tracers are well suited to constraining local-type primordial non-Gaussianity and the total neutrino mass, two of the key open questions motivating the transition to Stage-V surveys. By explicitly linking the limiting magnitudes of the LBG $u$-, $g$-, and $r$-dropout samples to the total survey time, spectroscopic efficiency, and resulting cosmological precision, we built an optimisation framework capable of identifying survey configurations that balance scientific return against the practical cost of observing time. This framework revealed a qualitatively different behaviour between the two high-redshift science cases: constraints on $\fnl$ saturate rapidly with survey time, as the relevant large scales are already close to cosmic-variance limited, whereas constraints on $M_\nu$ continue to improve significantly over several years of additional observation, reflecting the persistent value of increased shot-noise suppression at small scales for the neutrino-mass signal.

Taken together, these results illustrate a central theme of this paper: that the transition to Stage-V spectroscopic surveys need not simply scale up existing target selection strategies, but can instead benefit from selection approaches designed jointly with the survey's scientific goals, its instrumental capabilities, and its finite observing time. The comparison between DESI-like and MagMax tracers, and the survey-time optimisation performed for the high-redshift dark-time tracers, both exemplify this science-driven approach to survey design, in which the maximum magnitude of each sample is not fixed a priori, but instead derived from the scientific return it delivers.

Several aspects of this work call for further development. The MagMax star–galaxy separation and spectroscopic efficiency model would benefit from validation against larger and independent photometric and spectroscopic datasets beyond the COSMOS field used here. The cosmological forecasts presented in this paper rely on a linear Fisher-matrix formalism and, for several tracers, on simplified constant spectroscopic and bias models; extending these forecasts with full non-linear modelling, more realistic fibre-assignment simulations, and dedicated pilot spectroscopic observations — as already planned for DESI-II and other Stage-V precursors — will be essential to refine the survey designs proposed here. Finally, the WST Exposure Time Calculator used throughout this work to rescale DESI-based efficiencies could itself be directly validated against dedicated LBG spectroscopic pilot surveys, analogous to those already conducted for DESI and MSE. We leave these developments to future work, while emphasising that the framework presented here already provides a concrete and adaptable basis for the science-driven design of the WST cosmology survey, and more broadly for the coming generation of Stage-V spectroscopic experiments.\\

\noindent\textit{Primary author contribution statement:} CP contributed to the software, including its overall structure, LBG and QSO selection, WST survey design, and forecasts, as well as to the preparation of the paper. AR contributed to the paper's overall structure, target selection, WST survey normalisation and design, and paper preparation. AV contributed to the low-$z$ target selection (DESI-like and MagMax samples), star--galaxy separation for the MagMax sample, and WST survey design. WdA contributed to the MagMax selection and properties, the forecast methodology and results, and the preparation of the paper. BB contributed to quasar target selection, paper preparation, and proofreading. 
\section*{Acknowledgments}

This project has received funding from the European Union Horizon Europe Research and Innovation Action under grant agreement no. 101183153 -WST. Views and opinions expressed are however those of the author(s) only and do not necessarily reflect those of the European Union or the European Research Executive Agency (REA). Neither the European Union nor the REA can be held responsible for them. 
AR \& AV acknowledge support from the Swiss National Science Foundation (SNF) ‘Cosmology with 3D Maps of the Universe’ research grant $200020\_207379$. CP and CY acknowledge support from grant ANR-22-CE31-0009 for the HZ-3D-MAP project and from grant ANR-22-CE92-0037 for the DESI-Lya project. 
W. d’A. acknowledges support from the MICINN projects PID2019-111317GB-C32, PID2022-141079NB-C32 as well as the predoctoral program AGAUR-FI ajuts (2024 FI-1 00692) Joan Oro. Some parts of this work used the DiRAC Data Intensive service (CSD3) at the University of Cambridge, managed by the University of Cambridge University Information Services on behalf of the STFC DiRAC HPC Facility (\url{https://www.dirac.ac.uk}). The DiRAC component of CSD3 at Cambridge was funded by BEIS, UKRI and STFC capital funding and STFC operations grants. DiRAC is part of the UKRI Digital Research Infrastructure. This work has made use of CosmoHub, developed by PIC (maintained by IFAE and CIEMAT) in collaboration with ICE-CSIC. It received funding from the Spanish government (grant EQC2021-007479-P funded by MCIN/AEI/10.13039/501100011033), the EU NextGeneration/PRTR (PRTR-C17.I1), and the Generalitat de Catalunya.

The authors use ChatGPT Astra (OpenAI, \href{https://chatgpt.com}{https://chatgpt.com}) and Claude Opus (Anthropic, \href{https://claude.ai}{https://claude.ai}) for language editing, grammatical refinement, and assisting in debugging code syntax. AR thanks OpenAI for providing access to the ChatGPT for Academic Researchers program. Following the use of these tools, the authors reviewed, evaluated, and edited the material, and take full responsibility for the validity, integrity, and accuracy of both the manuscript text and the functional codebase. All scientific content, analysis, and conclusions are the sole responsibility of the authors. 

\section*{Data availability}

This work uses data obtained and processed as part of the CFHT Large Area U-band Deep Survey (CLAUDS), which is a collaboration between astronomers from Canada, France, and China. CLAUDS is based on observations obtained with MegaPrime/MegaCam,
a joint project of CFHT and CEA/Irfu, at the CFHT, which is operated by the National Research Council (NRC) of Canada, the Institut National des Sciences de l’Univers of the
Centre National de la Recherche Scientifique (CNRS) of France, and the University of Hawaii.

CLAUDS uses data obtained in part through the Telescope Access Program (TAP), which
has been funded by the National Astronomical Observatories, Chinese Academy of Sciences,
and the Special Fund for Astronomy from the Ministry of Finance of China. CLAUDS uses
data products from TERAPIX and the Canadian Astronomy Data Centre (CADC) and was carried out using resources from Compute Canada and Canadian Advanced Network For Astrophysical Research (CANFAR).

This paper is also based on data collected at the Subaru Telescope by the Hyper Suprime-Cam (HSC) collaboration and retrieved from the HSC data archive system, which
is operated by the Subaru Telescope and Astronomy Data Centre (ADC) at NAOJ. Their
data analysis was in part carried out with the cooperation of the Centre for Computational Astrophysics (CfCA), NAOJ. The HSC collaboration includes the astronomical communities of Japan and Taiwan, and Princeton University. The HSC instrumentation and software were developed by the National Astronomical Observatory of Japan (NAOJ), the Kavli Institute
for the Physics and Mathematics of the Universe (Kavli IPMU), the University of Tokyo, the High Energy Accelerator Research Organisation (KEK), the Academia Sinica Institute for Astronomy and Astrophysics in Taiwan (ASIAA), and Princeton University. Funding
was provided by the FIRST program from the Japanese Cabinet Office, the Ministry of
Education, Culture, Sports, Science and Technology (MEXT), the Japan Society for the Promotion of Science (JSPS), Japan Science and Technology Agency (JST), the Toray Science Foundation, NAOJ, Kavli IPMU, KEK, ASIAA, and Princeton University. The HSC collaboration members are honoured and grateful for the opportunity of observing the Universe from Maunakea, which has cultural, historical, and natural significance
in Hawaii.

\bibliographystyle{mnras}
\bibliography{reference} 
\section*{Affiliations}
\noindent
$^{1}$ Université Paris-Saclay, CEA, IRFU, 91191, Gif-sur-Yvette, France \\
$^{2}$ Laboratoire d’astrophysique, École Polytechnique Fédérale de Lausanne (EPFL) Observatoire, CH-1290 Versoix, Switzerland \\
$^{3}$ IFAE, The Barcelona Institute of Science and Technology, Campus UAB, 08193 Bellaterra Barcelona, Spain \\
$^{4}$ Institute for Astronomy, University of Edinburgh, Royal Observatory, Blackford Hill, Edinburgh, EH9 3HJ, U.K. \\
$^{5}$ Astronomy Centre, Department of Physics and Astronomy, University of Sussex, Falmer, Brighton BN1 9QH, UK \\
$^{6}$ Max Planck Institute for Extraterrestrial Physics, Giessenbachstrasse 1, 85748 Garching, Germany \\
$^{7}$ INAF – Osservatorio Astronomico di Brera, Via E. Bianchi 46, 23807 Merate (LC), Italy \\
$^{8}$ European Southern Observatory, Karl-Schwarzschild-Strasse 2, 85748 Garching bei München, Germany \\
$^{9}$ Dipartimento di Fisica e Astronomia, Universit\`a di Bologna, Via Gobetti 93/2, 40129 Bologna, Italy \\
$^{10}$ INAF – Osservatorio di Astrofisica e Scienza dello Spazio di Bologna, Via Piero Gobetti 93/3, I-40129 Bologna, Italy \\
$^{11}$ INAF – Osservatorio Astrofisico di Arcetri, Largo E. Fermi 5, 50125 Firenze, Italy \\
$^{12}$ Institute of Graduate Studies in Science, Istanbul University, Istanbul, Turkey \\
$^{13}$ Université Claude Bernard Lyon 1, IUF, IP2I Lyon, 4 rue Enrico Fermi, 69622 Villeurbanne, France \\
$^{14}$ National Center for Nuclear Research, Pasteura 7, 02-093 Warsaw, Poland \\
$^{15}$ INAF - Osservatorio Astronomico di Trieste, via Tiepolo 11, I-34131, Trieste, Italy \\
$^{16}$ Leibniz-Institut für Astrophysik Potsdam (AIP), An der Sternwarte 16, D-14482 Potsdam, Germany \\
$^{17}$ Leiden Observatory, Leiden University, P.O. Box 9513, 2300 RA Leiden, The Netherlands \\
$^{18}$ Center for Astrophysics and Space Science, NYUAD Research Institute, New York University Abu Dhabi, PO Box 129188, Abu Dhabi, UAE \\
$^{19}$ Faculty of Science, Department of Astronomy and Space Sciences, Istanbul University, Istanbul, Turkey \\
$^{20}$ Department of Space Engineering, Delft University of Technology, 2629 HS Delft, The Netherlands \\
$^{21}$ ILEWG LUNEX-EuroSpaceHub EuroMoonMars Earth-Space Innovation, Wassenaar, Leiden \& Noordwijk, The Netherlands \\
$^{22}$ Laboratoire d’Astrophysique de Marseille (UMR 6110), CNRS-Université de Provence, BP8, 13376 Marseille Cedex 12, France
\appendix

\section{Time-magnitude relation}
\label{app:exposuretime}

For a given target SNR in a given photometric band, we can determine the exposure time -- expressed as several nominal WST exposures -- required to reach that SNR for a source of a given magnitude. In the faint-object limit, the observation is sky-noise dominated, and the signal-to-noise ratio follows the standard CCD equation
\begin{equation}
   {\rm SNR}=\alpha_{\rm tel}\, I(m)\sqrt{t_{\rm exp}\,S_{\rm tel}},
   \label{eq:SNR}
\end{equation}
where $I(m) = I_0\times 10^{-0.4m}$ is the source flux ($I_0$ does not need to be specified explicitly here, since only flux ratios enter this work), and $\alpha_{\rm tel}$ is a coefficient encapsulating the relevant observing conditions. In practice, observations proceed in discrete units of a unitary exposure time $t_{1\,\rm exp}$. Inverting \eqreff{eq:SNR}, the exposure time required to reach a target SNR for a source of magnitude $m$ is therefore given by
\begin{equation}
    t_{\rm WST}(m\vert {\rm SNR})=\left\lceil \frac{1}{\,t_{1\,\rm exp}}\,\left(\frac{{\rm SNR}}{\sqrt{S_{\rm WST}}\,\alpha_{\rm WST}\,I(m)}\right)^2\right\rceil\,t_{1\,\rm exp},
\end{equation}
where $\lceil \cdot \rceil$ denotes the ceiling function, giving the integer number of exposures (passes) required to reach the desired SNR. This relation provides the exposure time required for a fixed target SNR; we now connect it to previous surveys, and specifically to DESI.

We write
\begin{equation}
    \alpha_{\rm WST} = \eta_{\rm WST\rightarrow DESI}\times\alpha_{\rm DESI},
\end{equation}
where $\eta_{\rm WST\rightarrow DESI}$ accounts for any difference in observing conditions between the two facilities. We expect $\eta_{\rm WST\rightarrow DESI} > 1$, owing to the generally better observing conditions anticipated for WST relative to DESI; for simplicity, however, we adopt $\eta_{\rm WST\rightarrow DESI} = 1$ throughout this work. In the WST exposure-time equation above, $\alpha_{\rm WST}$ is therefore the only unknown quantity, which we determine by rescaling the corresponding DESI coefficient $\alpha_{\rm DESI}$.

For a DESI reference tracer of magnitude $m_{\rm ref}$, the DESI pipeline applies an exposure time $t_{\rm ref\, DESI}$ to reach a target signal-to-noise ratio ${\rm SNR}_{\rm ref\, DESI}$, which fixes the value of $\alpha_{\rm DESI}({\rm SNR}_{\rm ref\, DESI}, t_{\rm ref\, DESI})$ by inverting the DESI exposure-time equation\footnote{Inverting the DESI exposure-time equation gives
\begin{equation}
    \alpha_{\rm DESI}({\rm SNR}_{\rm ref\, DESI}, t_{\rm ref\, DESI})=\frac{{\rm SNR}_{\rm ref\, DESI}}{I(m_{\rm ref})\sqrt{t_{\rm ref\, DESI}\,S_{\rm DESI}}}.
\end{equation}}.

\section{Linear galaxy bias of the MagMax sample with the \textit{Euclid} Flagship simulations}\label{app:magbias}

\begin{figure}
\centering
\includegraphics[width=1\linewidth]{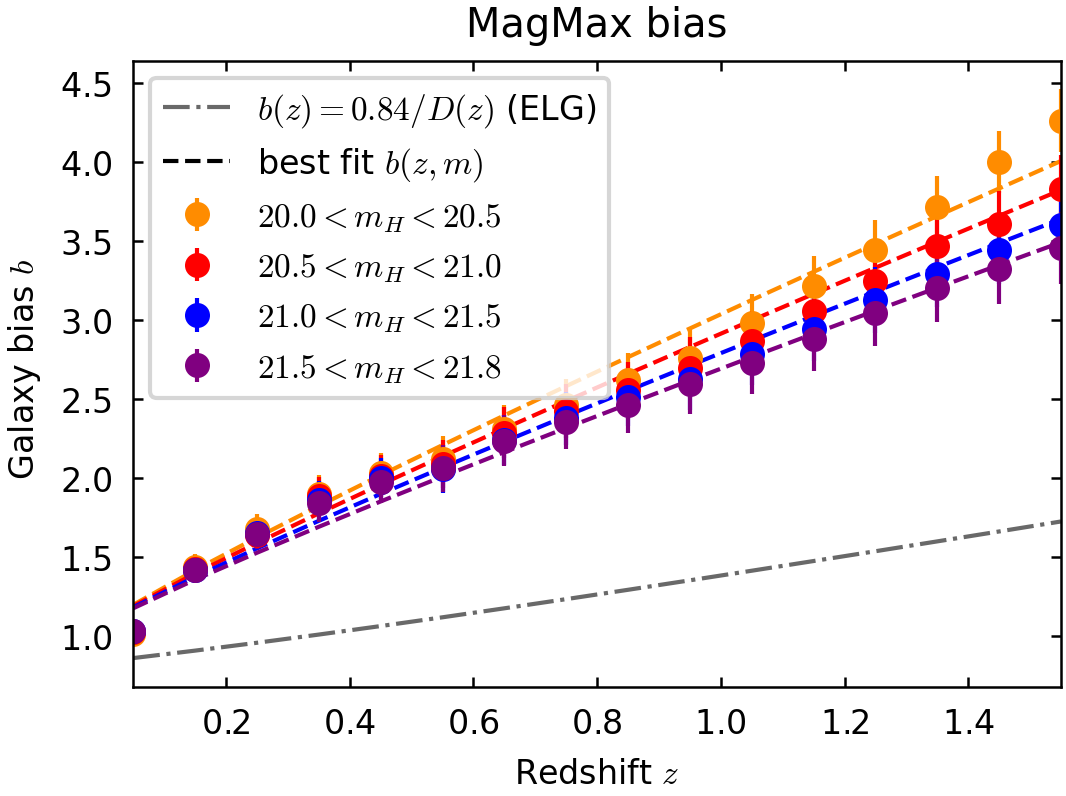}
\caption{Galaxy bias of the Magmax sample as a function of redshift for different Euclid-$H$ magnitude bins, together with the corresponding best-fitting model given by \eqreff{eq}. The DESI-ELG bias evolution is also shown for comparison.}
\label{fig:magmax_bias_flagship}
\end{figure}

To evaluate the forecasts, we need to account for both the redshift evolution of the galaxy bias and its dependence on tracer magnitude. We expect fainter galaxy samples to be less strongly biased towards overdense regions, and therefore anticipate a decrease in galaxy bias with increasing magnitude. For the DESI-like sample, we adopt an empirical redshift-evolution law calibrated at the maximum magnitude of the DESI selection. We extrapolate this relation to fainter magnitudes. While this provides a simple prescription for the forecasts, such an extrapolation is not fully justified and should be improved in future work.
For the Magmax sample, we cannot use the existing empirical bias relations as an anchor. Since the Euclid NIR data are not yet publicly available over a sufficiently large area, we instead estimate the galaxy bias using the realistic Flagship catalogue (version 2.1.10), described in detail in \cite{FS_2024}. The Euclid Flagship $N$-body dark-matter simulation consists of a $3600,h$Mpc-on-a-side simulation box containing approximately 4 trillion particles. It is one of the largest $N$-body simulations performed to date and covers a cosmological volume comparable to that expected to be surveyed by Euclid. The simulation resolves approximately 16 billion dark-matter halos, including those hosting the faintest galaxies targeted by Euclid. The halos are identified using the \textsc{Rockstar} halo finder \citep{Rockstar_2013}. A synthetic galaxy catalogue is subsequently generated using a combination of halo occupation distribution (HOD) and abundance matching (AM) techniques, together with empirical relations between galaxy properties \citep{FS_2024}.
We reproduce the Magmax sample selection using the \textsc{CosmoHub} platform \citep{Carretero_cosmo_hub}. The number of targets we get is in agreement with the data expectation of \figreff{fig:recap_density_all}. In principle, a similar procedure could be applied to the DESI-like tracers. However, the Flagship catalogue does not provide the DESI-like fibre magnitudes, which makes reproducing the corresponding selection considerably more challenging. In particular, the relation between total and fibre magnitudes is non-linear, as discussed in \cite{Verdier_2025}.
Even for the Magmax sample, we do not expect the Flagship catalogue to provide a sufficiently accurate description of galaxy bias to predict the bias of future selections with high fidelity. A data-driven measurement using Euclid observations will therefore be valuable in future work. Nevertheless, we consider the Flagship-based estimate sufficiently accurate for this first forecast study.
Once the catalogue has been selected, we divide the Magmax sample into redshift and Euclid-$H$-magnitude bins. For each bin, we measure the angular galaxy auto-correlation function, $w(\theta)$, over angular scales of $5$--$20$ arcmin. We then compare it with the corresponding non-linear dark-matter prediction, $w_{\rm m}(\theta)$, computed using \textsc{Halofit} \citep{halofit}. A similar procedure was used in \cite{desy6WZ} for DESI tracers and provides bias values consistent with the expected trends.
From the resulting measurements, ${b(z_{\rm eff},m_{\rm eff})}$, we determine the best-fitting coefficients of the phenomenological model
\begin{equation}
b(z,m) = a + b(c+m)z^d.
\end{equation}
This functional form was chosen following visual inspection of the measured bias evolution and provides a good description of the simulation results. We obtain
\begin{equation}
b_{\rm Mag}(z,,m_H)
= 1.1 + 0.25\,(27.8-m_H)\,z^{0.9}.
\label{eq}
\end{equation}
The resulting bias evolution is shown in \figreff{fig:magmax_bias_flagship}. The inferred biases are systematically higher than those of the ELG-like samples. This is expected given that the MagMax selection corresponds to a substantially brighter galaxy population than the ELG-like selection.

\section{Some forecasting details}
\label{app:forecast_details}

\subsection{Growth of structures}
\label{app:forecast_details_growth}

\begin{figure}
    \centering
    \includegraphics[width=\linewidth]{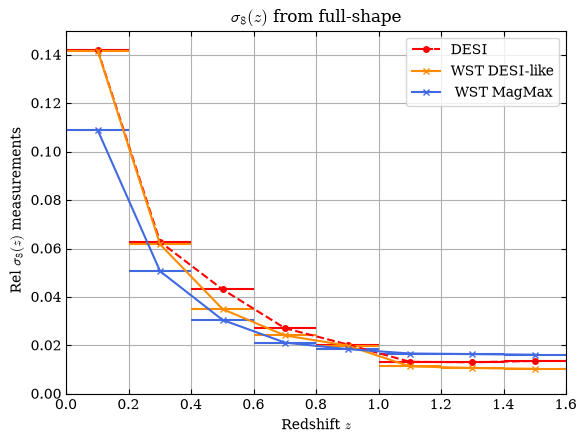}
    \caption{Errors on $\sigma_8(z)$ from full shape data without any external priors, for DESI, WST with DESI-like tracers, and WST MagMax sample. }
    \label{fig:rsd_fullshape}
\end{figure}

We follow the straightforward forecasting strategy of \cite{eboss_forecast} for the redshift-space distortion forecast. We rewrite the tracer power spectrum as
\begin{equation}
P_{\rm gg}(k,z)=\left(b_{\rm g}(z)\sigma_8(z)+f(z)\sigma_8(z)\mu^2\right)^2\frac{P_\text{m}(k,z=0)}{\sigma_8(z=0)^2}.
\end{equation}
We use as fit parameters $\ln[b_{\rm g}(z_{\rm eff})\sigma_8(z_{\rm eff})]$ and $\ln[f(z_{\rm eff})\sigma_8(z_{\rm eff})]$, from which we derive the associated relative errors on $b_{\rm g}(z_{\rm eff})\sigma_8(z_{\rm eff})$ and $f(z_{\rm eff})\sigma_8(z_{\rm eff})$. Besides the aforementioned parameters, all the other parameters are fixed, considered as fixing the 'template' power spectrum. 

We also use a more sophisticated full shape approach to get constraints on $\sigma_8$; see discussion in \secreff{sec:forecast_methodology}.

\begin{figure}
    \centering
    \includegraphics[width=1\linewidth]{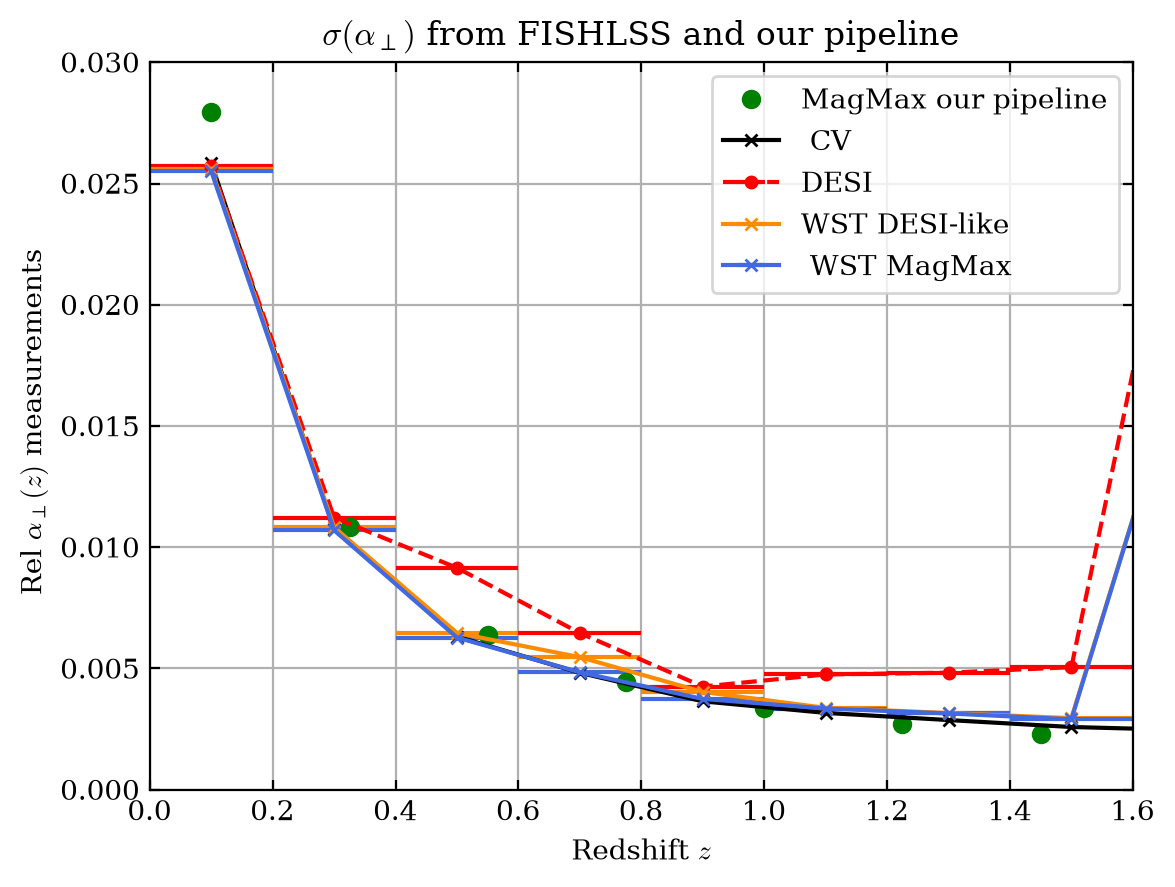}
    \caption{Uncertainties on $\alpha_\perp$ obtained with \texttt{FishLSS} for WST MagMax (blue) and DESI-like tracers (orange), DESI tracers (red), and the cosmic-variance limit (black), assuming a survey area of $18,000,\mathrm{deg}^2$. Uncertainties from the simplified prescription for MagMax are also reported in green. }
    \label{fig:comparison_codes}
\end{figure}

\subsection{Baryon Acoustic Oscillations (BAO) and dynamical dark energy (DDE)}

\paragraph*{AP forecasts and code comparison:} As detailed in \secreff{sec:forecast_methodology}, we use two forecasting approaches to evaluate the Fisher matrices for the AP parameters. The first follows the prescription of \cite{seo_improved_2007}, while the second uses the \texttt{FishLSS} code \cite{FISHLSS}. The latter accounts for non-linear effects and marginalises over the polynomial coefficients associated with the reconstruction procedure. \cite{FISHLSS} finds that the two approaches yield AP parameter uncertainties in agreement at the 10--15$\%$ level.
\figreff{fig:comparison_codes} shows the forecasted uncertainty on $\alpha_\perp$ obtained with \texttt{FishLSS} for WST Magmax and DESI-like tracers, DESI tracers, and the cosmic-variance limit, assuming a survey area of $18,000\,\mathrm{deg}^2$. We also show the WST Magmax forecast obtained using the prescription of \cite{seo_improved_2007}. The two approaches agree at the ($\sim10\%$) level for $\alpha_\perp$, consistent with the comparison presented in \cite{FISHLSS}. We find a similar level of agreement for $\alpha_\parallel$. Forecasts with \texttt{FishLSS} confirm our finding that in both low-$z$ tracer strategies -- WST-MagMax and WST-DESI-like-- our measurements are cosmic variance limited over the entire low-redshift range.

\paragraph*{Dynamical Dark Energy: } BAOs allow the relative measurement of the Hubble parameter and angular diameter distance in the combinations $H(z_{\rm eff})\,r_{\rm d}$ and $D_A(z_{\rm eff})/r_{\rm d}$, respectively rescaled by the sound horizon $r_{\rm d}$, through the AP parameters; see \eqreff{eq:AP}. Assuming that $r_{\rm d}$ is very precisely determined from CMB experiments, the relative errors on $D_A$ and $D_H$ are directly obtained from the forecasted relative errors on the AP parameters, computed using the Fisher matrix approximation from~\citep{seo_improved_2007}. This has been applied, for example, in \cite{eboss_forecast, FISHLSS, Doumerg2023forecast}. The Fisher matrix is then a $2\times 2$ matrix. Then we can study the expansion history of the Universe by constraining the DE equation of state
\begin{equation}
  w(z)=w_0+w_a\frac{z}{1+z}.
\end{equation}
For a flat $w_0--w_a\mathrm{CDM}$ cosmology, the angular diameter distance is given by
\begin{equation}
D_A(z)=\frac{c}{(1+z)\,H_0}
\int_0^z\frac{{\rm d}z'}{E(z';\,w_0,\,w_a)}
\end{equation}
while the Hubble parameter is
\begin{equation}
H(z)=H_0\,E(z;w_0,w_a)
\end{equation}
where
\begin{equation}
E^2(z)=\Omega_{\rm m}(1+z)^3+
(1-\Omega_{\rm m})(1+z)^{3(1+w_0+w_a)}
\exp\left[-3w_a\frac{z}{1+z}\right].
\end{equation}
We propagate our forecasts on $\alpha$s to forecasts on $w_0-w_a$ by  transforming our redshift-binned Fisher matrices $F_{\alpha_\parallel,\alpha_\perp}(z_i)$ into $F_{w_0,\,w_a,\,\Omega_{\rm m}} (z_i)$ introducing the Jacobians $J(z_l)=\left( \partial \theta_i/\partial \tilde{\theta}_j\right)_{z_l}$ of change of coordinates.
Then the total Fisher matrix is 
\begin{equation}
    {F_{w_0,\,w_a,\,\Omega_{\rm m}}}=\sum_{z_l} J_l^\top\,\cdot\,F_{\alpha_\parallel,\alpha_\perp}(z_l)\,\cdot\,J_l.
\end{equation}

\begin{table}
\caption{Fiducial values of cosmological parameters and their Gaussian prior from \citet{Planck2020cosmofull}.}
\centering
\begin{tabular}{lcc}
\hline
Parameter & Fiducial values & Prior standard deviation \\
\hline
$h$ & 0.677 & 0.0054 \\
$\omega_b$ & 0.02247 & $0.00015$ \\
$\omega_c$ & 0.1192 & $0.0012$ \\
$n_\text{s}$ & 0.9675 & $0.0042$ \\
$\ln(A_\text{s})$ & -13.073 & $0.015$ \\
$M_\nu$ (eV) & 0.06 & 0.5 \\
\hline
\end{tabular}

\label{tab:prior_fiducial}
\end{table}

\bsp	
\label{lastpage}
\end{document}